\documentclass[useAMS,usenatbib]{mn2e}

\usepackage{epsfig,graphicx,latexsym,amsmath,amssymb}
\usepackage{natbib}
\usepackage{hyperref}
\usepackage{breakurl}
\usepackage{mathrsfs}
\usepackage{psfrag}                              
\usepackage{url}

\usepackage{color}

\bibpunct[,]{(}{)}{;}{a}{}{,}

\title[Higher order moment models of dense stellar systems]
{Higher order moment models of dense stellar systems: \\
Applications to the modeling of the stellar velocity distribution function}

\author[Schneider, Amaro-Seoane \& Spurzem]
{Justus Schneider$^{1}$\thanks{e-mail:Justus@ari.uni-heidelberg.de},
Pau Amaro-Seoane$^{2,3}$ \& Rainer Spurzem$^{4,5,1}$\\
$^{1}$ Astronomisches Rechen-Institut, M{\"o}nchhofstra{\ss}e 12-14, 69120,
Zentrum f\"ur Astronomie, Universit\"at Heidelberg, Germany\\
$^{2}$ Max-Planck Institut f\"ur
Gravitationsphysik (Albert-Einstein-Institut), Am M{\"u}hlenberg 1,
D-14476 Potsdam, Germany\\
$^{3}$ Institut de Ci{\`e}ncies de l'Espai (CSIC-IEEC), Campus UAB, Torre C-5, parells, $2^{na}$ planta, ES-08193
Bellaterra, Barcelona, Spain \\
$^{4}$ National Astronomical Observatories of China, Chinese Academy of
Sciences, 20A Datun Lu, Chaoyang District, 100012, Beijing, China\\
$^{5}$ Kavli Institute for Astronomy and Astrophysics, Peking
University, China\\
}

\date{\today}

\pagerange{\pageref{firstpage}--\pageref{lastpage}} \pubyear{\today, submitted}

\begin{document}

\label{firstpage}

\maketitle

\begin{abstract}
Dense stellar systems such as globular clusters, galactic nuclei and nuclear
star clusters are ideal loci to study stellar dynamics due to the very high
densities reached, usually a million times higher than in the solar
neighborhood; they are unique laboratories to study processes related to
relaxation. 
There are a number of different techniques to model the global evolution of
such a system. We can roughly separate these approaches into two major groups; the
particle-based models, such as direct $N-$body and Monte Carlo models, and the
statistical models, in which we describe a system of a very large number of stars
through a one-particle phase space distribution function. In this approach we
assume that relaxation is the result of a large number of two-body gravitational
encounters with a net local effect. 
{We present two moment models that are based on the collisional 
Boltzmann equation. By taking moments of the Boltzmann equation one obtains 
an infinite set of differential moment equations where the equation for the moment of
order $n$ contains moments of order $n+1$.}
In our models we assume spherical symmetry but we do not require dynamical
equilibrium. 
{We truncate the infinite set of moment equations 
at order $n=4$ for the first model and at order $n=5$ for the second model. The
collisional terms on the right-hand side of the moment equations account for 
two-body relaxation and are computed by means of the Rosenbluth potentials. We 
complete the set of moment equations with closure relations which constrain the 
degree of anisotropy of our model by expressing moments of order $n+1$ by moments 
of order $n$.}
The accuracy of this approach relies on the number of moments included 
from the infinite series.
{Since both models include fourth order moments we can 
study mechanisms in more detail that increase or decrease the number 
of high velocity stars.}
{The resulting model allows us to derive a velocity distribution 
function, with unprecedented accuracy, compared to previous moment models.}

\end{abstract}


\section{Motivation}

Statistical continuum models such as Fokker-Planck (FP) 
{and 
moment} models separate the
treatment of the different astrophysical processes that control the evolution
of the system.  This allows us to isolate the effects of the distinct dynamical
mechanisms. 
{In particular statistical moment} models have provided us with
important contributions to the understanding of phenomena 
such as core collapse and gravothermal oscillations \citep{bettwieser1984}. 
These models decompose the local velocity distribution function into the 
different contributions of the moments, allowing us to study the influence of
the different moments on the evolution of star clusters and the impact of 
different dynamical mechanisms on the moments of the distribution function.  
This has a bearing in a number of crucial problems such as the contribution 
of high velocity stars to the evolution of star clusters, which we only can address 
by including fourth order moments. 

We present in this paper 
{two} {statistical moment models} 
for
dense, non-rotational and spherically symmetric stellar systems, such as
globular clusters (GCs) or nuclear star clusters (NCs). 
{The models 
include fourth order moments and thus allow us to study 
astrophysical scenarios that 
affect the number of high velocity stars.} The models
describe the evolution of a stellar system that slowly evolves due to the
effects of two-body relaxation.  Moment models have the advantage over 
particle-based techniques in that they are computationally much cheaper,
being based on the numerical integration of a relatively small set of
partial differential equations with just one variable, the radius $r$. The
numerical solution of the model equations is usually very fast as they are
equivalent to one-dimensional hydrodynamical equations. Since the system is
treated as a continuum, all macroscopic quantities (such as density, pressure
and energy flux) are smooth functions of radius $r$ and time $t$ and do not
suffer from the characteristic noise of particle-based approaches.

{Moment models began with  simple collisionless models 
and progressed to the anisotropic gaseous model
\citep{bettwieser1986,louis1991,Spurzem92,giersz1994,spurzem1995}. They have significantly 
contributed to the understanding of stellar dynamical systems by gradually 
adding new phenomena
such as two-body relaxation, three-body encounters, 
and energy transport processes in
stellar systems with a mass spectrum.} 

{Moment models could quite easily be coupled with hydrodynamical solvers to 
simulate the dynamical evolution of dense gas-star systems (DGSS) in galactic nuclei 
\citep{Langbein1990, AmaSpu2001, AmSpuJus2002, AS04, Spurzem2004}. In \citet{Langbein1990}
it was shown that gaseous models of dense star clusters can be regarded 
as a generalization of the Tolman-Oppenheimer Volkoff equation for relativistic 
anisotropic gases. Many years ago \cite{BisSuny1972}, \citet{Vilkoviski1975}, and 
\citet{Hara1978} have proposed DGSS as energy sources in galactic nuclei. Nowadays 
the idea is being reconsidered that supermassive stars are progenitors of the first 
supermassive black holes in galactic nuclei \citep{begelman2010}, and that galactic 
nuclei in their variety of appearances could be determined by the interplay of stellar 
and gas dynamics, including star formation and feedback \citep{ciotti2009, shin2010, 
ciotti2010}. These topics deserve further investigation with improved stellar dynamical 
modelling, as we provide it here with our new momentum model. Therefore we think a fresh
look at and improvement of the momentum model is timely and very useful. It should be 
noted that spherical symmetry yet has been a limitation of gaseous or momentum models
of star clusters. However, also here a generalization at least to axisymmetric models 
is possible by describing viscosity through two-body relaxation in analogy to heat 
conduction \citep[][unpublished Ph.D. thesis]{goodman1983a}. We have demonstrated that the 
aforementioned Goodman models can be used and solved numerically with sufficient accuracy 
in the case of direct solutions of the orbit averaged Fokker-Planck equation
\citep{einsel1999, kim2002, kim2004, kimsp2008, fiestassp2010}. There is no reason
to assume that also our momentum or gaseous model could not be extended to axial symmetry in the
future, using appropriate implicit hydrodynamic solvers. }

{
By extending the model with additional equations coupled with
collisional terms, we are in the position to address new
problems.
Thus }
we can investigate
accretion theory \citep{pau2004}, stellar collision, gas dynamics and 
coupling with the stellar system, including radiative transfer and turbulences,
the role of the loss-cone \citep{ASEtAl03,AS04,PauTesi04} and tidal fields
\citep{spurzem2005}.  Higher order moments are necessary to have a more
realistic description of the velocity distribution function and a more accurate
description of relaxation, reducing the number of approximations necessary
to the model. 

The numerical models used to study dynamical processes have to be constrained
by comparison with observations. In order to do so, both models and observations
must fulfill certain accuracy requirements. There are many methods for modeling GCs
which can be separated into particle based methods such as $N$-body or
Monte-Carlo simulations and continuum methods such as Fokker-Planck or moment
models (see next section).  In statistical moment models, we employ velocity
moments to characterize the local velocity distribution function. The $n$-th
moment of a velocity distribution $f(v)$ is defined as $\langle v^n \rangle =
\int v^nf(v) \,\, \mathrm{d}v$ (see also definition \eqref{eq:def_moment}). The
accuracy of these models is then limited by the order of the highest moment
included to describe the velocity distribution. A physical interpretation for
each moment up to the fourth order can be given. Since each stellar dynamical
process driving the evolution of a cluster has a different impact on the local
velocity distribution, this motivates us to construct a distribution function
that is able to reflect the effects of each of these processes properly 
so as not to lose information that influences the clusters evolution. The
velocity distribution can be written as a series expansion using a
\emph{truncated Gauss-Hermite series} \citep{gerhard1993, marel1993} to
illustrate the meaning of the first four moments:

\begin{equation}
\label{eq:gauss_hermite}
f(v_r) \propto \exp(-\frac{v_r-\bar{v}_r}{2\sigma})\left[1+\sum^4_{k=3}h_k H_k(v_r-\bar{v}_r)\right]
\end{equation}

\noindent        
$v_r$ might be the velocity in radial direction (or the line-of-sight velocity
which is the velocity measured in direction of an observer). $\bar{v}_r$,
$\sigma$, $h_3$ and $h_4$ are free parameters and will be explained in the
following. 
	
\begin{figure}
	\centering
	\psfrag{s}[c][c]{\scriptsize $\sigma=10\,\text{km}/\text{s}$}
	\psfrag{f}[c][c]{\scriptsize f($\text{v}_{\text{r}}$)}
	\psfrag{v}[c][c]{\scriptsize $\text{v}_{\text{r}}$}
	\includegraphics[width=7.5cm,clip]{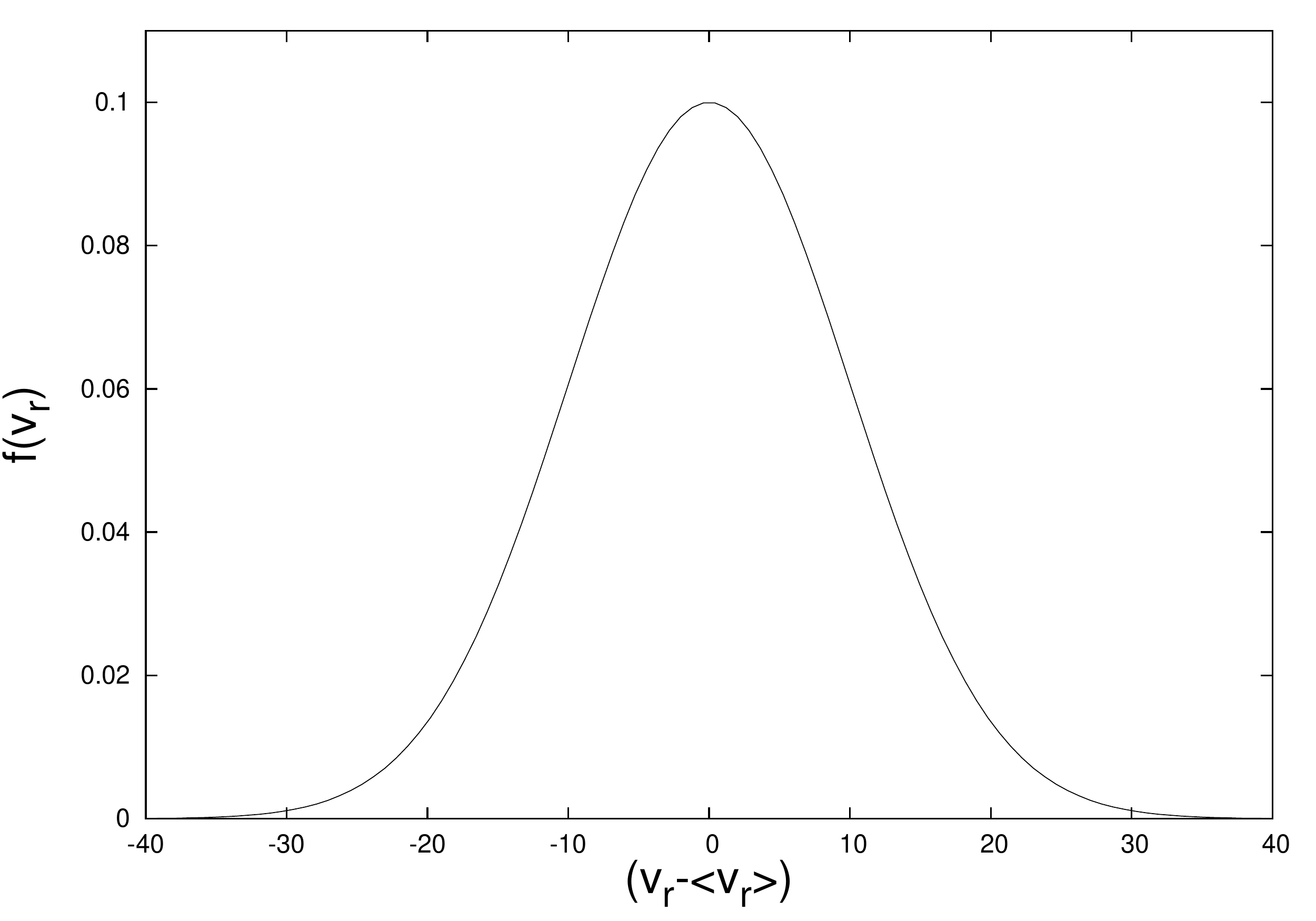}
	\includegraphics[width=7.5cm,clip]{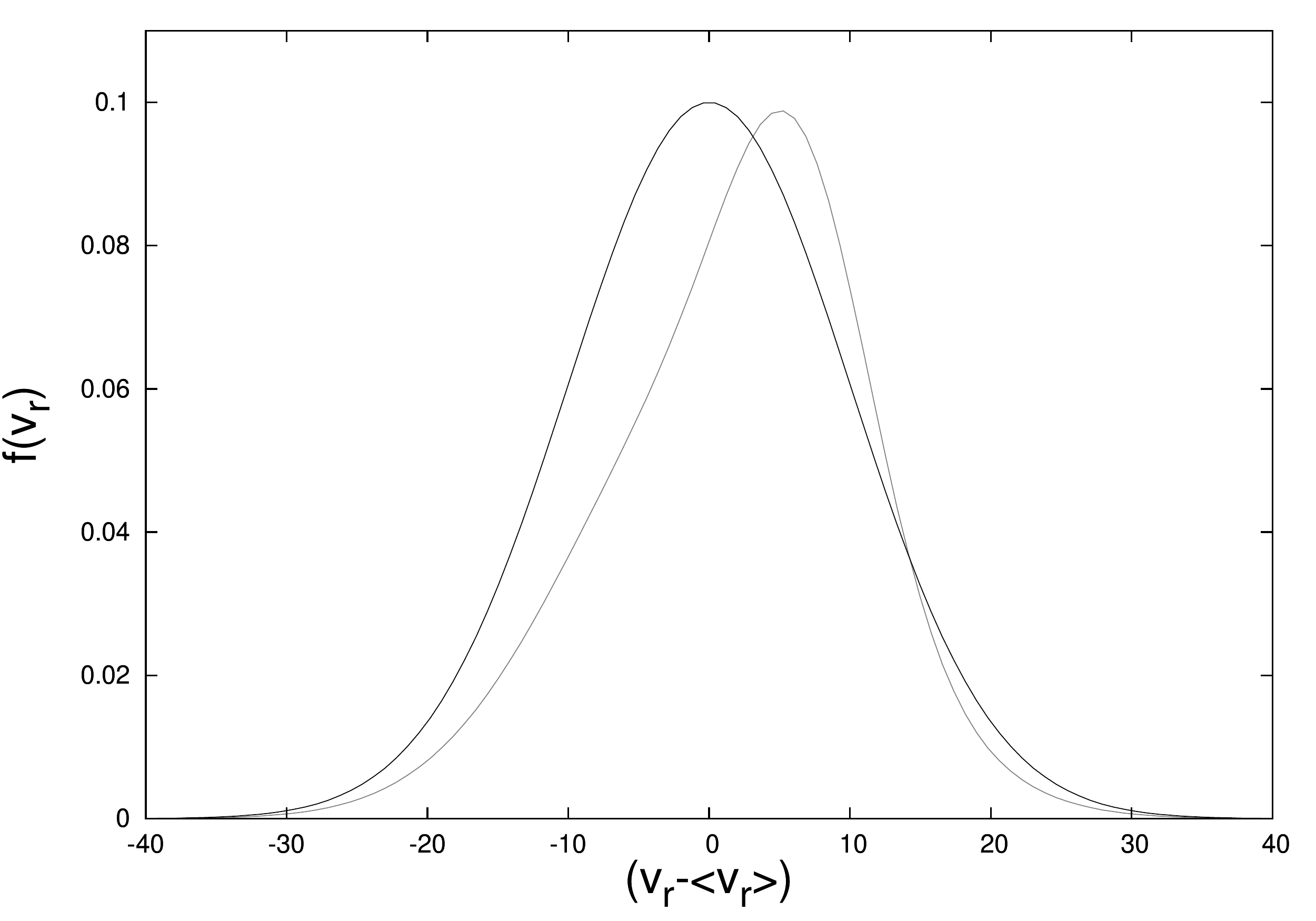}
	\includegraphics[width=7.5cm,clip]{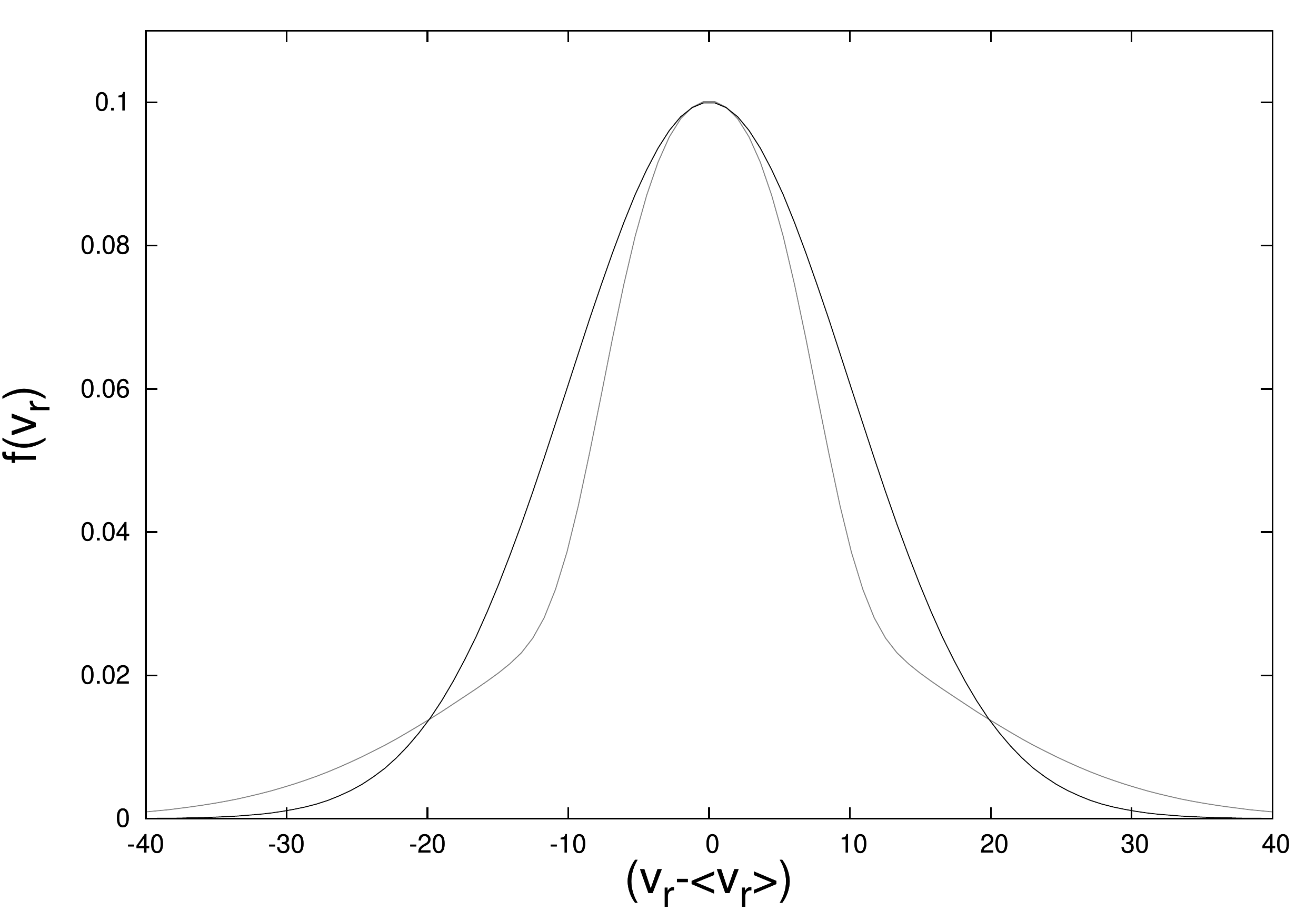}
	\includegraphics[width=7.5cm,clip]{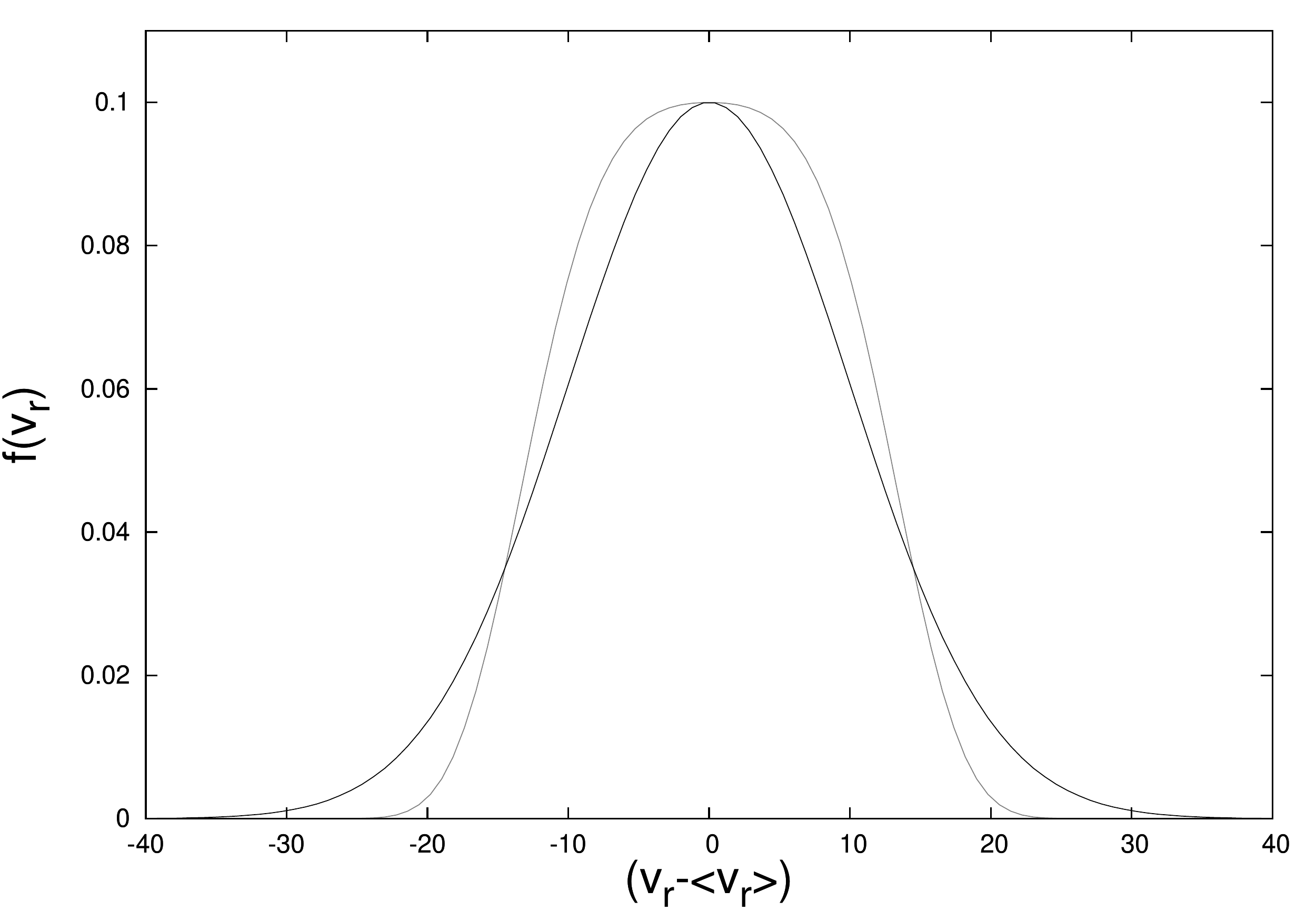}
	\caption{
These four plots show one-dimensional velocity distribution functions for
different cases. \emph{{top:}} Gaussian velocity distribution describing
thermodynamical equilibrium with a variance of
$\sigma=10\,\text{km}/\text{s}$. The Gaussian appears in the subsequent panels
for comparison (black). \emph{{upper middle}:} velocity distribution (grey) with a
skewness in positive $v_r$-direction indicating energy flow in $v_r$-direction.
\emph{{lower middle and bottom}:} two velocity distributions (grey) with an
excess and deficit of high velocity stars respectively as compared to
``thermodynamical equilibrium''
        }\label{fig:gauss_plots}
\end{figure}

        \begin{itemize}
		\item \emph{0th moment:} \\
		The zeroth moment of a velocity distribution is 1 due to normalization. 
		\item \emph{1st moment:} \\
		The first moment of a velocity distribution is the mean velocity $\bar{v}_r$ and denotes the bulk mass transport velocity. 
		\item \emph{2nd moment:} \\
		The second moment of a velocity distribution is the variance $\sigma$ and is equal to the velocity dispersion. It determines the width of  $f(v_r)$ and thus the scattering of stellar velocities around the mean velocity $\bar{v}_r$. If $f(v_r)$ is fully determined by $\bar{v}_r$ and $\sigma$ and $h_3=h_4=0$ it is a Gaussian ({top pannel} in figure \ref{fig:gauss_plots}) corresponding to thermal equilibrium. Then the symmetry of the one-dimensional velocity distribution $f(v_r)$ to $\bar{v}_r$ reflects isotropy. 
		\item \emph{3rd moment:}\\
		The third moment, denotes the transport of random kinetic energy and depends on $h_3$. If the third moment of the velocity distribution does not vanish, implying that $h_3\ne0$, then the shape of the velocity distribution is a skewed Gaussian (figure \ref{fig:gauss_plots}, {upper middle pannel}). The asymmetry indicates the direction of the energy flux, and the uneven distribution of velocities in different directions denotes anisotropy.
		\item \emph{4th moment:}\\
		The fourth moment is a measure of the excess or deficiency of particles/stars with high velocities as compared to thermodynamical equilibrium, and depends on the value of $h_4$. An excess of particles with high velocities results in thicker wings of the velocity distribution and a more pointed maximum (figure \ref{fig:gauss_plots}, {lower middle pannel}). A deficiency of high velocities causes a broader shape around the mean and thinner wings of the velocity distribution (figure \ref{fig:gauss_plots}, {bottom pannel}). 
	\end{itemize}

Third and fourth order moments therefore denote deviations from thermodynamical
equilibrium. Modeling processes that lead to the transport of random kinetic
energy in a cluster or that strongly affect the high velocity wings of the
distribution suggest the use of a model that includes 4th order moments. These
processes are, for example, the ``evaporation'' of high velocity stars from the
cluster, which reduces the number of high velocity stars. On the other hand,
binaries and a mass spectrum transfer kinetic energy 
{between different} stellar
components and thereby produce high velocity stars. These high velocity stars then transfer
their excess energy to their environment in subsequent distant two-body
encounters which can lead to a transport of kinetic energy between different
regions in the GC. 

Neglecting third and fourth order moments in these cases results in a loss of
information by failing to fully model the effect of the processes they
represent on the evolution of the cluster.

\section{Particle-based techniques vs statistical methods}
\label{chapter: models}
	
The methods for studying star clusters can be divided into two types;
statistical continuum models, such as Fokker-Planck (FP), or moment models and
particle-based techniques, such as direct $N$-body models and Monte Carlo.  They
have different advantages and deliver complementary information about the
processes and mechanisms that drive the evolution of star clusters.

\subsection{Direct integration techniques}

By using direct $N$-body we integrate Newton's equations of motion. In
principle all gravitational dynamics phenomena are naturally
included in the integration. Thus, this method is not subject to any
approximations nor restricted to any assumptions, such as spherical symmetry.
In contrast to statistical methods, it does not require additional physics in
order to include gravitational interactions between pairs, triples (binary-star
interactions) or quadruples (binary-binary interactions) as they are 
inherent to the model. Including a mass spectrum or tidal field is also, in
principle, straightforward. On the other hand, direct-summation methods of this
type are
computationally expensive, and as a consequence it is not possible to
realistically model a stellar cluster with a typical number of $10^{7-8}$
stars.  This is due to the fact that the computation of all pairwise
interactions of a system consisting of $N$ particles scales with $N^{2-3}$.
Using modern hardware we are severely limited to integrations of at most a few 
$10^6$ particles for a very short time, typically a few dynamical times. 
Another drawback of direct $N$-body is that it suffers from noise, as an 
individual N-body calculation in star cluster dynamics have exponential 
instabilities; nevertheless, the results can be used in a statistical average
\citep[e.g.][]{miller1964, giersz1994a}. 

There exist many schemes for integrating Newton's gravitational equation, 
some of them are faster and more effective than others. Among these
we should mention the \emph{Euler scheme} or an improvement of this, the
\emph{leapfrog scheme} \citep[e.g.][]{hut1995}. We can gain more accuracy by
the \emph{divided difference scheme} or the \emph{Hermite scheme}
\citep{makino1992, aarseth1999}, which is used in the NBODY6 and NBODY6++ codes
for the orbit integration. Additionally, various $N$-body codes incorporate a
number of approaches which are necessary for maintaining adequate accuracy
and efficiency over many dynamical times; these include the use of many individual
time steps, computation of forces from near neighbors and distant stars with
different frequencies, special treatments of compact pairs (binaries) and other
few-body configurations \citep{mikkola1990, mikkola1993}. Direct $N$-body
simulation is a powerful tool for realistically simulating a wide range of
astrophysically interesting scenarios such as black holes in galactic nuclei or
GCs, binaries of massive black holes in (rotating) clusters
\citep{ASF06,Amaro-SeoaneEtAl09a,pau2010} or binary black hole mergers in
galactic nuclei \citep{berentzen2009}. 

\subsection{The Monte Carlo approach}

Other powerful particle-based techniques are the Monte Carlo (MC) methods, in
which relaxation is treated using the Fokker-Planck approximation. These
methods rely also on the assumptions that the system is spherically
symmetric and that the gravitational potential can be separated into two parts.
The advantage of MC is that it is orders of magnitude faster than direct
$N-$body, yet it is still slower than statistical methods and also suffers
from numerical noise.

Spitzer and collaborators pioneered the MC scheme in a series of papers, such
as \cite{spitzer1971a, spitzer1971b, spitzershapiro1972, spitzerthuan1972,
spitzerchevalier1973, spitzershull1975a, spitzershull1975b, spitzermathieu1980}
The initial models were soon improved by Shapiro and his collaborators
\citep{shapiromarchant1978, marchantshapiro1979, marchantshapiro1980,
duncanshapiro1982, shapiro1985}. MC, being particle-based, follows the
individual stellar orbits and allows us to model processes occurring on both
relaxation and crossing time scales. Spitzer's method was used to explore a
variety of important phenomena, including mass segregation, anisotropy of the
velocity distribution, tidal shocking, and the role of primordial binary stars,
to mention a few.

The second MC approach was devised by \citet{henon1971a, henon1971b, henon1972,
henon1975} and later improved by \citet{sto1982, sto1986}. In contrast to the
models of Spitzer, H{\'e}non's models assumed dynamical equilibrium; the
distribution function must also depend only on isolated integrals of motion. It
is worth mentioning that it was the first scheme to break through the impasse
of core collapse \citep{henon1975}. The algorithm was further improved by
\citet{sto1985} by including processes such as the formation of binaries by
two- and three-body encounters, mass loss from stellar evolution and tidal
shocking. 

\cite{giersz1998,giersz2001} in a series of papers modeled $\omega$Cen
\citep{giersz2003}, M4 \citep{heggie2008}, M67 \citep{giersz2008} and NGC 6397
\citep{giersz2009} with MC techniques. In these papers several additional
improvements were also included, as two-body relaxation, most kinds of three-
and four-body interactions involving of primordial binaries and those formed
dynamically, the Galactic tide and the internal evolution of both single and
binary stars.  MC techniques can be coupled with continuum models to describe
the stochastic process of binary formation energy generation and movement
\citep{spurzemgiersz1996, giersz2000, gierszspurzem2003}. 
{This has been successfully used to examine the gravitational radiation from binary black holes
in star clusters \citep{downing2009}.}

\cite{joshi2000, joshi2001, fregeau2003, fregeau2007} developed a MC technique
based on a modified version of H{\'e}non's algorithm for solving the
Fokker-Planck equation. Their scheme 
{includes} a mass spectrum, stellar
evolution, and primordial binary interactions and the direct integration of
binary scattering interactions.  The H{\'e}non-type MC approach has been used by
M.~Freitag, who developed another MC code with the special purpose of studying
semi-Keplerian systems. Applying this code he extensively studied the structure of 
galactic nuclei containing a central MBH \citep{FreitagThesis00,FB02b,FASK06}.

\subsection{A statistical model: The Fokker-Planck technique}

Fokker-Planck (FP) models are based on the direct numerical solution of the
orbit-averaged FP equation. \citet{cohn1979, cohn1980} pioneered a direct
numerical finite-difference solution of the 1-dimensional FP equation (for a
phase space distribution function: $f=f(E)$). Similar methods had been
developed for a fixed potential by \citet{ipser1977} and by \citet{cohn1978},
and since then different FP codes have been written independently by
\citet{inagaki1984} and by \citet{chernoff1990}. Whereas Cohn's formulation
assumes spherical symmetry, codes which can handle a rotating cluster have been
devised by \citet{goodman1983a} and \citet{einsel1996}. \citet{takahashi1995,
takahashi1996, takahashi1997} has developed FP models for GCs, based on the
numerical solution of the orbit averaged 2D FP equation (i.e. solving the FP
equation for the distribution $f=f(E, J^2)$) as a function of energy and
angular momentum, and thus accounting for anisotropy. 

\citet{drukier1999} followed with results from another 2D FP code based on
the original idea of \citet{cohn1979}. There have been several
comparative studies \citep{giersz1994a, giersz1994b, giersz1997, takahashi1995,
giersz1994, spurzem1995, freitag2006, khalisi2007} showing that for isolated,
non-rotating star clusters the results of FP simulations are generally in good
agreement with those of $N$-body simulations.  However, when a tidal boundary
is included, discrepancies between $N$-body and FP models occur. 

Also, \cite{einsel1999} found that rotating GCs collapse faster than
non-rotating ones with a 2D FP technique that had a distribution function
depending on the $z$-component of the angular momentum, $f=f(E, J_z)$.
\cite{kim2002} improved the approach by including an energy source due to
formation and hardening of three-body binaries. These two studies only
investigated single-mass models. Later, \cite{kim2004} extended this method to
multi-mass systems, finding interesting results concerning segregation of mass
and angular velocity with heavy stars in the cluster. \citet{fiestas2006} have
modeled rotating globular clusters and \citet{fiestas2010} included a star
accreting black hole with a loss cone. Comparative studies for rotating star
clusters between FP and $N$-body methods have been done as well by
\citet{boily2000, boilyspurzem2000, ardi2005, ernst2007, kimsp2008}. They
produced fairly similar results, though there were small discrepancies in the
core-collapse time.

\subsection{Advantages and disadvantages of statistical models as compared to
direct-summation techniques}

>From the three different techniques, direct $N$-body models appear as the most
realistic model. However, as mentioned before, it suffers from exponential
instabilities; small deviations in the initial conditions result in 
exponential divergence of the phase space distribution of the particles of the
system \citep{miller1964,giersz1994a}. These instabilities make it difficult to
compare a realistic model of GCs to observational data.  Statistical models
produce averaged physical quantities and are better suited for comparison
with observations. 

Also, as $N$-body models are not restricted by boundary conditions such as
spherical symmetry, they can be applied to the widest range of stellar
dynamical systems and study them under the most diverse scenarios.  This relies
on a microscopic description of dynamical processes and translates into a
complexity that requires a massive computational effort.  As a consequence we 
depend on the development of hardware to push the number of particles that 
we can integrate forward. On the other hand, statistical continuum models 
which are based on a comparatively small set of differential equations are 
computationally cheap.

These algorithms have also the important property that the contribution of
various dynamical processes to the overall evolution of a star cluster can be
isolated. This is so because the different mechanisms have to be included
separately by additional terms in the model equations. Therefore, it is
possible to identify each mechanism and its effect. 

{The downside of statistical moment models is that they are 
subject to a large number of approximations. Some of these approximations 
are inherent in the approach, such as the description of the phase space 
distribution function by a finite number of its velocity moments. 
} 
Additional approximations consist of the limitation in the number of processes included such as
two-body relaxation, star-binary deflections, binary-binary encounters or
anisotropy.  Such processes are natural in $N$-body models. 
The bottom line is
that in order to build up a detailed understanding of stellar dynamical systems
we need the different properties of particle-based {\em and} statistical
models.

\section{Self-gravitating, conducting gas spheres}

In the previous section we have given an overview on the different numerical tools to
address stellar dynamics including relaxation. Now that we have highlighted the
advantages of statistical methods, we introduce an interesting alternative to
FP 
. More than 35
years ago \citet{hachisu1978} and \citet{lynden-Bell1980} proposed transport
process in a self-gravitating, conducting gas sphere as a way to mimic two-body
stellar relaxation. Later, \citet{bettwieser1983, bettwieser1984,
bettwieser1986, heggie1984} \citet{heggie1989} and \cite{louis1991} implemented
anisotropy and \citet{giersz1994} and \cite{spurzem1995} added a multi-mass
distribution and improved the detailed form of the conductivities to have
better accuracy. 
{The resulting model is often called the \emph{anisotropic 
gaseous model} (AGM).} 
This allows us to compare with $N-$body models to calibrate
the approach.  \citet{pau2004} addressed the accretion of stars on to a massive
black hole by adding collisional terms corresponding to loss-cone physics as
well as tidal effects and \citet{spurzem2005} investigated the evolution and
dissolution of star clusters under the combined influence of internal
relaxation and external tidal fields.

In this approach, we emulate spherically symmetric systems as a continuum;
relaxation is treated as a diffusive process in phase space using the FP
equation.  We employ the {local approximation} to simplify the FP equation by
neglecting the diffusion in position. The idea behind this is that an
encounter takes place in a volume that is much smaller than the dimensions of
the whole system.  We model energy transfer by a local heat flux equation with
an appropriately tailored conductivity.

The basis of the equations of the model is the FP equation which describes the
time evolution of the probability density function.
Using spherical polar coordinates, the Boltzmann equation takes the form:

\begin{equation}\label{seboltzmanneq}	
        \frac{\partial f}{\partial t} + v_{\rm{r}} \frac{\partial f}{\partial r} + 
	\dot{v}_{\rm{r}} \frac{\partial f}{\partial v_{\rm{r}}} + 
	\dot{v}_{\rm{\phi}} \frac{\partial f}{\partial v_{\rm{\phi}}} +
	\dot{v}_{\rm{\theta}} \frac{\partial f}{\partial v_{\rm{\theta}}} 
	= \left(\frac{\delta f}{\delta t}\right)_{FP}
\end{equation}

\noindent
In the last equation the right-hand side denotes that collisions are given in
terms of the FP approximation. Due to symmetry, we can define a tangential
velocity $v_t^2=v_{\rm{\phi}}^2+v_{\rm{\theta}}^2$, so that we have two
velocities $v_t$ and $v_r$ to describe the system. The ``centralized'' moments are
defined by multiplying the velocity distribution $f$ with powers of $v_t$ and
$(v_r-\bar{v}_r)$ and integrating over velocity space. 

The term ``centralized'' means that the moments are defined with respect to the
mean velocity components $\bar{v}_r=\langle v_r \rangle = u$ and
$\bar{v}_t=\langle v_t \rangle = 0$, because we assume spherical symmetry. The
order of a moment is defined by $n+m$ where $n$ and $m$ are the powers of
velocities in the definition of moments, i.e. $\int
(v_r-\bar{v}_r)^nv_t^m\,f\,\, \mathrm{d}^3v$.  The moments defined this way
correspond to the density of stars, $\rho$, the bulk velocity, $u$, the radial
and tangential pressures, $p_r$ and $p_t$, and the radial and tangential
kinetic energy fluxes, $F_r$ and $F_t$. In order to obtain the set of
differential moment equations, we multiply equation \eqref{seboltzmanneq} with
powers of $v_t$ and $(v_r-\bar{v}_r)$ and integrate it in velocity space. After
some recasting the integrals can be substituted by the moments. Up to second
order the moment equations are the continuity equation, the Euler equation
(force) and radial and tangential energy equations:

\begin{equation}
\begin{split}
\label{moment equations}
    & \frac{\partial \rho}{\partial t} + \frac{1}{r^2}\frac{\partial}{\partial r}(r^2 u \rho) = 0 \\
    & \frac{\partial u}{\partial t} + u \,\frac{\partial u}{\partial r} + \frac{G M_r}{r^2} + \frac{1}{\rho} \frac{\partial p_r}{\partial r} + 2 \frac{p_r - p_t}{\rho r} = 0 \\ 
    & \frac{\partial p_r}{\partial t} + \frac{1}{r^2}\frac{\partial}{\partial r}(r^2 u p_r) + 2 p_r \,\frac{\partial u}{\partial r} + \frac{1}{r^2}\frac{\partial}{\partial r}(r^2 F_r) - \frac{2 F_t}{r} \\
    & = -\frac{3}{5} \frac{p_r-p_t}{\lambda_A t_{\mathrm{rx}}} + \left(\frac{\delta p_r}{\delta t}\right)_{\rm{bin}3}\\
    & \frac{\partial p_t}{\partial t} + \frac{1}{r^2}\frac{\partial}{\partial r}(r^2 u p_t ) + \frac{2 p_r u}{r} + \frac{1}{2 r^2}\frac{\partial}{\partial r}(r^2 F_t) + \frac{F_t}{r} = \\
    &
\frac{3}{10} \frac{p_r-p_t}{\lambda_A t_{\mathrm{rx}}} + \left(\frac{\delta p_t}{\delta t}\right)_{\rm{bin}3}
\end{split}
\end{equation}
                
\noindent
Here $\lambda_A$ is a numerical constant related to the time-scale of collisional
anisotropy decay, necessary to describe the relaxation effects on cluster
evolution. It should become unity when describing GCs using higher moment
models. However, this can only be confirmed by simulations. The value of
$\lambda_A$ is discussed in \citet{giersz1994} and is chosen by calibrating
with $N-$body simulations.  The authors found that $\lambda_A=0.1$ is a
physically realistic value within the half-mass radius for all numbers of
particles.

The two terms on the right-hand sides of the equations for radial and
tangential energy equations are the collisional terms. The fist term accounts
for relaxation from uncorrelated two-body encounters and can be derived from
the FP equation. The second term, which is marked with ``bin3'', refers to
star-binary encounters. Close 3-body or star-binary encounters generate kinetic
energy. If the energy generation is high enough, this mechanism can reverse
core collapse. 

The radial and tangential pressure, $p_r$ and $p_t$ are related to the random
velocity dispersions; $p_r= \rho \, \sigma_r^2$ and $p_t= \rho \, \sigma_t^2$.
They are linked to observable quantities in stellar clusters such as the radial
velocity dispersion. The average velocity dispersion is $\sigma^2 = (\sigma_r^2
+ 2\sigma_t^2)/3$, where the factor 2 comes from the fact that there are two
tangential directions. The radial energy flux of random kinetic energy is
$F=(F_r + F_t)/2$. We can see this by adding the two-moment equations for
radial and tangential pressure to obtain the gas-dynamical equation for the
energy density. The velocities for energy transport are defined by

\begin{equation}\begin{split}
& v_r = \frac{F_r}{3p_r} + u \\
& v_t = \frac{F_t}{2p_t} + u 
\end{split}\end{equation}

\noindent

{In the case of \emph{weak isotropy}, $p_r = p_t$ everywhere and hence $F_r =
\frac{3}{2}F_t$, so that $v_r = v_t$. Therefore, the transport velocities for
radial and tangential random kinetic energy are equal.}

In order to close the set of moment equations \eqref{moment equations} three
more equations are set up, a mass relation and two equations accounting for
heat flux. The mass relation defines the mass $M_r$ contained in a sphere of
radius $r$, 

\begin{equation} \label{Poisson}
\frac{\partial M_r}{\partial r} = 4 \pi  r^2 \rho_M
\end{equation}

{
\noindent
where $\rho_M = M \cdot \rho$ is the mass density and $M$ the mass of the stellar component.} We thus obtain a set of gas dynamical equations \eqref{moment equations}
coupled with the Poisson equation \eqref{Poisson}. Since the moment equations
of order $n$ obtained from the Boltzmann equation contain moments of the order
$n+1$, we need closure relations connecting the moments of order $n+1$ with
lower-order moments. 
{This is achieved}
with the \emph{heat conduction closure},
a phenomenological approach obtained in an analogous way to gas dynamics. It is
motivated by the resemblance between a star consisting of a large number of
atoms and a star cluster with large number of stars not only on the simple
level of the virial theorem but also due to similarities in heat transport,
energy generation and core-halo evolution. It was used by
\citet{lynden-Bell1980}, initially restricted to isotropic systems. In
this approximation we assume that heat transport is proportional to the
temperature gradient 

\begin{equation}\label{heat flux} 
F = - \kappa \frac{\partial
T}{\partial r} = - \Lambda \frac{\partial \sigma^2}{\partial r}
\end{equation} 

This equation describes the heat flux in gases and liquids and for this reason
the models using this closure are also called \emph{conducting gas sphere
models}. Even though the use of equation \eqref{heat flux} is based on the
assumption of small mean free paths for the particles, which is certainly
questionable for stellar dynamical systems, models like the AGM agree with
other modeling methods (e.g. $N$-Body, FP) 
{\citep{giersz1994, spurzem1995}}

In the classical approach $\Lambda \propto \rho\bar{\lambda}^2/\tau$, where
$\bar{\lambda}$ is the mean free path and $\tau$ the collisional time. Choosing
the Jeans length $\lambda_J^2 = \sigma^2/(4 \pi G \rho)$ for $\bar{\lambda}^2$
and the standard Chandrasekhar local relaxation time $t_{\rm{rx}} \propto
\sigma^3/\rho$ \citep{chandrasekhar1942} for $\tau$, we obtain the conductivity
$\Lambda \propto \rho/\sigma$. More precisely, the conductivity takes the form
found in \citet{lynden-Bell1980}: 

\begin{equation} \Lambda = \frac{3 C G m \rho N}{\sigma},  
\end{equation} 

\noindent
where C is a dimensionless numerical constant of order unity. By means of the
velocities of energy transport the heat flux equation can be recast to find the
two closure relations in the anisotropic case 

\begin{equation} 
v_r-u+ \frac{\lambda}{4 \pi G \rho t_{\rm{rx}}} \frac{\partial \sigma^2}{\partial r} =
0 \qquad v_r = v_t, 
\end{equation} 

\noindent
where 

\begin{equation} 
\lambda = \frac{27 \sqrt{\pi}}{10} C 
\end{equation} 

It should be emphasized that $\lambda$ is a free parameter that has to be
determined by comparison with other models such as $N$-body \citep{giersz1994},
Louis' fluid dynamical model \citep{louis1991} or FP models. In the isotropic
limit, $\lambda$ is just a scaling scaling factor, but when taking into account
anisotropy, $\lambda$ prescribes the relative speed of two processes: the decay
of anisotropy and the heat flow between warm and cold regions. With increasing
$\lambda$ heat flows faster, so there is less time for gravitational encounters
to destroy anisotropy. A larger $\lambda$ thus results in stronger anisotropy. 

\section{Higher moment models} 
\label{sec:moment_model}
{In this section we present a new higher order moment model. We derive the model equations which }
consist of differential
equations for the velocity moments of the phase space distribution function, a
Poisson equation and three equations to close the system of equations. 
We first compute the left-hand sides of the differential moment equation and
then use a polynomial ansatz for the phase space distribution function to
obtain the right-hand sides.  We define two models, \emph{model a} and
\emph{model b}, which differ in the number of differential (moment) equations
and their closure relation.

\subsection{Left-hand sides} 
\label{subsec:lhs}

Without collisions, the Boltzmann equation takes the form of a
conservation equation ($\mathrm{d} f/\mathrm{d} t = 0$) and describes the
advective rate of change of the phase space distribution function $f$. If we
follow the trajectory of a particle in a system described by the collisionless
Boltzmann equation, the number density in phase space around the particle does
not change. This implies that flow in phase space is incompressible.  
{It becomes compressible when collisions are introduced with FP
 terms on the right hand side of the Boltzmann equation.}

Assuming that the stellar system is spherically symmetric, we can use spherical
coordinates when we write the collisional Boltzmann equation,

\begin{equation}\label{eq:boltzmann_spherical}
 \frac{\partial f}{\partial t}+ v_r \frac{\partial f}{\partial r}+\dot{v}_r\frac{\partial f}{\partial v_r}+\dot{v}_{\theta}\frac{\partial f}{\partial v_{\theta}}+\dot{v}_{\phi}\frac{\partial f}{\partial v_{\phi}} = \left(\frac{\delta f}{\delta t}\right)_{\text{enc}}
\end{equation}

Using the Lagrangian of a particle in a spherical symmetric potential $\Phi(r,t)$, we have that

\begin{equation}
 \mathcal{L} = \frac{1}{2}(\dot{r}^2+r^2\dot{\theta}^2+r^2\sin^2\theta \,\dot{\phi}^2)-\Phi(r,t)
\end{equation}

We then apply the Euler-Lagrange equations to the Lagrangian, to derive the equations of motion

\begin{eqnarray}\label{eq:eq_of_motion}
 \dot{v}_r&=&-\frac{\partial \Phi}{\partial r}+\frac{v^2_{\theta}+v^2_{\phi}}{r}\nonumber \\
 \dot{v}_{\theta}&=& -\frac{v_r v_{\theta} }{r}+\frac{v^2_{\phi}}{r \tan \theta} \\
 \dot{v}_{\phi}&=& -\frac{v_rv_{\phi}}{r}-\frac{v_{\theta}v_{\phi}}{r \tan \theta}\nonumber
\end{eqnarray}

After substituting equation \eqref{eq:eq_of_motion} into the Boltzmann equation
\eqref{eq:boltzmann_spherical}, we use the approach of spherical symmetry to
define the tangential velocity $v_t = \sqrt{v^2_{\theta}+v^2_{\phi}}$ and
obtain

\begin{equation}\label{eq:boltzmann_spherical_vel}
 \frac{\partial f}{\partial t}+ v_r \frac{\partial f}{\partial r}+\left(\frac{v^2_t}{r}-\frac{\partial \Phi}{\partial r}\right)\frac{\partial f}{\partial v_r}-\frac{v_r v_t}{r}\frac{\partial f}{\partial v_t} = \left(\frac{\delta f}{\delta t}\right)_{\text{enc}}
\end{equation}

We now define the velocity moments of the distribution function
$f=f(r,v_r,v_t,t)$ by multiplying it by powers of $v_r$ and $v_t$ and
integrating over velocity space,

\begin{equation}\label{eq:def_moment}
 [n,m]=\int \textnormal{d}^3v\,f \,v^n_r\,v^m_t=2 \pi 
 \int^{\infty}_{0}\textnormal{d}v_t \int^{\infty}_{-\infty}\textnormal{d}v_r\,f\, v^n_r\,v^{m+1}_t
\end{equation}

{Again, the }
order of a moment is defined as $k=n+m$. To obtain the differential
equations for the moments $[n,m]$ we multiply equation
\eqref{eq:boltzmann_spherical_vel} with powers of $v_r$ and $v_t$ and integrate
over velocity space. After some recasting we can substitute the integrals by
$[n,m]$ which yields

\begin{equation}\begin{split}\label{eq:moment_dgl}
&\frac{\partial}{\partial t}[n,m]+\frac{\partial}{\partial r}[n+1,m]+\frac{m+2}{r}[n+1,m]\\
&-\frac{n}{r}[n-1,m+2]+n[n-1,m]\frac{\partial \Phi}{\partial r}=\left(\frac{\delta}{\delta t}[n,m]\right)_{\text{enc}}
\end{split}\end{equation}
We now want to find a differential equation equivalent to equation \eqref{eq:moment_dgl} for centralized moments. The centralized velocity moments are defined with respect to their mean velocity. Due to the assumed spherical symmetry of the system the mean velocities of the tangential components $\bar{v}_{\theta}=\bar{v}_{\phi}=0$ vanish. The mean velocity is only given by the radial velocity component 
\begin{equation}
\bar{v}_r = [1,0]=u=2 \pi \int^{\infty}_{0}\textnormal{d}v_t \int^{\infty}_{-\infty}\textnormal{d}v_r\,f\, v_r \,v_t
\end{equation}
We hence obtain the definition for centralized moments by substituting $v_r$ in equation \eqref{eq:def_moment} with $(v_r-\bar{v}_r)$. Furthermore, the centralized moments can be expressed in terms of the moments $[n,m]$ and are defined as
\begin{eqnarray}\label{eq:central_moments_def}
 \langle n,m \rangle &=& \int \textnormal{d}^3v\, (v_r-\bar{v}_r)^n\,v^m_t \,f\nonumber \\
&=& 2 \pi \int^{\infty}_{0}\textnormal{d}v_t \int^{\infty}_{-\infty}\textnormal{d}v_r\, (v_r-\bar{v}_r)^n\,v^{m+1}_t\,  f\\
&=& \sum^{n}_{k=0} \binom{n}{k} \, (-1)^{n-k}\, [1,0]^{n-k}\, [k,m]\nonumber
\end{eqnarray} 
It is evident from the second line of \eqref{eq:central_moments_def} that the first centralized moment $\langle 1,0 \rangle=0$. 

We adopt the following notation for the centralized moments:

\begin{equation}
\begin{split}
\rho & = \langle0,0\rangle  \qquad \langle1,0\rangle =0\\
p_r & = \langle 2,0 \rangle \qquad 2p_t = \langle 0,2 \rangle\\
F_r & = \langle 3,0 \rangle \qquad F_t= \langle 1,2 \rangle\\
\kappa_r & = \langle 4,0 \rangle \qquad \kappa_{rt}= \langle 2,2 \rangle \qquad \kappa_t= \langle 0,4 \rangle\\
G_r & = \langle 5,0 \rangle \qquad G_{rt}= \langle 3,2 \rangle \qquad G_t= \langle 1,4 \rangle\\
H_r & = \langle 6,0 \rangle \qquad H_{r,t}= \langle 4,2 \rangle \qquad H_{t,r}= \langle 2,4 \rangle \qquad H_t = \langle 0,6 \rangle
\end{split}
\end{equation}

\noindent
{Again, } 
$\rho$ is the particle density, $p_r$ and $p_t$ are the radial and tangential
pressure and are related to the radial and tangential velocity dispersion
$\sigma_r = p_r/\rho$ and $\sigma_t = p_t/\rho$, and $F_r$ and $F_t$ denote the
radial and tangential energy flux.

We obtain a linear system of equations which can be solved for the moments
$[n,m]$ by computing all centralized moments $\langle n , m \rangle$ up to
order $n+m=6$ using equation \eqref{eq:central_moments_def}:

\begin{equation}
\begin{split}
\label{eq:transform}
	& [2,0] = p_r + \rho u^2 \\ 
	& [0,2] = 2 p_t \\
	& [3,0] = \rho u^3 + 3 u p_r + F_r \\
	& [1,2] = 2 u p_t + F_t \\
	& [4,0] = \rho u^4 + 6 u^2 p_r + 4 u F_r + \kappa_r\\
	& [2,2] = 2 u^2 p_t + 2 u F_t + \kappa_{rt} \\ 
	& [0,4] = \kappa_t \\
	& [5,0] = \rho u^5 + 10 u^3 p_r+ 10 u^2 F_r + 5 u \kappa_r + G_r \\ 
	& [3,2] = 2 u^3 p_t + 3 u^2 F_t + 3 u \kappa_{rt} + G_{rt} \\
	& [1,4] = u \kappa_t + G_t \\
	& [6,0] = \rho u^6 + 15 u^4 p_r + 20 u^3 F_r + 15 u^2 \kappa_r + 6 u G_r + H_r \\
	& [4,2] = 2 u^4 p_t + 4 u^3 F_t + 6 u^2 \kappa_{rt} + 4 u G_{rt} + H_{rt} \\
	& [2,4] = u^2 \kappa_t + 2 u G_t + H_{tr} \\
	& [0,6] = H_t 
\end{split}
\end{equation}

To obtain the differential equations for the centralized moments, we substitute
the transformation from equation \eqref{eq:transform} into equation \eqref{eq:moment_dgl} and then successively use differential equations for lower-order moments to simplify the differential equations for higher orders. We divide the
differential moment equations into three sets, defined as follows

\begin{description}
\item[Set I :]
\begin{equation}\begin{split}\label{moment_equations_a}
 &\frac{\partial \rho}{\partial t}+ \textnormal{div}(\rho u) =\left(\frac{\delta \rho}{\delta t}\right)_{\text{enc}}\\
 &\frac{\partial \rho u}{\partial t}+ \textnormal{div}(\rho u^2)+\frac{\partial p_r}{\partial r}+\frac{2}{r}(p_r-p_t)+\rho\frac{\partial \Phi}{\partial r}= \left(\frac{\delta \rho u}{\delta t}\right)_{\text{enc}}\\
 &\frac{\partial p_r}{\partial t}+ \textnormal{div}(F_r+up_r)+2 p_r \frac{\partial u}{\partial r}-\frac{2}{r}F_t=\left(\frac{\delta p_r}{\delta t}\right)_{\text{enc}}\\ &2\frac{\partial p_t}{\partial t}+ \textnormal{div}(F_t+2 up_t)+\frac{2}{r}(F_t+2u p_t)=2 \left(\frac{\delta p_t}{\delta t}\right)_{\text{enc}}\\
\end{split}\end{equation}

\item[Set II :]
\begin{equation}\begin{split}\label{moment_equations_b}
&\frac{\partial F_r}{\partial t}+ \textnormal{div}(\kappa_r+u F_r)+3 F_r \frac{\partial u}{\partial r}\\
&-3\frac{p_r}{\rho}\textnormal{div}p_r-\frac{3}{r}(\kappa_{rt}-\frac{2p_r p_t}{\rho})=\left(\frac{\delta F_r}{\delta t}\right)_{\text{enc}}\\ 
&\frac{\partial F_t}{\partial t}+ \textnormal{div}(\kappa_{rt}+u F_t)+ F_t \frac{\partial u}{\partial r}-\frac{2 p_t}{\rho}\textnormal{div}p_r\\
&-\frac{1}{r}(\kappa_t-2\kappa_{rt}-2uF_t-4\frac{p_t^2}{\rho})=\left(\frac{\delta F_t}{\delta t}\right)_{\text{enc}}\\
& \frac{\partial \kappa_r}{\partial t}+ \textnormal{div}(G_r+u\kappa_r)+4\kappa_r\frac{\partial u}{\partial r}- 4\frac{F_r}{\rho}\textnormal{div}p_r\\
&-\frac{4}{r}(G_{rt}-\frac{2p_t F_r}{\rho})=\left(\frac{\delta \kappa_r}{\delta t}\right)_{\text{enc}}\\
&\frac{\partial \kappa_{rt}}{\partial t}+ \textnormal{div}(G_{rt}+u\kappa_{rt})+2\kappa_{rt}\frac{\partial u}{\partial r}- 2\frac{F_t}{\rho}\textnormal{div}p_r\\
&+\frac{2}{r}(G_{rt}-G_t+u\kappa_{rt}+2\frac{p_t F_t}{\rho})=\left(\frac{\delta \kappa_{rt}}{\delta t}\right)_{\text{enc}}\\
& \frac{\partial \kappa_t}{\partial t}+ \textnormal{div}(G_t+u\kappa_t)+\frac{4}{r}(G_t+u\kappa_t)=\left(\frac{\delta \kappa_t}{\delta t}\right)_{\text{enc}}\\
\end{split}\end{equation}

\item[Set III :]
\begin{equation}\begin{split}\label{moment_equations_c}
& \frac{\partial G_r}{\partial t}+ \text{div}(H_r+uG_r)+5G_r\frac{\partial u}{\partial r}-5\frac{\kappa_r}{\rho}\text{div}p_r\\
&-\frac{5}{r}(H_{rt}-2\frac{p_t\kappa_r}{\rho})=\left(\frac{\delta G_r}{\delta t}\right)_{\text{enc}}\\
&\frac{\partial G_{rt}}{\partial t}+ \text{div}(H_{rt}+uG_{rt})+3G_{rt}\frac{\partial u}{\partial r}-3\frac{\kappa_{rt}}{\rho}\text{div}p_r\\
&+\frac{1}{r}(2H_{rt}-3H_{tr}+2uG_{rt}+6\frac{p_t\kappa_{rt}}{\rho})=\left(\frac{\delta G_{rt}}{\delta t}\right)_{\text{enc}}\\
& \frac{\partial G_t}{\partial t}+ \text{div}(H_{tr}+uG_t)+G_t \frac{\partial u}{\partial r}-\frac{\kappa_t}{\rho}\text{div}p_r\\
&-\frac{1}{r}(H_t-4H_{tr}-4uG_t-2\frac{p_t\kappa_t}{\rho})=\left(\frac{\delta G_t}{\delta t}\right)_{\text{enc}}\\
\end{split}\end{equation}

\end{description}

Note that the divergence operator in spherical symmetry reduces to:
\begin{equation}
 \text{div}=\frac{1}{r^2}\frac{\partial }{\partial r}\, r^2
\end{equation}

We now define two models with different accuracies,

\begin{description}

\item[Model a] -- including \eqref{moment_equations_a} and \eqref{moment_equations_b}

\item[Model b] -- including all, \eqref{moment_equations_a},
\eqref{moment_equations_b} and \eqref{moment_equations_c}

\end{description}

The potential $\Phi(r,t)$ is determined by the
fraction of cluster mass $M_r(t)$ contained at radius $r$

\begin{equation}
	\Phi= -\frac{GM_r}{r}
\end{equation}

\noindent
$\Phi$ obeys the Poisson equation $\Delta \Phi= 4 \pi \rho_{\text{M}}$, where
$\rho_{\text{M}}=M \rho$ is the mass density of the cluster. This leads to the
equation for $M_r$

\begin{equation} \label{Poisson}
\frac{\partial M_r}{\partial r} = 4 \pi  r^2 \rho_{\text{M}} 
\end{equation}

We note that the moment equations of order $n$ contain moments of order $n+1$. To
close the system of equations we need closure equation where moments of order
$n+1$ are expressed with lower-order moments.  We derive these relations in the
next section.

\subsection{{FP collision terms}} 

We now compute the right-hand sides of the differential moment equations
\eqref{moment_equations_a}, \eqref{moment_equations_b} and
\eqref{moment_equations_c}, i.e. the collisional terms.  Our starting point is
the collisional Boltzmann equation \eqref{eq:boltzmann_spherical}. We have to
find an expression for the term $(\delta f / \delta t)_{\text{enc}}$. This can
be done by approximating it with the Fokker-Planck equation, which requires
that the evolution of the stellar system is driven by uncorrelated distant
encounters.

The Fokker-Planck equation is a diffusion equation that describes the diffusion
of the phase space distribution function in position and velocity space. We
assume that the volume in which a stellar encounter takes place is small when
compared to the volume of the whole system. As a consequence, we can assume
that during an encounter only the velocity of the particle is modified, {\em
but not the position}. We thus neglect the diffusion of the phase space
distribution function in position space. This approach is usually referred to as the ``local
approximation''. Therefore, the right-hand side of equation
\eqref{eq:boltzmann_spherical} is

\begin{equation}\begin{split}
 \left(\frac{\delta f}{\delta t}\right)_{\text{enc}}= &-\sum^3_{i=1}\left[\frac{\partial}{\partial v_i}\left(f(\vec{x},\vec{v})D(\Delta v_i)\right)\right] \\
&+\frac{1}{2} \sum^3_{i,j=1}\bigg[\frac{\partial^2}{\partial v_i \partial v_j} \left(f(\vec{x},\vec{v})D(\Delta v_i \Delta v_j)\right)\bigg]
\end{split}\end{equation}

$D(\Delta v_i)$ and $D(\Delta v_i \Delta v_j)$ are the diffusion coefficients which depend on position and velocity coordinates. They determine the diffusion of the phase space distribution function in velocity space and describe the average change of the $i$-th component of velocity per unit time due to stellar
collisions. This is expressed by their dependence on the change of the $i$th
velocity component $\Delta v_i$. Note that there are no diffusion coefficients
that depend on $\Delta x_i$ as we are using the local approximation.  The
diffusion coefficients are \citep{rosenbluth1957}:

\begin{equation}\begin{split}
 D(\Delta v_i) &= 4 \pi G^2 m_f \ln \Lambda \frac{\partial }{\partial v_i} h(\vec{v})\\
 D(\Delta v_i \Delta v_i) &= 4 \pi G^2 m_f \ln \Lambda \frac{\partial^2 }{\partial v_i \partial v_j} g(\vec{v})
\end{split}\end{equation}
{$\ln \Lambda$ is the Coulomb logarithm, where $\Lambda$ is the ratio between the upper and lower cut-off impact parameter $b$ in a stellar collision.}
$h(\vec{v})$ and $g(\vec{v})$ are called the Rosenbluth potentials which are given by
\begin{equation}\begin{split}\label{eq:rosenbluth_potentials}
h(\vec{v})&=(m+m_f)\int \frac{f(\vec{v_f})}{|\vec{v}-\vec{v}_f|}\text{d}^3\vec{v}_f\\
g(\vec{v})&=m_f\int f(\vec{v_f})|\vec{v}-\vec{v}_f|\text{d}^3\vec{v}_f
\end{split}\end{equation}

\noindent
{Thus,}
$m$ denotes the mass of a test star that moves through a distribution 
$f(V_f,\mu_f)$ of field stars with a mass $m_f$. The FP equation then takes the form:

\begin{equation}\begin{split}\label{eq:FPgh}
 \left(\frac{\delta f}{\delta t}\right)_{\text{enc}}= &-4 \pi G^2 m_f \ln \Lambda \bigg[ \sum^3_{i=1} \frac{\partial}{\partial v_i}\left(f(\vec{v})\frac{\partial h}{\partial v_i}\right)\\
&-\frac{1}{2} \sum^3_{i,j=1} \frac{\partial^2}{\partial v_i \partial v_j} \left(f(\vec{v})\frac{\partial^2 g}{\partial v_i \partial v_j}\right)\bigg]
\end{split}\end{equation}

To compute this expression we need to know the phase space distribution
function $f(\vec{r},\vec{v},t)$. We approximate the distribution function by a
series expansion which accounts for the spherical symmetry of the system
\citep[see e.g.][]{rosenbluth1957,larson1970}. The expansion coefficients 
are expressed in terms of the velocity moments needed to compute the
collisional terms of the moment equations. We then compute
the phase space distribution function $f$ to calculate
the Rosenbluth potentials $h$ and $g$, the right-hand side of the FP
equation and thus the collisional terms of the Boltzmann equation.

\subsubsection{Construction of the distribution function}
\label{section:constr distr}

The phase space distribution function in spherical symmetry only depends on
$r$, $v_r$, $v^2_{\theta}+v^2_{\phi}$, and $t$, i.e.\
$f=f(r,v_r,v^2_{\theta}+v^2_{\phi},t)$, which implies that the system is axially
symmetric in the velocity space with axes $v_r$, $v_{\theta}$ and $v_{\phi}$.
The velocity components can be written in spherical coordinates

\begin{eqnarray}\label{velocities}
 v_r-\bar{v}_r &=& V\, \cos \theta' \nonumber \\
 v_{\theta} &=& V\,\sin \theta' \, \cos \phi' \\
 v_{\phi} &=& V\, \sin \theta' \, \sin \phi', \nonumber 
\end{eqnarray}

\noindent
where $V$ is the modulus of $\vec{v}$ , $\theta'$ the angle between $\vec{v}$
and the radial direction, and $\phi'$ the angle which defines the orientation
of the tangential component of $\vec{v}$ in the $v_{\theta}v_{\phi}$-plane.
Note that our model describes non-rotational spherically symmetric systems and
thus the mean tangential velocities are $\bar{v}_{\theta}=\bar{v}_{\phi}=0$.
Thus, equations \eqref{velocities} denote the components of the radial and
tangential velocities with respect to their means. As the phase space
distribution function $f$ only depends on $v^2_{\theta}+v^2_{\phi}$, it is
independent on the angle $\phi'$. We can henceforth omit the prime from angle
$\theta'$. 

Since we are operating in velocity space, in the following we refer to
$f$ as the (local) velocity distribution function (VDF). Substituting $\mu=\cos
\theta$ yields

\begin{equation}\label{spherical_axial}
 v_r-\bar{v}_r=V\mu, \qquad  v^2_t=v^2_{\theta}+v^2_{\phi}=V^2(1-\mu^2)
\end{equation}

These coordinates are appropriate for a series expansion of the VDF in Legendre
polynomials \citep{larson1970}:

\begin{equation}\begin{split}
 f(V,\mu)&= g(V) + \sum_{l=0}^{\infty} a_l(V) P_l(\mu)\\
         &=\sum^{\infty}_{l=0}A_l(V)P_l(\mu)
\end{split}\end{equation}

\noindent
where $A_0(V)=g(V)+a_0(V)$ and $A_l(V)=a_l(V)$ for $l\ge1$.
In this expansion 

\begin{displaymath}
 g(V) = \rho \frac{1}{\sqrt{2 \pi} \sigma^3} \exp(-\frac{V^2}{\sigma^2})
\end{displaymath}

Thus $g(V)V^2$ is the Maxwell-Boltzmann (MB) VDF. $P_l(\mu)$ are the Legendre
polynomials, and the functions $a_i(V)$ are defined by

\begin{displaymath}
 a_i(V) = g(V) \sum_{j=0}^{l_{max}} c_{ij} V^j
\end{displaymath}
where $l_{max}$ denotes the highest order of the Legendre polynomials $P_l(\mu)$ 
in the expansion of the VDF.
Due to axial symmetry in velocity space the VDF can only depend on powers of
$v_t$ and $v_r$. Using equations \eqref{spherical_axial} and fully expanding
the VDF we find the following constraints for the coefficients $c_{nm}$:

\begin{itemize}
 \item $n\le m$
 \item $n$ and $m$ are either both even or both odd
\end{itemize}
otherwise $c_{ij}= 0$. 

We obtain for \emph{model b} a VDF which extends to order $l=5$ in the Legendre Polynomials $P_l(\mu)$ which reads

\begin{equation}\begin{split}\label{expansion}
 f(V,\mu) =& \,g(V)+ g(V) (c_{00}+c_{02}V^2+c_{04}V^4)P_0(\mu)\\
&+g(V)(c_{11}V+c_{13}V^3+c_{15}V^5)P_1(\mu)\\
&+g(V)(c_{22}V^2+c_{24}V^4)P_2(\mu)\\
&+g(V)(c_{33}V^3+c_{35}V^5)P_3(\mu)+g(V)c_{44}V^4P_4(V)\\
&+g(V)V^5 c_{55}P_5(\mu)
\end{split}\end{equation}

\noindent
The VDF for \emph{model a} only extends to order $l=4$ and can be obtained
from equation \eqref{expansion} by setting all coefficients $c_{ij}$ with $j>4$
to zero. 

We can now calculate the coefficients $c_{ij}$ using the definition of the
centralized moments from equation \eqref{eq:central_moments_def}. However, we
first have to transform equation \eqref{eq:central_moments_def} to the new
coordinate system $(V,\mu)$. The volume element $\text{d}^3v$ in these
coordinates is written as:

\begin{equation}
 \textnormal{d}^3v = V^2\,\rm{d}V \, \rm{d}(\cos\theta) \,\rm{d}\phi = V^2\,\rm{d}V \, \rm{d}\mu \,\rm{d}\phi \quad \textnormal{where} \quad \mu=\cos\theta
\end{equation}

Thus we obtain for the centralized moments:

\begin{eqnarray}
 \langle n,m \rangle &=& \int \textnormal{d}^3v f\,(v_r-u)^n v_t^m \nonumber \\
&=& \int V^2\textnormal{d}V \,\textnormal{d}\mu \,\textnormal{d}\phi \,f\,(v_r-u)^n v_t^m \nonumber \\
&=& 2 \pi \int V^2\textnormal{d}V \textnormal{d}\mu\, f\,V^n \mu^n (V^2(1-\mu^2))^{m/2} \nonumber \\
&=& 2 \pi \int \textnormal{d}V \,\textnormal{d}\mu \,f\,V^{2+n+m} \mu^n (1-\mu^2)^{m/2}
\end{eqnarray}

We can obtain a linear system of equations to be solved for the coefficients
$c_{ij}$ by computing the different moments via this equation 
{with the expansion of the VDF from equation \eqref{expansion}.}
It must be noted that:

\begin{itemize}
\item The first centralized moment vanishes, since $\langle v_r-\bar{v}_r\rangle=\bar{v}_r- \bar{v}_r =0$, i.e.\:
\begin{equation}
 \langle 1,0 \rangle = 0 
\end{equation}
\item Since there are two tangential directions we add a factor 2 in the definition below
\begin{equation}
 2 p_t = \langle 0,2 \rangle
\end{equation}
\end{itemize}
We obtain for the VDF of \emph{model a} the coefficients $c_{ij}$:	
\begin{equation}\begin{split}\label{l4}
&c_{00}=\frac{27}{8}-\frac{7 (p_r+2p_t)}{4 \rho  \sigma ^2}+\frac{\kappa _r+2\kappa_{rt}+\kappa_t}{8 \rho  \sigma ^4}\\
&c_{02}=-\frac{7}{4 \sigma ^2}+\frac{p_r+2 p_t}{\rho  \sigma ^4}-\frac{\kappa _r+2\kappa_{rt}+\kappa_t}{12 \rho  \sigma ^6}\\
&c_{04}=\frac{1}{8 \sigma ^4}-\frac{p_r+2 p_t}{12 \rho  \sigma ^6}+\frac{\kappa _r+2\kappa_{rt}+\kappa_t}{120 \rho  \sigma ^8}\\
&c_{11}=-\frac{F_r+F_t}{2 \rho  \sigma ^4}\\
&c_{13}=\frac{F_r+F_t}{10 \rho  \sigma ^6}\\
&c_{22}=\frac{3 (p_r-p_t)}{2 \rho  \sigma ^4}-\frac{2\kappa_r+\kappa _{\text{rt}}-\kappa_t}{12\rho  \sigma ^6}\\
&c_{24}=-\frac{p_r-p_t}{6 \rho  \sigma ^6}+\frac{2\kappa_r+\kappa _{\text{rt}}-\kappa_t}{84
\rho  \sigma ^8}\\
&c_{33}=\frac{F_r-\frac{3}{2}F_t}{15 \rho  \sigma ^6}\\
&c_{44}=\frac{\frac{1}{3}\kappa _r-\kappa _{\text{rt}}+\frac{1}{8}\kappa_t}{35 \rho  \sigma ^8}
\end{split}\end{equation}

Since the coefficients $c_{ij}$ with $i=0,1$ only depend on sums of moments, we
can find a definition for total moments (see section \ref{total_moments}).

The role of anisotropy comes into the open when going to higher-order
coefficients, like $c_{2j} \propto (p_r-p_t) \propto a$, where $a =
1-{p_t}/{p_r}$ is the anisotropy parameter. We envisage a system as
isotropic in a ``weak'' sense if $a= 0$ everywhere. Strong isotropy holds if
the distribution has the strict dependence
$f=f(r,(v_r-\bar{v}_r)^2+(v^2_{\theta}+v^2_{\phi})^2,t)=f(r,V,t)$ on the
modulus of the velocity, which results in $F_r=F_t=0$ for the radial and
tangential energy flux, i.e. spherical symmetry in velocity space.  
We now compute the moments of fifth order with the VDF of \emph{model a}.
In this case the expansion in Legendre polynomials $P_l(\mu)$ expands up 
to $l=4$ 
{and the fifth order moments are}

\begin{equation}\label{GrGrtGt}
 G_r=10\sigma^2 F_r, \qquad G_{rt}=2\sigma^2F_r+3\sigma^2F_t, 
\qquad G_t=8\sigma^2 F_t
\end{equation}
{As we saw before, the system of differential moment equations 
\eqref{moment_equations_a} and \eqref{moment_equations_b} combined with 
the mass relation \eqref{Poisson} was not complete. We can now
close it by including the three relations in equation \eqref{GrGrtGt}. 
In these equations the fifth-order moments $G$ are expressed through 
lower-order moments. We now have a set of equations 
(\eqref{moment_equations_a}, \eqref{moment_equations_b}, \eqref{Poisson} and 
\eqref{GrGrtGt}) that is numerically solvable. This set describes our 
\emph{model a}}.

Combining the three relations we have

\begin{equation}\label{c55}
 G_r-5G_{rt}+\frac{15}{8}G_t=0
\end{equation}
We will see that the left-hand side of equation \eqref{c55} appears in the 
coefficient $c_{55}$ when the coefficients $c_{ij}$ of the VDF of 
\emph{model b} are computed. It then becomes clear that equation \eqref{c55} 
is a result of setting $c_{55}=0$ since $c_{44}$ is the highest coefficient 
of the VDF of \emph{model a}.

For the VDF of \emph{model b} we find
\begin{equation}\begin{split}
c_{00}=&\frac{27}{8}-\frac{7 (p_r+2p_t)}{4 \rho  \sigma ^2}+\frac{ (\kappa _r+2\kappa_{rt}+\kappa_t)}{8 \rho  \sigma ^4}\\
c_{02}=&-\frac{7}{4 \sigma ^2}+\frac{(p_r+2p_t)}{\rho  \sigma ^4}-\frac{(\kappa_r+2\kappa_{rt}+\kappa_t)}{12 \rho  \sigma ^6}\\
c_{04}=&\frac{1}{8 \sigma ^4}-\frac{(p_r+2p_t)}{12 \rho  \sigma^6}+\frac{(\kappa _r+2\kappa_{rt}+\kappa_t)}{120 \rho  \sigma ^8}\\
c_{11}=&-\frac{9 (F_r+F_t)}{4 \rho  \sigma ^4}+\frac{G_r+2G_{rt}+G_t}{8 \rho  \sigma ^6}\\
c_{13}=&\frac{4 (F_r+F_t)}{5 \rho  \sigma ^6}-\frac{G_r+2G_{rt}+G_t}{20 \rho  \sigma ^8}\\
c_{15}=&-\frac{F_r+F_t}{20 \rho  \sigma ^8}+\frac{G_r+2G_{rt}+G_t}{280 \rho  \sigma ^{10}}\\
c_{22}=&\frac{3 (p_r-p_t)}{2 \rho  \sigma ^4}-\frac{(2\kappa_r+\kappa _{\text{rt}}-\kappa_t)}{12 \rho  \sigma ^6}\\
c_{24}=&-\frac{(p_r-p_t)}{6 \rho \sigma ^6}+\frac{(2\kappa_r+\kappa _{\text{rt}}-\kappa_t)}{84 \rho  \sigma ^8}\\
c_{33}=&\frac{11 (F_r-\frac{3}{2}F_t)}{30 \rho  \sigma ^6}-\frac{G_r-\frac{1}{2}G_{rt}-\frac{3}{2}G_t}{30 \rho  \sigma ^8}\\
c_{35}=&-\frac{F_r-\frac{3}{2}F_t}{30 \rho  \sigma ^8}+\frac{G_r-\frac{1}{2}G_{rt}-\frac{3}{2}G_t}{270 \rho  \sigma ^{10}}\\
c_{44}=&\frac{\frac{1}{3}\kappa _r-\kappa _{\text{rt}}+\frac{1}{8}\kappa_t}{35 \rho  \sigma ^8}\\
c_{55}=&\frac{G_r-5G_{rt}+\frac{15}{8}G_t}{945 \rho  \sigma ^{10}}\\
\end{split}\end{equation}

We have the same dependencies of the coefficients $c_{ij}$ on the sum of
moments and relations that determine the degree of anisotropy, such as
$p_r-p_t$, $F_r-\frac{3}{2} F_t$.  As we predicted before, we can
obtain relation \eqref{c55} by setting $c_{55}=0$.  Similarly, the relation
$\frac{1}{3}\kappa_r-\kappa_{rt}+\frac{1}{8}\kappa_t=0$ obtained by calculating
the fourth order moments with the VDF used in \citet{spurzem1995} can be found in the
coefficient $c_{44}$ of the VDF for $l=4$ and $l=5$ again.

Computing the moments of order $n+m=6$ leads to the four equations:
\begin{equation}\begin{split}\label{Hrelations}
 	&H_r = 15 \rho \sigma^6 -45 \sigma^4 p_r+15 \sigma^2 \kappa_r\\
 	&H_{rt}= 6 \rho \sigma^6 - 12 \sigma^4 p_r - 6 \sigma^4 p_t + 2 \sigma^2 \kappa_r + 6 \sigma^2 \kappa_{rt}\\  
	&H_{tr} = 8 \rho \sigma^6 - 8 \sigma^4 p_r - 16 \sigma^4 p_t + 8 \sigma^2 \kappa_{rt} + \sigma^2 \kappa_t\\
 	&H_t = 48 \rho \sigma^6 - 144 \sigma^4 p_t + 18 \sigma^2 \kappa_t
\end{split}\end{equation}

These equations are the closure relations for \emph{model b}. 
{The complete set of equations of \emph{model b}
consists therefore of equations \eqref{moment_equations_a}, 
\eqref{moment_equations_b}, \eqref{moment_equations_c}, \eqref{Poisson} 
and \eqref{Hrelations}.}

 By means of equation \eqref{Hrelations} we also find the relation
 
\begin{equation}
 \frac{8}{15} H_r - 4 H_{rt} + 3 H_{tr} - \frac{1}{6}H_t = 0, 
\end{equation}
where the left hand side of this equation appears in the coefficient 
$c_{66}$ if we take the Legendre expansion of the VDF up to $l=6$.

\section{Weak isotropy, total moments and Rosenbluth potentials}

{In this section we identify different degrees of isotropy. 
They are specified by anisotropy parameters that can be found in the 
coefficients $c_{ij}$ of the VDF. }
{We start our discussion with the conducting gas sphere model 
of \citet{giersz1994, spurzem1996}. In these two studies the authors use a 
VDF of second order $l=2$ in Legendre polynomials $P_l(\mu)$ in order to 
compute the collisional terms of their model equations.}
\begin{equation}\label{eq:GS_vdf}
		f(V,\mu) = g(V) P_0(\mu)+ \frac{p_r-p_t}{2 \rho\sigma ^2}g(V) P_2(\mu)
	\end{equation}

\noindent {As we explained before, the definition of \emph{weak isotropy}
is}

\begin{equation}\begin{split}\label{weak_isotr}
	& p_r=p_t \\
	& F_r= \frac{3}{2} F_t,
\end{split}\end{equation}

\noindent
{This concept of isotropy includes second and third order moments.  
In the case of weak isotropy, the VDF becomes the MB distribution 
$g(V)$ since $P_0(\mu)=1$.}
{To generalize the definition of \emph{weak isotropy} we retake 
the MB VDF.  This VDF describes thermal equilibrium 
and is defined as}

	\begin{equation}\label{eq:vdf_l=2}
		f(V,\mu) = g(V) = \rho \frac{1}{\sqrt{2 \pi} \sigma^3} e^{-\frac{V^2}{\sigma^2}}
	\end{equation}
	We then compute the two moments of second order,
	\begin{equation}\begin{split}\label{pco}
	& p_r = \rho \sigma^2 \\
	& 2 p_t = 2 \rho \sigma^2 \quad
	\end{split}\end{equation}
	The factor 2 in front of $p_t$ accounts for two tangential directions. 
        We recover the known isotropy condition by dividing the second equation 
	by two and then subtracting the two resulting equations, 
	\begin{equation}\label{eq:p_isotropy}
		p_r-p_t = 0
	\end{equation}
	
        {For a spherical symmetric stellar system this relation 
	describes the highest degree of isotropy. We define the lowest-order 
	anisotropy parameter as}
	\begin{equation}
		a_p=p_r-p_t. 
	\end{equation}
	
        {Thus, computing second order moments with 
	a zeroth order VDF produces the two relations in equation 
	\eqref{pco}. This can be used to derive the isotropy condition 
	in equation \eqref{eq:p_isotropy} which appears as an 
	anisotropy parameter $a_p$ in the second order VDF of equation 
	\eqref{eq:GS_vdf}.}

        We can now recover an expression for the velocity dispersion  
        $\sigma$ by simply adding the two equations in \eqref{pco} and 
        solving for $\rho \sigma^2$:
	\begin{equation}
		\rho \sigma^2 = \frac{p_r+2 p_t}{3}
	\end{equation}

The random kinetic energy $e$ is defined as $e = (p_r+2 p_t)/2$; then, applying
the isotropic condition $p'=p_r=p_t$, we find that
$e=\frac{3}{2}\,p'=\frac{3}{2}\,\rho \sigma^2$. This is the equipartition
theorem for $f=3$ degrees of freedom, which states that in thermal equilibrium
at a temperature $T$ every degree of freedom contains the same amount of
average energy $e_i= \frac{1}{2} k_{\mathrm{B}} T \hat{=} \frac{1}{2} \rho
\sigma^2$. 

{In order to find isotropy relations for higher-order 
moments, we use a second- and fourth-order VDF. The fourth-order VDF was 
computed in the previous section. The second-order VDF is }

\begin{equation}\begin{split}
f(V,\mu) = &g(V)+ g(V) (c_{00}+c_{02}V^2)P_0(\mu)\\
   &+g(V)c_{11}V P_1(\mu)+g(V)c_{22}V^2P_2(\mu)
\end{split}\end{equation}
	
We determine the coefficients $c_{ij}$ and then compute the fourth order moments $\kappa$:

\begin{equation}\begin{split}\label{kappaco}
& \kappa_r = -3 \rho \sigma^4 + 6 \sigma^2 p_r \\
& \kappa_{rt} = -2\rho \sigma^4 + 2 \sigma^2 p_r +2 \sigma^2 p_t \\
& \kappa_{t} = -8 \rho \sigma^4 + 16 \sigma^2 p_t
\end{split}\end{equation}

{We can assume that these relations constrain our degree of anisotropy, 
since the information contained in higher-order moments can be expressed by lower-order 
moments. Similarly, we use now these relations to compute isotropy conditions 
that reappear as linear combinations of the $\kappa$'s in the coefficients $c_{22}$, 
$c_{24}$ and $c_{44}$ of the fourth-order VDF. These linear combinations should vanish 
in case of isotropy and thus can be identified as the anisotropy parameters of fourth 
order.}

{For the linear combination of $\kappa$'s in the coefficient
$c_{44}$ we directly find by inserting equation \eqref{kappaco}}

\begin{equation}
	 \kappa_r- 3 \kappa_{rt} + \frac{3}{8} \kappa_t = 0
\end{equation}
{For the linear combination of the $\kappa$'s in $c_{22}$ and 
$c_{24}$ we obtain in the same way}
\begin{equation}	
	2 \kappa_r + \kappa_{rt} - \kappa_t = 0 \qquad \Leftrightarrow \qquad p_r = p_t
\end{equation}
{If the conditions for weak isotropy defined in equation 
\eqref{weak_isotr} hold as well, the fourth-order VDF only depends on the 
Legendre polynomials $P_0(\mu)$ and $P_1(\mu)$ and sums of moments $p$, 
$F$ and $\kappa$.}

{We thus conclude that linear combinations of moments in 
the coefficients $c_{ij}$ for $2\le i$ describe anisotropy parameters. 
The order of an anisotropy parameter is equal to the order of moments it 
consists of. The degree of isotropy is hence determined by the lowest-order 
anisotropy parameters that vanish. We therefore introduce a new definition for 
weak isotropy by requiring that all anisotropy parameters vanish, which
corresponds to demanding that the VDF 
depends only on the Legendre polynomials $P_0(\mu)$ and $P_1(\mu)$.}

	\subsection{Total moments}\label{total_moments}
	
We are now in position to define the total centralized moments of the 
VDF, since this has been totally determined.
	\begin{equation}\begin{split}\label{vmoments}	
		\langle v^n \rangle &= \langle \left((v_r-\bar{v}_r)^2+ v_{\phi}^2 + v_{\theta}^2 \right)^{\frac{n}{2}} \rangle\\
				    &= \langle \left( \mu^2 V^2 + V^2 (1- \mu^2) \right)^{\frac{n}{2}} \rangle = \langle V^n \rangle \\
				    &= 2 \pi \int_0^{\infty}  \int_{-1}^1\, \, V^{n+2}\, \,f(\mu,V)\, \, \mathrm{d}\mu \, \, \mathrm{d}V, 
	\end{split}\end{equation}
	since $\bar{v}_{\theta}=\bar{v}_{\phi}=0$. 
        With the help of \eqref{vmoments} we calculate the even moments, which we define as $p$, $\kappa$ and $H$,
	\begin{equation}\begin{split}\label{randommoments}
		 p &= \langle V^2 \rangle = p_r + 2 p_t \\
		 \kappa &= \langle V^4 \rangle= \kappa_r + \kappa_{rt} + \kappa_t \\
		 H &= \langle V^6 \rangle=	105 (\rho \sigma^6 - \sigma^4 (p_r+2 p_t)) + 21 \sigma^2 (\kappa_r + \kappa_{rt} + \kappa_t) \\
		   &= H_r + 3 (H_{rt} + H_{tr}) + H_t 
	\end{split}\end{equation}
	
\noindent
In the last line we have employed the relations given by \eqref{Hrelations}. 
With $p$ and $\kappa$ and VDF up to order $l=4$ we find
	\begin{equation}\begin{split}
		\langle V^1 \rangle &= \sigma \frac{1}{\sqrt{2 \pi}}\left(\frac{9}{2}\rho-\frac{3}{10}\rho\frac{\kappa}{p^2}\right)\\
		\langle V^3 \rangle &= \sigma \sqrt{\frac{2}{\pi}} \left(\frac{5}{3} p + \frac{3}{5} \rho \frac{\kappa}{p} \right)\\
		\langle V^5 \rangle &= \sigma  \sqrt{\frac{8}{\pi}} \left(-\frac{7}{3} \frac{p^2}{\rho} + 3 \kappa \right),
	\end{split}\end{equation}
which is independent of the uneven moments $F_r$, $F_t$, 
$G_r$, $G_{rt}$ and $G_t$. We thus define the uneven total moments 
	\begin{equation}\begin{split}
		F &= \frac{1}{2}(F_r + F_t)\\
		G &= G_r + 2 G_{rt} + G_t
	\end{split}\end{equation}
{The factor $1/2$ in the definition of $F$ is chosen 
in order to obtain consistency with the physical interpretation of 
$F$. This becomes clear when we add the two differential equations 
for the radial and tangential pressure $p_r$ and $p_t$ in equation 
\eqref{moment_equations_a}, where we find that $F=\frac{1}{2}(F_r + F_t)$ 
corresponds to the radial flux of random kinetic energy.} 

With these definitions our the coefficients of the VDF $c_{ij}$ for 
$i=0,1$ now only depend on total moments. 	

\subsection{Rosenbluth Potentials}

After having calculated the expansion coefficients for the VDF $f(V,\mu)$ in 
section \ref{section:constr distr}, we now can calculate
the Rosenbluth potentials, given by

\begin{equation}\begin{split}\label{eq:rosenbluth_hg}
 h(V,\mu)&=(m+m_f)\,\int\limits_{0}^{2\pi}\int\limits_{-1}^{1}\int\limits_{0}^{V} \,\frac{f(V_f,\mu_f)}{|\vec{v}-\vec{v}_f|} \,V^2_f \,\text{d}V_f\, \text{d}\mu_f \,\text{d}\phi  \\
 g(V,\mu)&=m_f\, \int\limits_{0}^{2\pi}\int\limits_{-1}^{1}\int\limits_{0}^{V} \,f(V_f,\mu_f) |\vec{v}-\vec{v}_f|\, V^2_f \, \text{d}V_f\, \textnormal{d}\mu_f \,\text{d}\phi
\end{split}\end{equation}
So as to integrate for $h(V,\mu)$ we can make a multi-pole expansion, i.e. 
\begin{equation}\label{eq:trick_1}
 \frac{1}{|\vec{v}-\vec{v}_f|}= \sum^{\infty}_{l=0}\sum^l_{m=-l} \frac{v^l_{<}}{v^{l+1}_{>}} \, \frac{4 \pi}{2l+1} Y^{\star}_{l,m}(\theta,\phi) Y_{l,m}(\theta_f,\phi_f) 
\end{equation}
 where 
\begin{equation} 
\begin{split}
v_{<} & = min(v,v') \\
v_{>} & = max(v,v'),
\end{split}
\end{equation}

\noindent
and the spherical harmonics are defined in the usual way,

\begin{equation}
\begin{split}
 Y_{l,m}(\theta,\phi)    & = \sqrt{\frac{2l+1}{4 \pi} \frac{(l-m)!}{(l+m)!}} \,P^m_l(\cos(\theta))\, e^{im\phi}\\
 Y_{l,-|m|}(\theta,\phi) & =(-1)^{m}\,Y^{\star}_{l,|m|}(\theta,\phi),
\end{split}
\end{equation}
with the associated Legendre polynomials $P^m_l(\cos \theta)$. 

We use $\mu =\cos \theta$ and insert equation \eqref{eq:trick_1} into
the Rosenbluth potential $h(V,\mu)$ of equation \eqref{eq:rosenbluth_hg}. After
integrating over $\phi$, the associated Legendre polynomials are reduced to
Legendre polynomials $P_l(\mu)$ and we hence can apply the orthogonality relation

\begin{equation}\label{eq:orthogonality}
 \int^{1}_{-1} P_l(\mu_f)P_k(\mu_f) \textnormal{d}\mu_f = \delta_{kl} \, \frac{2}{2l+1}
\end{equation}

To compare our results with the lower-order estimation of
\citet{spurzem1995}, we adopt their notation for the integrals over $V$,

\begin{equation}\begin{split}\label{eq:I_n,K_n}
 I_n&=\int^V_0V^n_f\,g(V_f)\, \text{d}V_f\\
 K_n&=\int^{\infty}_VV^n_f\,g(V_f)\, \text{d}V_f
\end{split}\end{equation}

With a VDF of order $l=5$ we obtain the Rosenbluth potential $h(V,\mu)$:

\begin{equation}
\begin{split}\
&\frac{h(V,\mu)}{4 \pi(m+m_f)}=\\
&\bigg[\Big(\frac{I_2}{V}+K_1\Big)(1+c_{00})+\Big(\frac{I_4}{V}+K_3\Big)c_{02}+\Big(\frac{I_6}{V}+K_5\Big)c_{04}\bigg]P_0(\mu)\\
&+\bigg[\Big(\frac{I_4}{3 V^2}+\frac{1}{3}V K_1\Big)c_{11} +\Big(\frac{I_6}{3 V^2}+\frac{1}{3}V K_3\Big)c_{13}+\\
& \Big(\frac{I_8}{3 V^2}+\frac{1}{3}V K_5\Big)c_{1,5}\bigg] P_1(\mu)
+\bigg[\Big(\frac{I_6}{5 V^3}+\frac{1}{5}  V^2 K_1\Big) c_{22}+\\
&\Big(\frac{I_8}{5 V^3}+\frac{1}{5}  V^2 K_3\Big) c_{24}\bigg]P_2(\mu)
+\bigg[\Big(\frac{I_8}{7 V^4}+\frac{1}{7}  V^3 K_1\Big) c_{33}+\\
&\Big(\frac{I_{10}}{7 V^4}+\frac{1}{7}  V^3 K_3\Big) c_{35}+\frac{I_{10} }{7 V^4}+\frac{1}{7} V^3 K_3 \bigg]P_3(\mu)\\
&+\bigg[\Big(\frac{I_{10}}{9 V^5}+\frac{1}{9}  V^4 K_1\Big) c_{44}\bigg]P_4(\mu)+\\
&\bigg[\frac{I_{12}}{11 V^6}+\frac{1}{11}  V^5 K_1\bigg] c_{55} P_5(\mu),
\end{split}
\end{equation}
{If we set all coefficients $c_{ij}=0$ with the exception of $c_{22}$, we recover the Rosenbluth potential $h(V,\mu)$ from \citet{giersz1994}. This confirms the correctness of our result. Moreover, we obtain the Rosenbluth potential $h(V,\mu)$ for order $l=4$ by setting the coefficients $c_{15}=c_{55}=0$.}

To calculate $g(V,\mu)$ we write
\begin{equation}\label{eq:trick_2}
 |\vec{v}-\vec{v}_f|=\frac{(|\vec{v}-\vec{v}_f|)^2}{|\vec{v}-\vec{v}_f|} = \frac{(V^2+V^2_f-2VV_f\cos\chi)}{|\vec{v}-\vec{v}_f|},
\end{equation} 
where $\chi$ is the angle between the vectors $\vec{v}$ and $\vec{v}_f$. This can be rewritten in terms of the angles $\theta$, $\phi$, $\theta_f$, $\phi_f$ with the general formula for Legendre polynomials
\begin{equation}\label{eq:cos}
 P_l(\cos\chi)= \sum^{l}_{m=-l} \frac{(l-|m|)!}{(l+|m|)!}\,P^{|m|}_l(\cos\theta)P^{|m|}_l(\cos\theta_f)e^{-im(\phi-\phi_f)}
\end{equation}
Setting $l=1$ we can substitute $\cos \chi$ in equation \eqref{eq:trick_2} with
$P_l(cos\chi)$ from equation \eqref{eq:cos}and insert the result
into the Rosenbluth potential $g(V,\mu)$ of equation \eqref{eq:rosenbluth_hg}.
This leads to an expression for $g(V,\mu)$ depending on products of Legendre
and associated Legendre polynomials. After carrying out the integration over
$\phi$ we use relations between the Legendre and associated Legendre
polynomials that reduce the products and enable us to apply the orthogonality
relation \eqref{eq:orthogonality}. We can now write the result in the notation
of \citet{spurzem1995} to verify that their lower order potential $g(V,\mu)$
is contained in our result by using a VDF up to order $l=5$;

\begin{equation}\begin{split}
&\frac{g(V,\mu)}{4\pi m_f}=\\
&\bigg[\Big(V I_2+\frac{1}{3 V}I_4+\frac{V^2}{3} K_1+K_3\Big)(1+c_{00})+\\
&\Big(V I_4+\frac{1}{3 V}I_6+\frac{V^2}{3} K_3+K_5\Big) c_{02}+\\
&\Big(V I_6+\frac{1}{3 V}I_8+\frac{V^2}{3} K_5+K_7\Big) c_{04}\bigg]P_0(\mu)\\
&+\bigg[\Big(-\frac{1}{3} I_4+\frac{I_6}{15 V^2}+\frac{1}{15} V^3 K_1-\frac{1}{3} V K_3\Big) c_{11}+\\
&\Big(-\frac{1}{3} I_6+\frac{I_8}{15 V^2}+\frac{1}{15} V^3 K_3-\frac{1}{3} V K_5\Big) c_{13}\\
&+\Big(-\frac{1}{3} I+\frac{I_10}{15 V^2}+\frac{1}{15} V^3 K_5-\frac{1}{3} V K_7\Big) c_{15}\bigg]P_1(\mu)\\
&+\bigg[\Big(-\frac{I_6}{15 V}+\frac{I_8}{35 V^3}+\frac{1}{35} V^4 K_1-\frac{1}{15} V^2 K_3\Big) c_{22}\\
&+\Big(-\frac{I_8}{15 V}+\frac{I_{10}}{35V^3}+\frac{1}{35} V^4 K_3-\frac{1}{15} V^2 K_5\Big) c_{24}\bigg]P_2(\mu)\\
&+\bigg[\Big(-\frac{I_8}{35 V^2}+\frac{I_{10}}{63 V^4}+\frac{V^5}{63} K_1-\frac{V^3}{35} K_3\Big) c_{33}+\\
&\Big(-\frac{I_{10}}{35 V^2}+\frac{I_{12}}{63 V^4}+\frac{V^5}{63} K_3-\frac{V^3}{35} K_5\Big) c_{35}\bigg]P_3(\mu)\\
&+\bigg[\Big(-\frac{I_{10}}{63 V^3}+\frac{I_{12}}{99 V^5}+\frac{V^6}{99} K_1-\frac{V^4}{63} K_3 \Big)c_{44}\bigg]P_4(\mu)\\
&+\bigg[-\frac{I_{12}}{99 V^4}+\frac{I_{14}}{143 V^6}+\frac{V^7}{143} K_1-\frac{V^5}{99} K_3\bigg]c_{55}\,P_5(\mu)
\end{split}\end{equation}
{We again can set all coefficients $c_{ij}=0$ with the exception of $c_{22}$. 
We find that this leads to the second-order result of \citet{spurzem1995}, which corroborates
the correctness of our result for $g(V,\mu)$. The fourth-order Rosenbluth potential $g(V,\mu)$ 
can be recovered by setting the coefficients $c_{15}=c_{55}=0$.}

Eventually we carry out the integration over $V$ for both Rosenbluth 
potentials $h(V,\mu)$ and $g(V,\mu)$ which is needed for the further 
computation of the right-hand sides of the moment equations. 

\section{Collision terms}\label{section: collision terms}
With the coordinates in velocity space $V$ and $\mu$ the Fokker-Planck 
equation \eqref{eq:FPgh} transforms to \citep[from][]{rosenbluth1957}

\begin{align}
\label{rhsFP}
&\frac{1}{\Gamma}\bigg(\frac{\delta f(V,\mu)}{\delta t}\bigg)_{\text{enc}} =\\ \nonumber
&-\frac{1}{V^2}\frac{\partial}{\partial V}\bigg(f(V,\mu) V^2 \frac{\partial h(V,\mu)}{\partial V}V\bigg)\\ \nonumber
&-\frac{1}{V^2}\frac{\partial }{\partial \mu}\bigg(f(V,\mu) \bigg(1-\mu^2\bigg) \frac{\partial h(V,\mu )}{\partial \mu} \bigg)\\ \nonumber
&+\frac{1}{2 V^2}\frac{\partial^2}{\partial V^2}\bigg(f(V,\mu) V^2 \frac{\partial^2 g(V,\mu)}{\partial V^2}\bigg)\\ \nonumber
&+\frac{1}{2 V^2}\frac{\partial^2}{\partial \mu^2}\bigg(f(V,\mu) \bigg(\frac{(1-\mu ^2)^2}{V^2}\frac{\partial^2 g(V,\mu)}{\partial \mu^2}\\ \nonumber
&+\frac{(1-\mu ^2)}{V}\frac{\partial g(V,\mu)}{\partial V}-\mu \frac{(1-\mu ^2)}{V^2}\frac{\partial g(V,\mu)}{\partial \mu}\bigg)\bigg)\\ \nonumber
&+\frac{1}{V^2} \frac{\partial^2}{\partial V \partial \mu}\bigg(f(V,\mu ) (1-\mu ^2)\bigg(\frac{\partial^2 g(V,\mu)}{\partial V \partial \mu}-\frac{1}{V} \frac{\partial g(V,\mu)}{\partial \mu}\bigg) \bigg)\\ \nonumber
&+\frac{1}{2 V^2} \frac{\partial}{\partial V}\bigg(f(V,\mu) \bigg(-\frac{(1-\mu ^2)}{V} \frac{\partial^2 g(V,\mu)}{\partial\mu^2}-2 \frac{\partial g(V,\mu)}{\partial V}\\ \nonumber
&+2 \frac{\mu }{V} \frac{\partial g(V,\mu)}{\partial \mu} \bigg)\bigg) \\ \nonumber
&+\frac{1}{2 V^2} \frac{\partial }{\partial \mu}\bigg(f(V,\mu) \bigg(\mu  \frac{(1-\mu ^2)}{V^2} \frac{\partial^2 g(V,\mu)}{\partial \mu^2}\\ \nonumber
&+2 \frac{\mu }{V} \frac{\partial g(V,\mu)}{\partial V}+2 \frac{(1-\mu ^2)}{V}\frac{\partial^2 g(V,\mu)}{\partial V \partial \mu}-\frac{2}{V^2} \frac{\partial g(V,\mu)}{\partial \mu}\bigg)\bigg),
\end{align}

\noindent
where $\Gamma=4 \pi G^2 m_f \ln \Lambda$ and $\ln \Lambda$ is the 
Coulomb logarithm.To obtain the collision terms of the moment equations
 \eqref{moment_equations_a}, \eqref{moment_equations_b} and 
\eqref{moment_equations_c}, we multiply the FP equation with powers of 
velocity components and integrate over velocity space
\begin{equation}\begin{split}
&\left(\frac{\delta \langle n,m \rangle}{\delta t}\right)_{\text{enc}} = \int \textnormal{d}^3v\,\left(\frac{\delta f(V,\mu)}{\delta t}\right)_{\text{enc}} (v_r-\bar{v}_r)^n\,v^m_t \\
&= 2 \pi \int V^2\textnormal{d}V \textnormal{d}\mu\, \left(\frac{\delta f(V,\mu)}{\delta t}\right)_{\text{enc}}\,V^n \mu^n (V^2(1-\mu^2))^{m/2} \\
&= 2 \pi \int \textnormal{d}V \,\textnormal{d}\mu \,\left(\frac{\delta f(V,\mu)}{\delta t}\right)_{\text{enc}}\,V^{2+n+m} \mu^n (1-\mu^2)^{m/2}
\end{split}\end{equation}

\noindent
For a single-mass model $m_f=m$ and some collisional terms must vanish.
These are the particle density $\rho$, due to
particle/mass conservation, the collision term of the bulk velocity $u$ (or
$\rho u$ since internal collisions do not disturb the motion of the barycenter)
and the collisional term for the energy density defined as $e=(p_r+2p_t)/2$, due
to energy conservation;

\begin{equation}\begin{split}
 &\left(\frac{\delta \rho}{\delta t}\right)_{\text{enc}} = 0 , \quad \left(\frac{\delta u}{\delta t}\right)_{\text{enc}}=0 , \\
 &\left(\frac{\delta e}{\delta t}\right)_{\text{enc}}  =\left(\frac{\delta p_r}{\delta t}\right)_{\text{enc}}  +2\left(\frac{\delta p_t}{\delta t}\right)_{\text{enc}} = 0,
\end{split}\end{equation}

\noindent
as expected, which proves that our calculations are right.
We define the
anisotropy parameters, which appear in the coefficients $c_{ij}$ as

\begin{equation}\begin{split}\label{set:anisotropy_param} 
	&a_p = p_r-p_t \\
	&a_F = 2F_r-3 F_t \\
	&a_{\kappa1} = 2 \kappa_r + \kappa_{rt} -\kappa_t \\
	&a_{\kappa2} = 8\kappa_r - 24\kappa_{rt} +3\kappa_t\\
	&a_{G1} = 2 G_r -G_{rt} -3 G_t\\
	&a_{G2} = 8 G_r - 40 G_{rt} +15 G_t
\end{split}\end{equation}
and use the total moments
\begin{equation}\begin{split}\label{set:total_moments}
	& F= \frac{1}{2} (F_r+F_t)\\
	& \kappa= \kappa_r + 2\kappa_{rt} +\kappa_t\\
	& G = G_r + 2G_{rt} + G_t
\end{split}\end{equation}
We then give the collisional terms for the two models \emph{a} and 
\emph{b} in appendix \ref{app:collisional_terms}. 

\section{The velocity distribution function}
\label{sec:distribution_funktion} 

In this section we investigate the influence of moments and anisotropy 
parameters on the VDF. For that, we use a VDF with moments up to fifth 
order. We express the VDF with the total moments $F$,$\kappa$ and
$G$ and anisotropy parameters $a_p$, $a_F$, $a_{\kappa1,2}$ and $a_{G1,2}$,

\begin{equation}\begin{split}\label{distr}
&f(V,\mu)=g(V)\\
&+g(V)\bigg(-\frac{15}{8}+\frac{5 V^2 }{4 \sigma ^2} -\frac{V^4 }{8 \sigma ^4}\\
&+\frac{\kappa  }{8 \rho  \sigma ^4}-\frac{V^2 \kappa}{12 \rho  \sigma ^6}+\frac{V^4 \kappa }{120 \rho  \sigma ^8}\bigg)\,\,P_0(\mu)\\
	&+g(V)\bigg(-\frac{9 V F}{2 \rho  \sigma ^4}+\frac{8 V^3 F}{5 \rho  \sigma ^6}-\frac{V^5 F}{10 \rho  \sigma ^8}\\
&+\frac{V G}{8 \rho  \sigma ^6}-\frac{V^3 G}{20 \rho  \sigma ^8}+\frac{V^5 G}{280 \rho  \sigma ^{10}}\bigg)\,\,P_1(\mu)\\
	&+g(V)\bigg(\frac{3 V^2 a_p}{2 \sigma ^2}-\frac{V^4  a_p}{6 \sigma ^4}-\frac{V^2 a_{\text{$\kappa $1}}}{12 \rho  \sigma^6}+\frac{V^4  a_{\text{$\kappa $1}}}{84 \rho  \sigma ^8}\bigg)\,\,P_2(\mu)\\
	&+g(V)\bigg(\frac{11 V^3 a_F}{60 \rho  \sigma ^6}-\frac{V^5 a_F }{60 \rho  \sigma ^8}-\frac{V^3 a_{G1}}{30 \rho  \sigma ^8}+\frac{V^5 a_{G1}}{540 \rho  \sigma ^{10}}\bigg)\,\,P_3(\mu)\\
	&+g(V)\frac{V^4 a_{\text{$\kappa $2}}}{840 \rho  \sigma ^8}\,\,P_4(\mu)+g(V)\frac{V^5 a_{\text{G2}}}{7560 \rho  \sigma ^{10}}\,\,P_5(\mu)
\end{split}\end{equation}

In order to obtain the MB VDF in the case of thermal equilibrium, 
$g(V)V^2$, we have to multiply $f(V,\mu)$ with $V^2$. In the figures, 
the $V$-axis denotes the modulus of the velocity and the $\mu$-axis 
the direction of the velocity vector. When $\mu=0$, the radial velocity 
component is $v_r=\mu V = 0$ and the tangential velocity component is 
$v_t=\sqrt{V^2(1-\mu^2)}=V$ and vice-versa for $\mu=1$.  The $z$-axis 
indicates the phase space probability density $f(V,\mu)$.  If not stated 
otherwise, we choose $\sigma=10 \text{km}\,\text{s}^{-1}\text{pc}^{-3}$ 
and $\kappa=150 \,000 \text{km}^4 \text{s}^{-4}\text{pc}^{-3}$. We 
normalize $f(V,\mu)$ by setting the particle density $\rho=1\text{pc}^
{-3}$, and then $\int f(V,\mu) V^2 \mathrm{d} V \mathrm{d} \mu = 1$. 
We set to zero the values of the moments $F$ and $G$ and the anisotropy 
parameters $a_p$, $a_F$, $a_{\kappa1,2}$ and $a_{G1,2}$.  To emphasize 
the effects of moments and anisotropy parameters, we choose very high 
and low values for these quantities in some plots.  This results in 
negative values of the distribution function which is unphysical but 
reflects the polynomial ansatz of the truncated series expansion of 
the VDF.  We explore the parameter space to analyze their influence 
on the VDF.

  \begin{table*}
	\centering
  \begin{tabular}{|c  c | c | c | c | c | c | c | c | c | c |}
    \hline
           &  & $a_p$ & $F$ & $a_F$ & $\kappa$ & $a_{\kappa1}$ & $a_{\kappa2}$ & $G$ & $a_{G1}$ & $a_{G2}$ \\ 
           & & $\left[\frac{\displaystyle\text{km}^2}{\displaystyle\text{s}^2\text{pc}^{3}}\right]$&$\left[\frac{\displaystyle\text{km}^3}{\displaystyle\text{s}^3\text{pc}^{3}}\right]$&$\left[\frac{\displaystyle\text{km}^3}{\displaystyle\text{s}^3\text{pc}^{3}}\right]$&$\left[\frac{\displaystyle\text{km}^4}{\displaystyle\text{s}^4\text{pc}^{3}}\right]$&$\left[\frac{\displaystyle\text{km}^4}{\displaystyle\text{s}^4\text{pc}^{3}}\right]$&$\left[\frac{\displaystyle\text{km}^4}{\displaystyle\text{s}^4\text{pc}^{3}}\right]$&$\left[\frac{\displaystyle\text{km}^5}{\displaystyle\text{s}^5\text{pc}^{3}}\right]$&$\left[\frac{\displaystyle\text{km}^5}{\displaystyle\text{s}^5\text{pc}^{3}}\right]$&$\left[\frac{\displaystyle\text{km}^5}{\displaystyle\text{s}^5\text{pc}^{3}}\right]$ \\ \hline
   FIG. 2. & \emph{left} & 0 & 0 & 0 & {0} & 0 & 0 & 0 & 0 & 0 \\ 
           & \emph{right}& 0 & 0 & 0 & $1.5\cdot10^5$ & 0 & 0 & 0 & 0 & 0 \\ \hline 
   FIG. 3. & \emph{left} & {$-5$} & 0 & 0 & $1.5\cdot10^5$ & 0 & 0 & 0 & 0 & 0 \\ 
           & \emph{right}& {$3.3$} & 0 & 0 & $1.5\cdot10^5$ & 0 & 0 & 0 & 0 & 0 \\ \hline 
   FIG. 4. & \emph{left} & 0 & {$-2.5 \cdot 10^2$} & 0 & $1.5\cdot10^5$ & 0 & 0 & 0 & 0 & 0 \\ 
           & \emph{right}& 0 & {$3.5 \cdot 10^2$} & 0 & $1.5\cdot10^5$ & 0 & 0 & 0 & 0 & 0 \\ \hline 
   FIG. 5. & \emph{left} & 0 & $1.0 \cdot 10$ & {$-1.3 \cdot 10^3$} & $1.5\cdot10^5$ & 0 & 0 & 0 & 0 & 0 \\ 
           & \emph{right}& 0 & $1.0 \cdot 10^2$ & {$2.0 \cdot 10^3$} & $1.5\cdot10^5$ & 0 & 0 & 0 & 0 & 0 \\ \hline 
   FIG. 6. & \emph{left} & 0 & 0 & 0 & {$8.0\cdot10^4$} & 0 & 0 & 0 & 0 & 0 \\ 
           & \emph{right}& 0 & 0 & 0 & {$2.0\cdot10^5$} & 0 & 0 & 0 & 0 & 0 \\ \hline 
   FIG. 7. & \emph{left} & 0 & 0 & 0 & $1.5\cdot10^5$ & {$-7.5 \cdot 10^4$} & 0 & 0 & 0 & 0 \\ 
           & \emph{right}& 0 & 0 & 0 & $1.5\cdot10^5$ & {$1.35 \cdot 10^5$} & 0 & 0 & 0 & 0 \\ \hline 
   FIG. 8. & \emph{left} & 0 & 0 & 0 & $1.5\cdot10^5$ & 0 & {$-4.5 \cdot 10^5$} & 0 & 0 & 0 \\ 
           & \emph{right}& 0 & 0 & 0 & $1.5\cdot10^5$ & 0 & {$6.0 \cdot 10^5$} & 0 & 0 & 0 \\ \hline 
   FIG. 9. & \emph{left} & 0 & 0 & 0 & $1.5\cdot10^5$ & 0 & 0 & {$-8.0 \cdot 10^5$} & 0 & 0 \\ 
           & \emph{right}& 0 & 0 & 0 & $1.5\cdot10^5$ & 0 & 0 & {$1.0 \cdot 10^6$} & 0 & 0 \\ \hline 
   FIG. 10.& \emph{left} & 0 & 0 & 0 & $1.5\cdot10^5$ & 0 & 0 & $2.0 \cdot 10^2$ & {$-1.14 \cdot 10^6$} & 0 \\ 
           & \emph{right}& 0 & 0 & 0 & $1.5\cdot10^5$ & 0 & 0 & $2.0 \cdot 10^2$ & {$2.0 \cdot 10^6$} & 0 \\ \hline 
   FIG. 11.& \emph{left} & 0 & 0 & 0 & $1.5\cdot10^5$ & 0 & 0 & $2.0 \cdot 10$ & 0 & {$2.0 \cdot 10^{-3}$} \\ 
           & \emph{right}& 0 & 0 & 0 & $1.5\cdot10^5$ & 0 & 0 & $2.0 \cdot 10$ & 0 & {$1.2 \cdot 10^{-3}$} \\ \hline 
  \end{tabular}
  \caption{The table gives an overview over the different values of 
the total moments and anisotropy parameters in each plot. The values 
for the density $\rho= 1\, \text{pc}^{-3}$ and the velocity dispersion 
$\sigma = 10\,\, \text{km}\, \text{s}^{-1}\,\text{pc}^{-3}$ are constant 
over all plots and therefore do not appear in the table. 
\label{tab:parameter_set}}
  \end{table*}

In order not to clutter the figures with the values for the set of 
parameters listed in equations \eqref{set:anisotropy_param} and 
\eqref{set:total_moments} the values for the plots are collected 
in table \ref{tab:parameter_set}. The parameters that change 
the VDF with respect to the MB distribution are denoted
in the each plot.

\begin{figure*}
		\centering
			\includegraphics[width=8cm,clip]{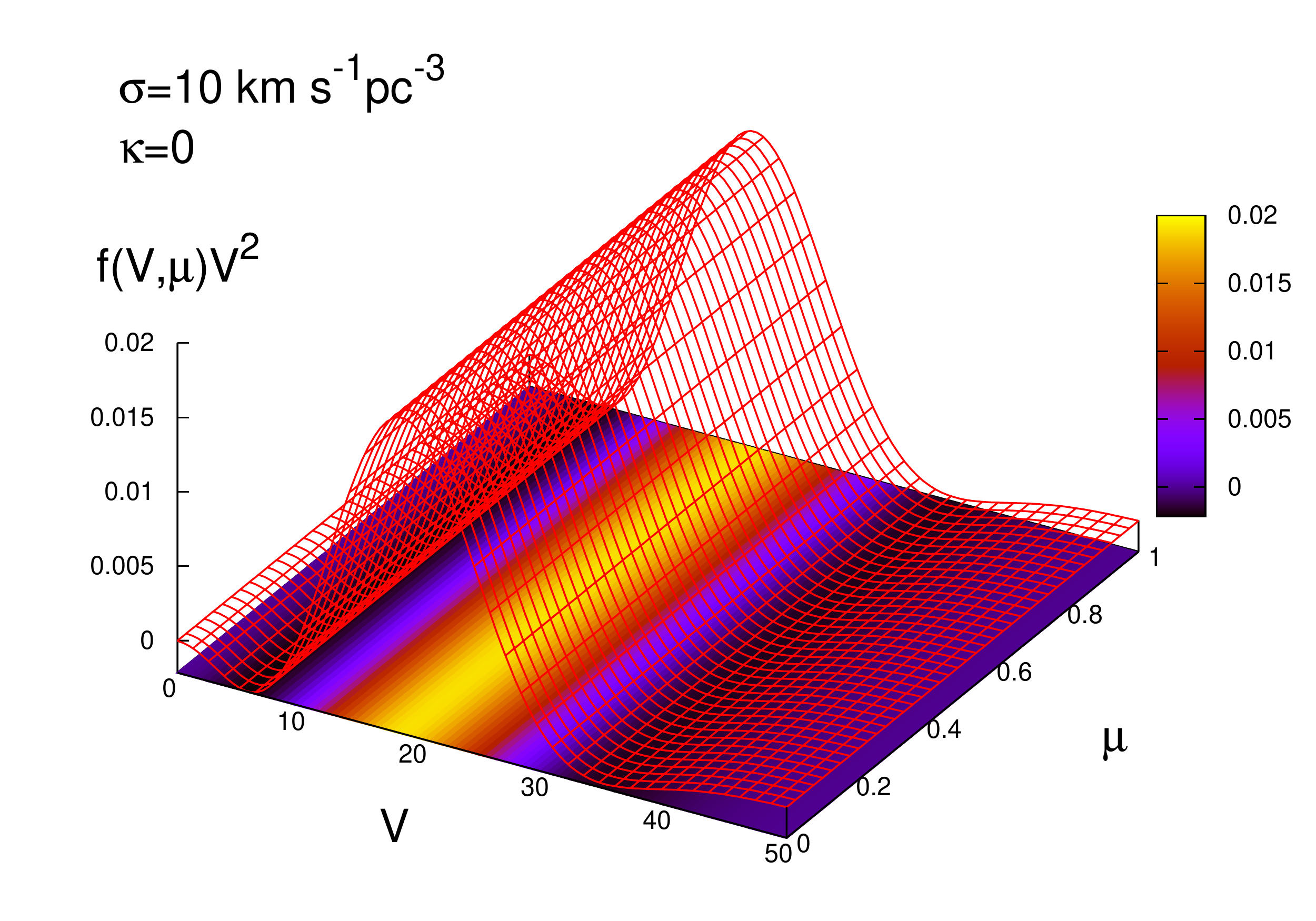}
       			\includegraphics[width=8cm,clip]{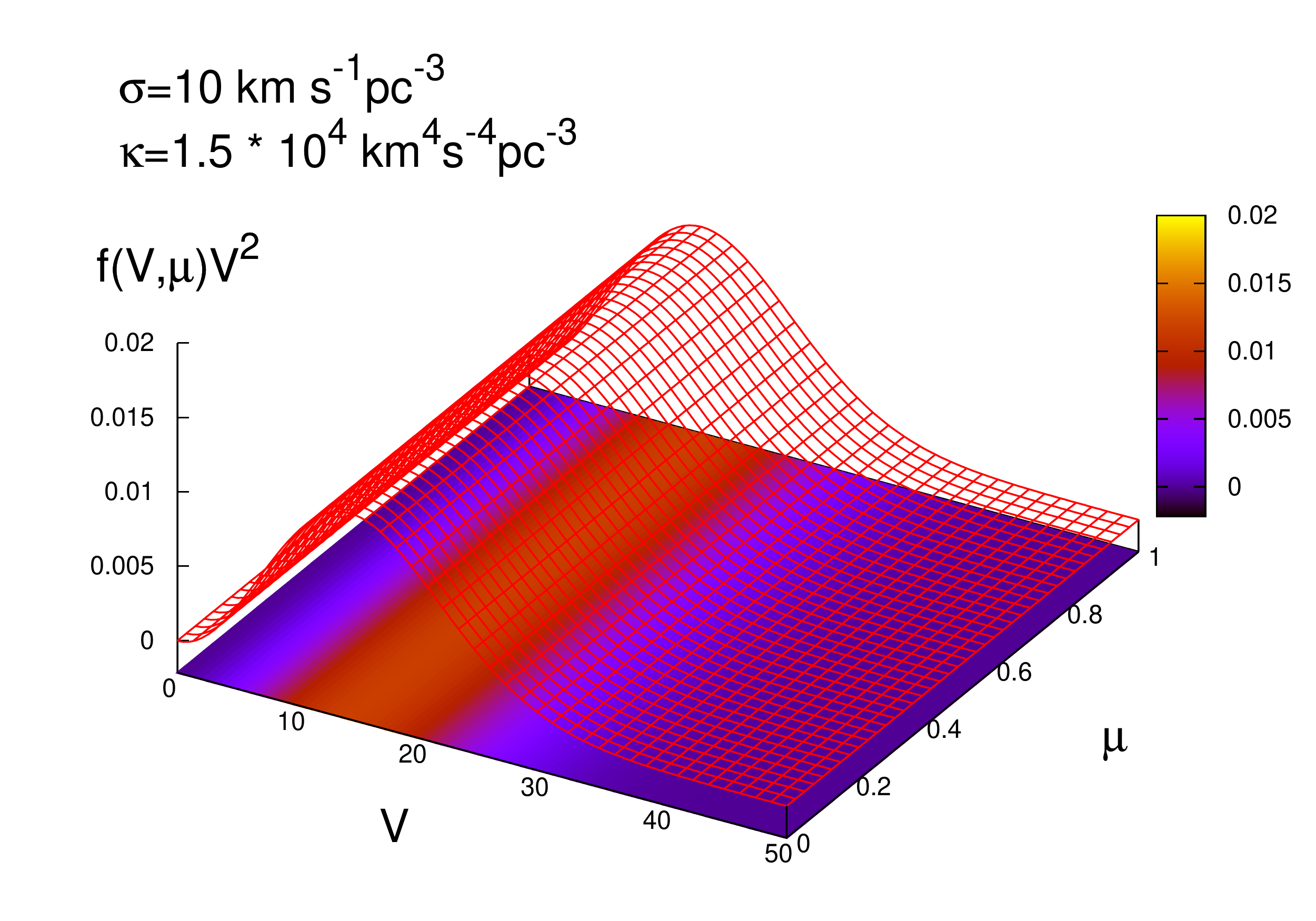}
       		\caption{\emph{left:} VDF where $\sigma=10 \text{km}\,
		\text{s}^{-1}\text{pc}^{-3}$, the remaining moments and 
		anisotropy parameters are set zero. \emph{right:} VDF where 
		$\sigma=10 \text{km}\,\text{s}^{-1}\text{pc}^{-3}$, $\kappa= 
		15 \rho \sigma^4$, the remaining moments and anisotropy 
		parameters are set zero. The right plot shows the MB 
		distribution in thermal equilibrium}
		\label{isotropy}
	\end{figure*}

Figure \ref{isotropy} displays two plots of the VDF $f(V,\mu)V^2$.  
In the left plot $\kappa=0$ and it is clear due to the shape that this 
is not the MB VDF for thermal equilibrium. To choose the right value 
for $\kappa$ compute it by means of equation \eqref{vmoments} using 
the MB distribution $g(V)V^2$ this yields\footnote[1]{This is also the 
reason why \citet{louis1990} defined $\kappa'=\kappa/15$ in his model. 
Then thermodynamical equilibrium or isotropy yields $\kappa'=\rho \sigma^4$. 
Nevertheless, this collides with equations \eqref{randommoments}, where the 
total moments were computed giving a more natural definition.}
\begin{equation}\label{15}
	\kappa=15 \rho \sigma^4
\end{equation}
For given $\sigma$ and $\rho$ this is the value we have to choose for 
$\kappa$ which is in our case $\kappa=150\,000\mathrm{km}^4\mathrm{s}
^{-4}\text{pc}^{-3}$. Then we obtain the MB distribution as can be 
seen in the right plot of figure \ref{isotropy} where this value was 
used.

	\begin{figure*}
		\centering
       		\includegraphics[width=8cm,clip]{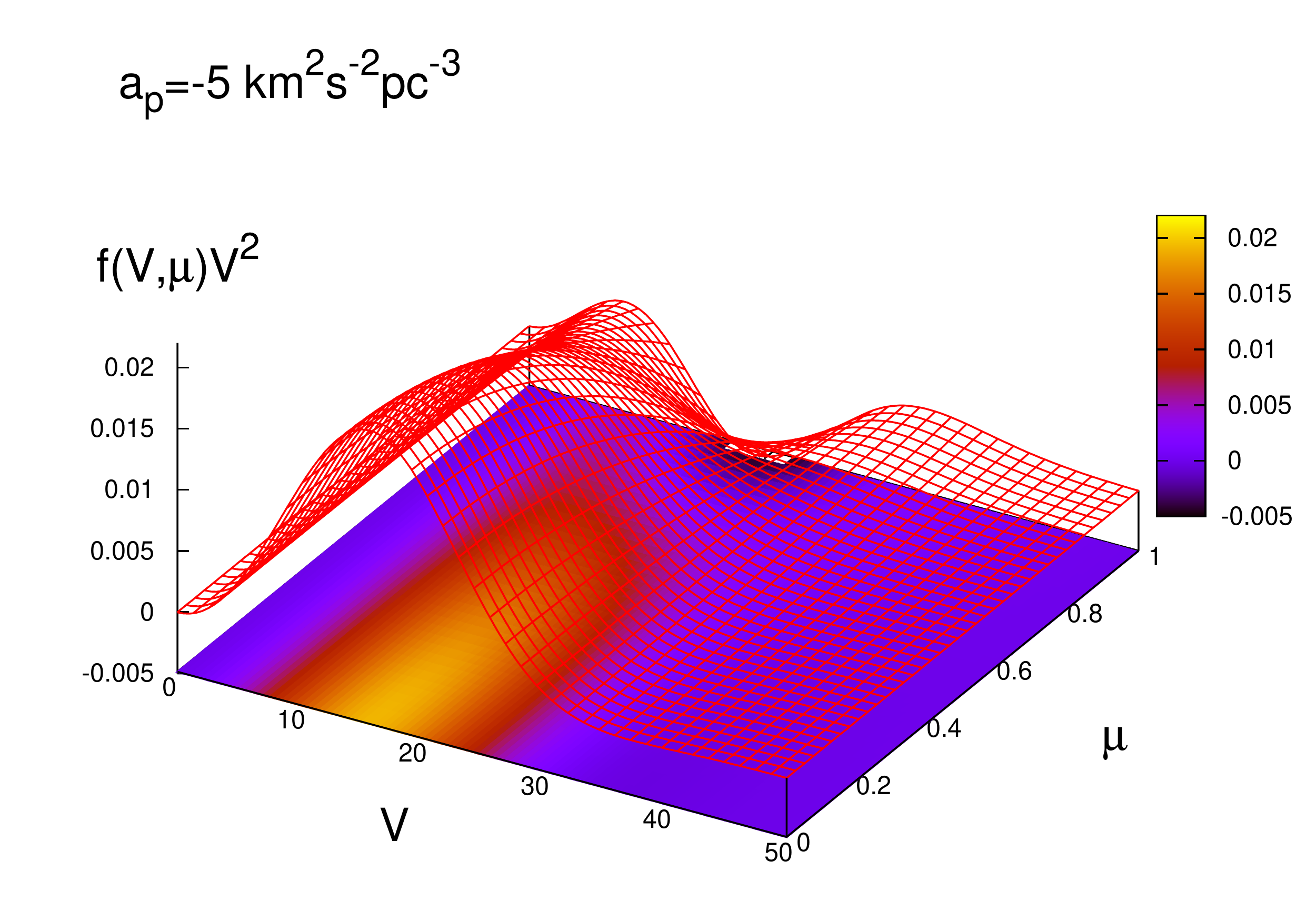}
       		\includegraphics[width=8cm,clip]{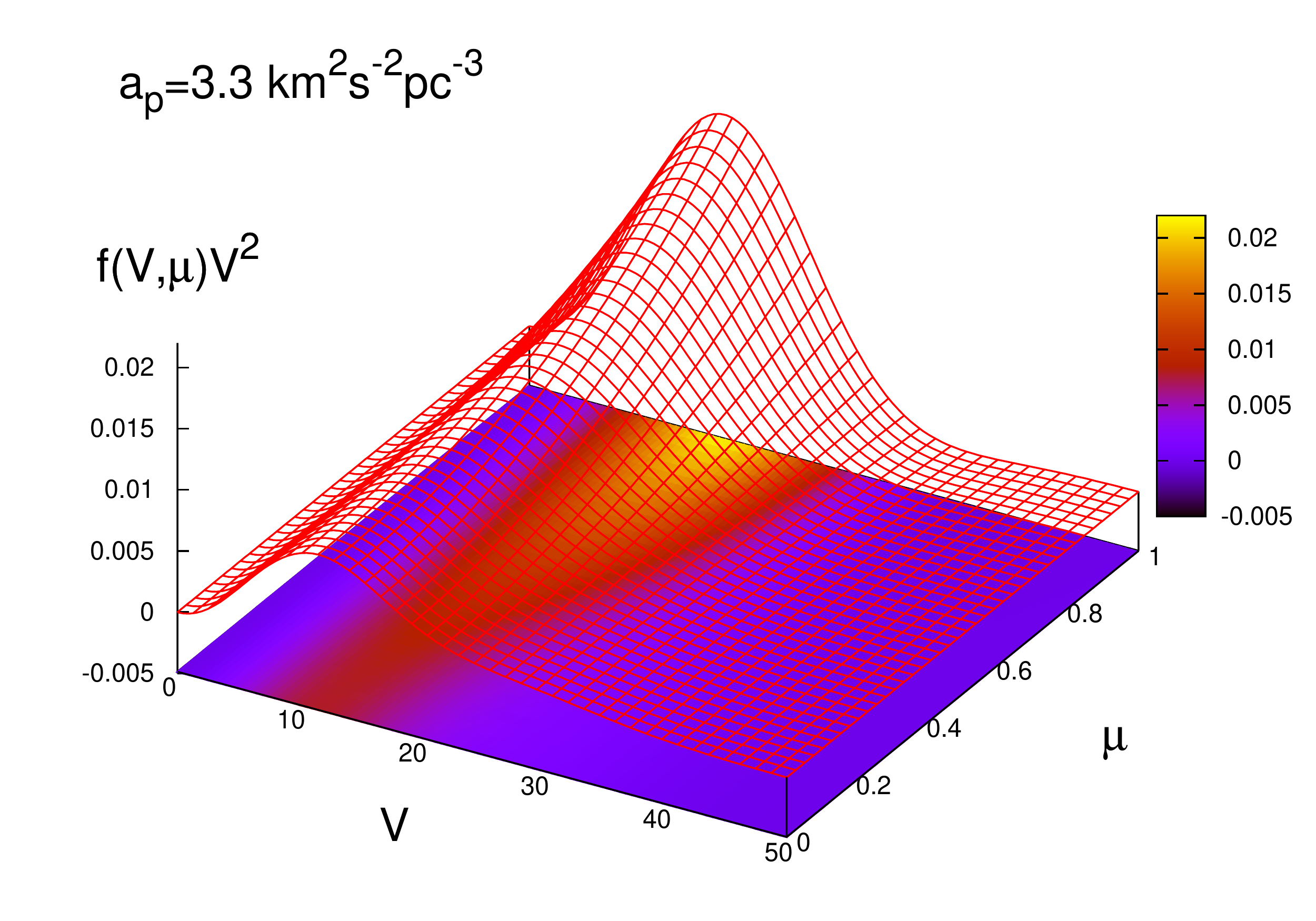}
       		\caption{VDF with two different values for the anisotropy 
		$a_p$. \emph{left:} Shows the effect of negative anisotropy 
		$a_p$ on a MB distribution. \emph{right:} Shows the effect 
		of positive anisotropy $a_p$ on a MB distribution.
}
		\label{ap}
	\end{figure*}
	\begin{figure*}
		\centering
       		\includegraphics[width=8cm,clip]{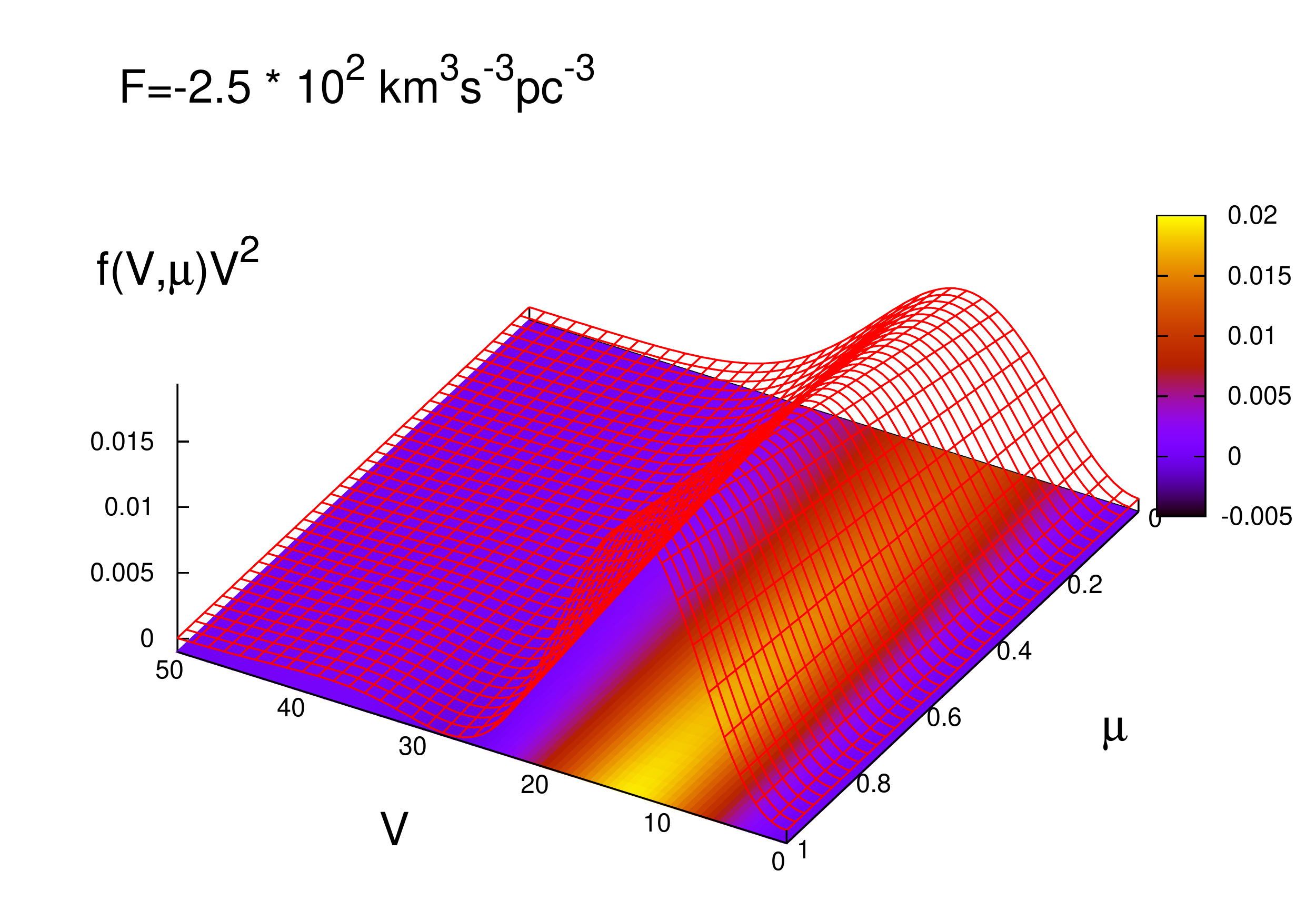}
       		\includegraphics[width=8cm,clip]{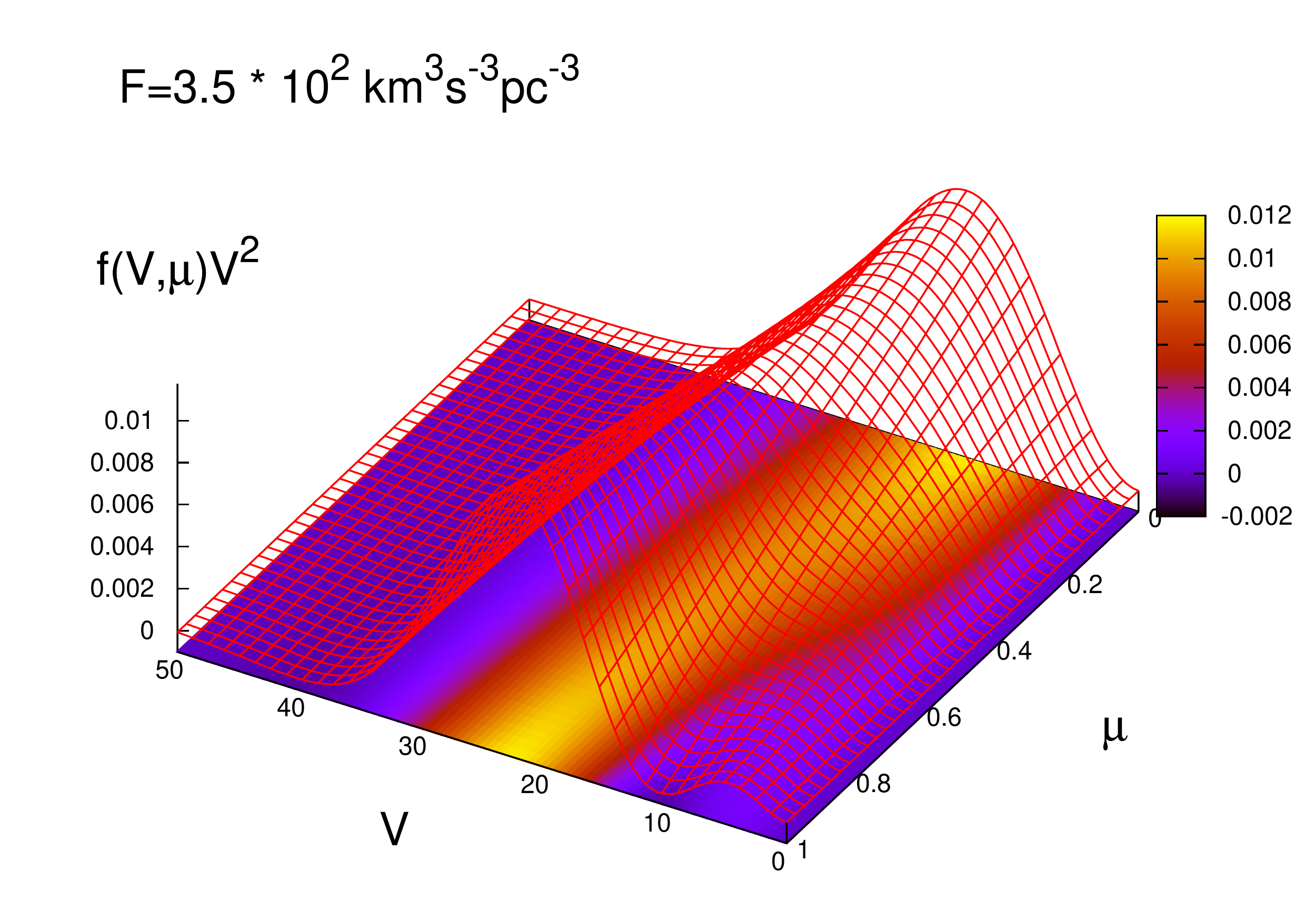}
       		\caption{VDF with two different values for the third order 
		total moment $F$ corresponding to energy flux. \emph{left:} 
		Shows the effect of negative $F$ on a MB distribution. 
		\emph{right:} The effect of positive $F$ on a MB distribution}
		\label{F}
	\end{figure*}
	\begin{figure*}
		\centering
       		\includegraphics[width=8cm,clip]{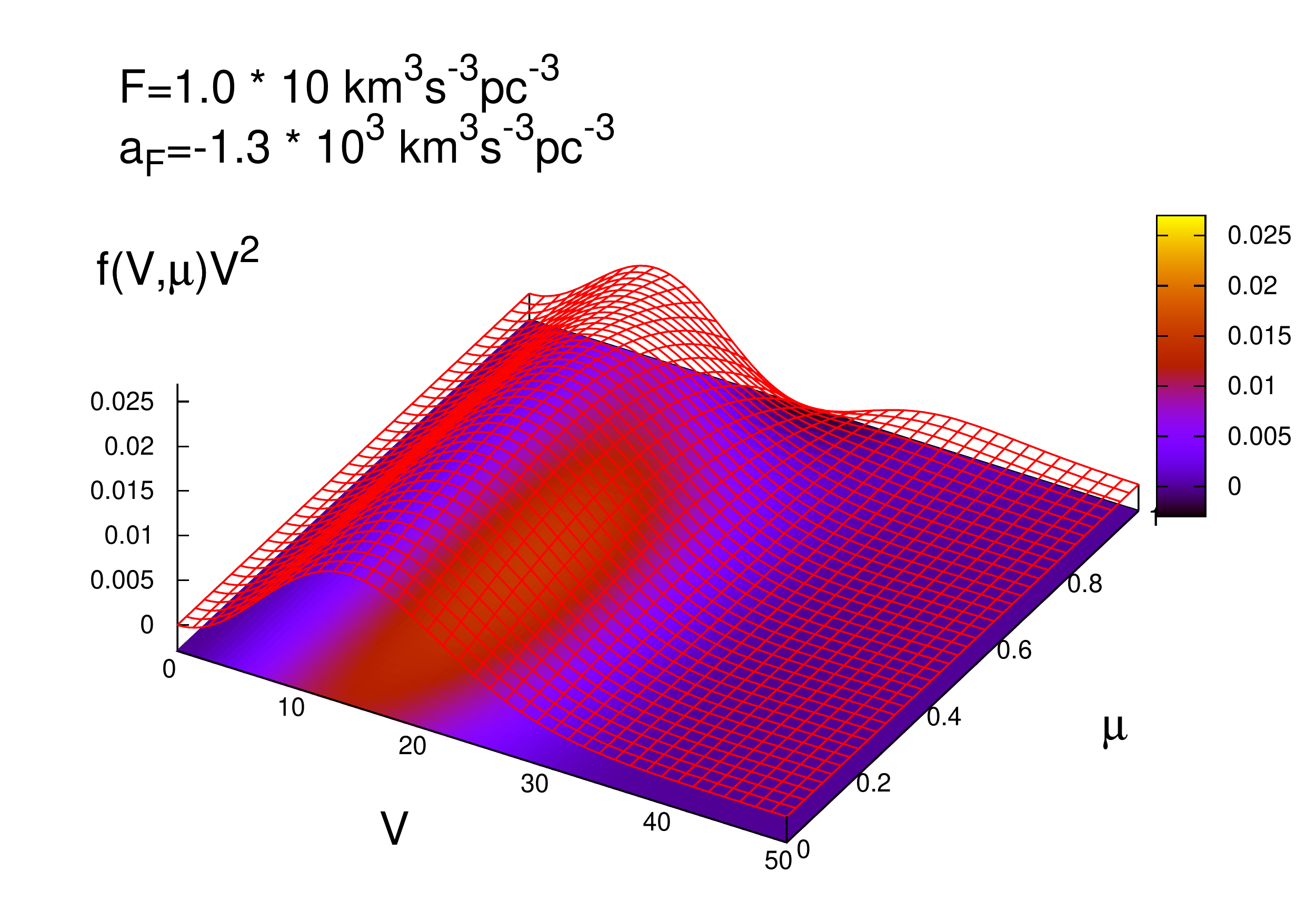}
		\includegraphics[width=8cm,clip]{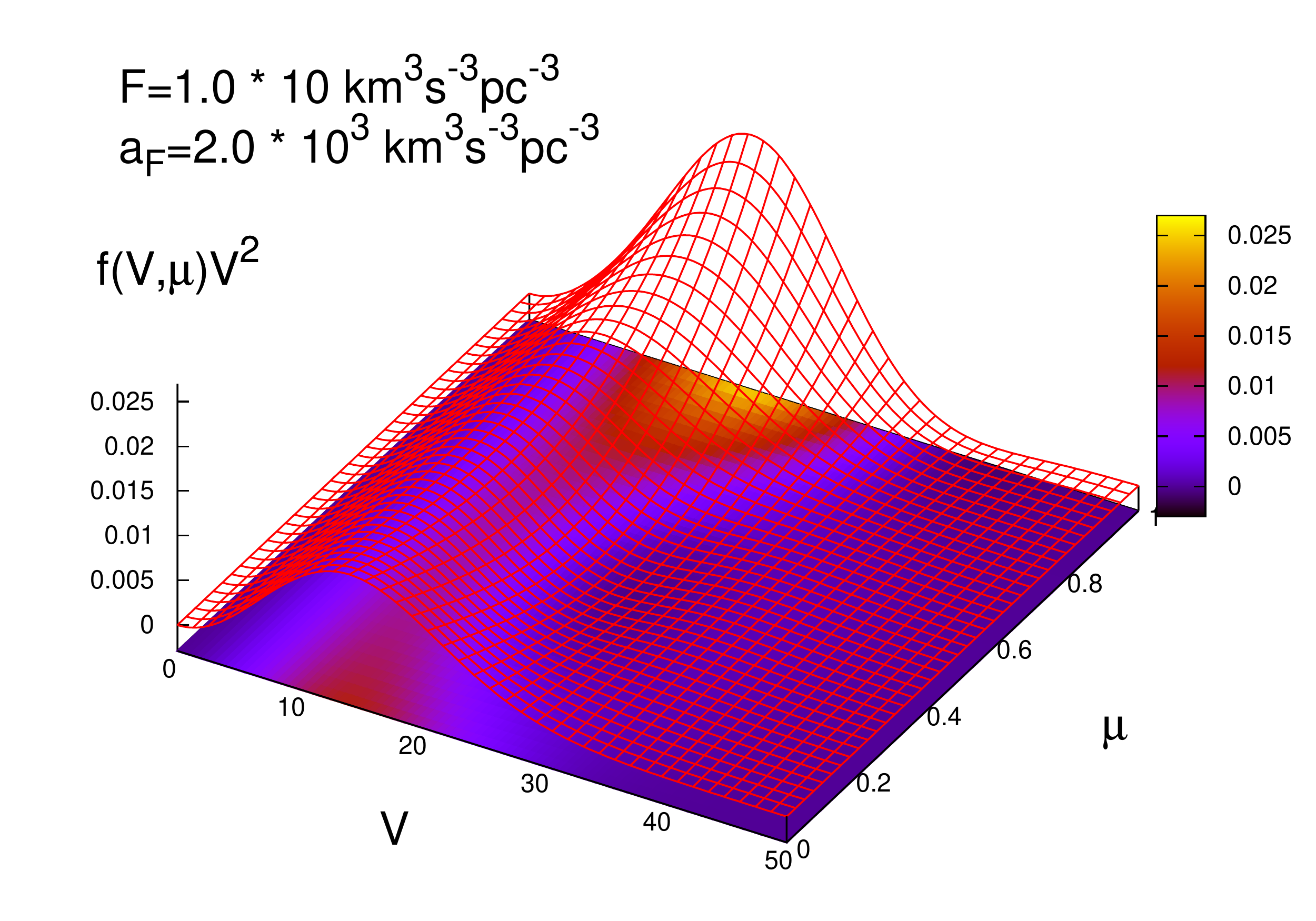}
       		\caption{VDF with two different values for the third order 
		anisotropy $a_F$. \emph{left:} Shows the effect of negative 
		anisotropy $a_F$ on the VDF. \emph{right:} The effect of 
		positive anisotropy $a_F$ on the VDF.}
		\label{aF}
	\end{figure*}
	\begin{figure*}
		\centering
       		\includegraphics[width=8cm,clip]{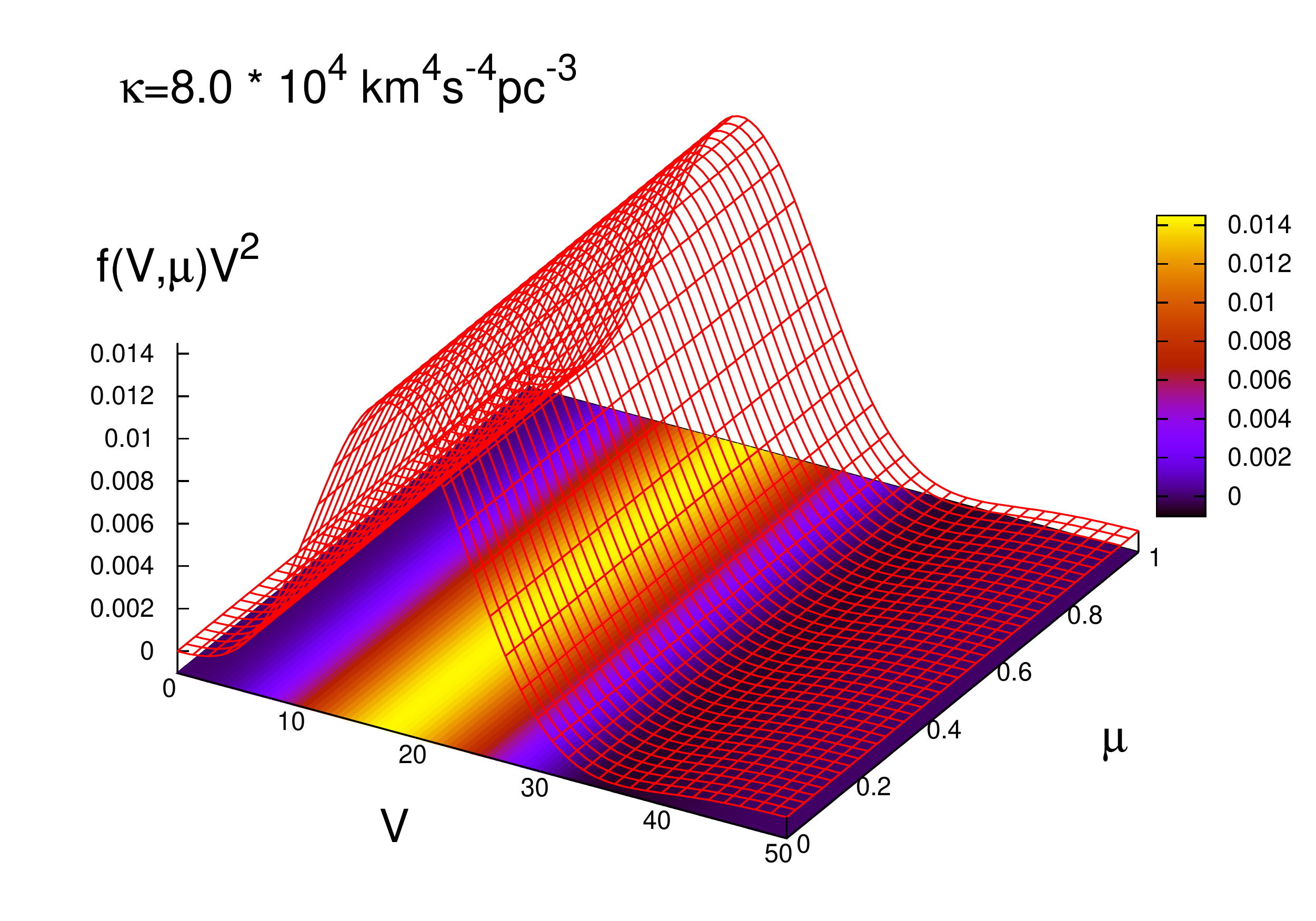}
       		\includegraphics[width=8cm,clip]{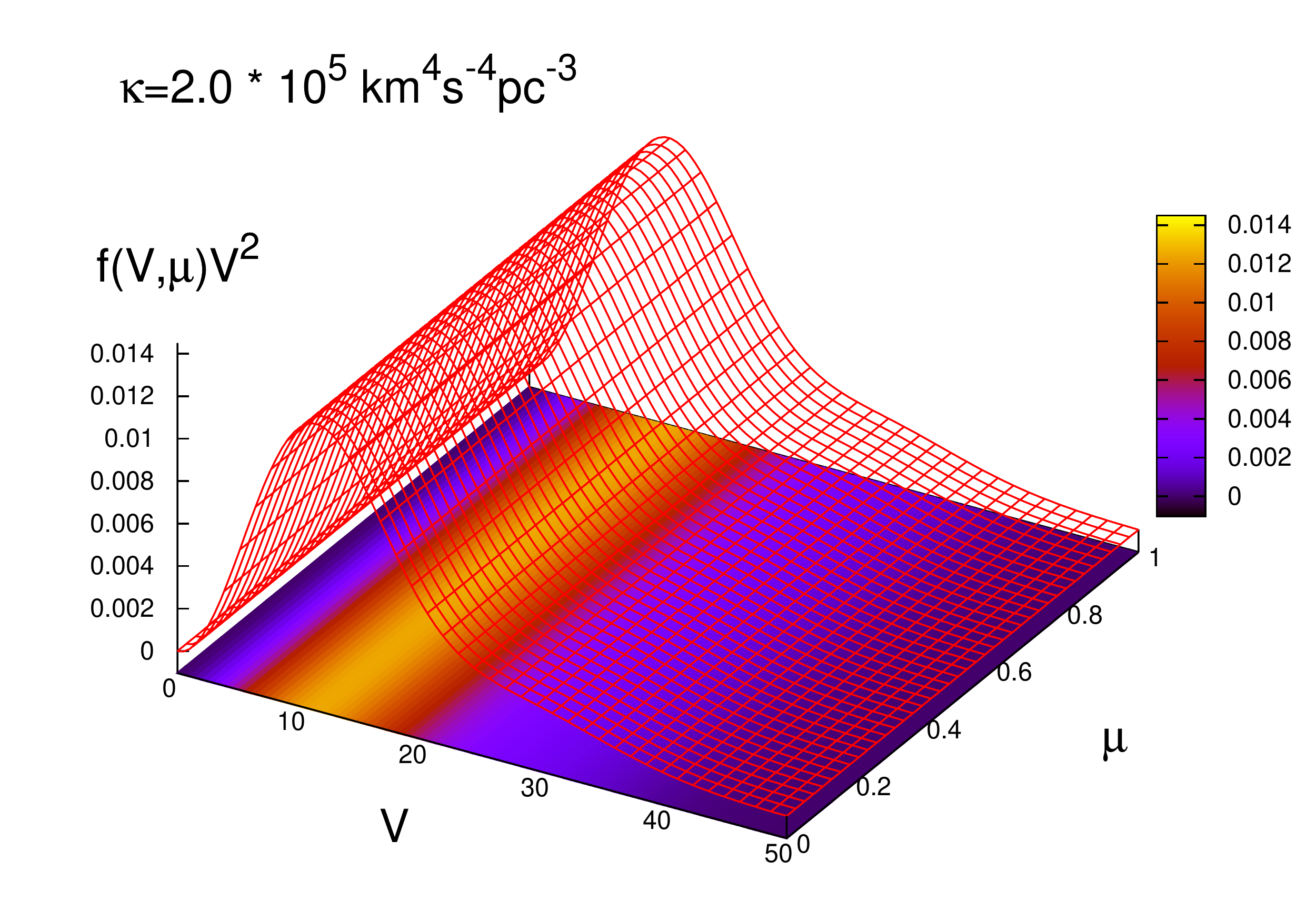}
       		\caption{VDF with two different values for the total 
		moments $\kappa$. \emph{left:} VDF with lower value 
		for $\kappa$ with respect to the MB distribution. 
		\emph{right:} VDF with higher value for $\kappa$ 
		with respect to the MB distribution.}
		\label{kappa}
	\end{figure*}
	\begin{figure*}
		\centering
       		\includegraphics[width=8cm,clip]{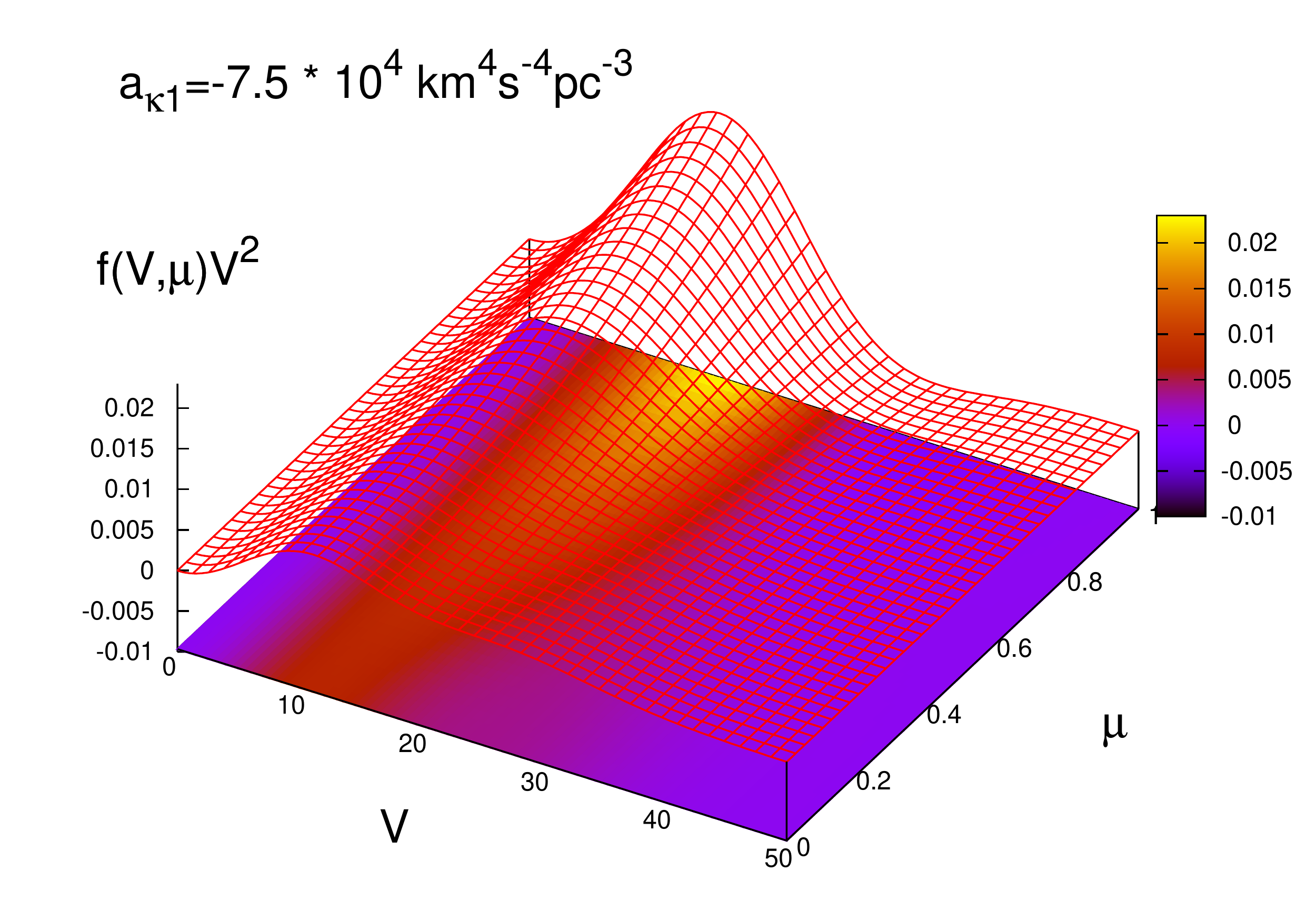}
       		\includegraphics[width=8cm,clip]{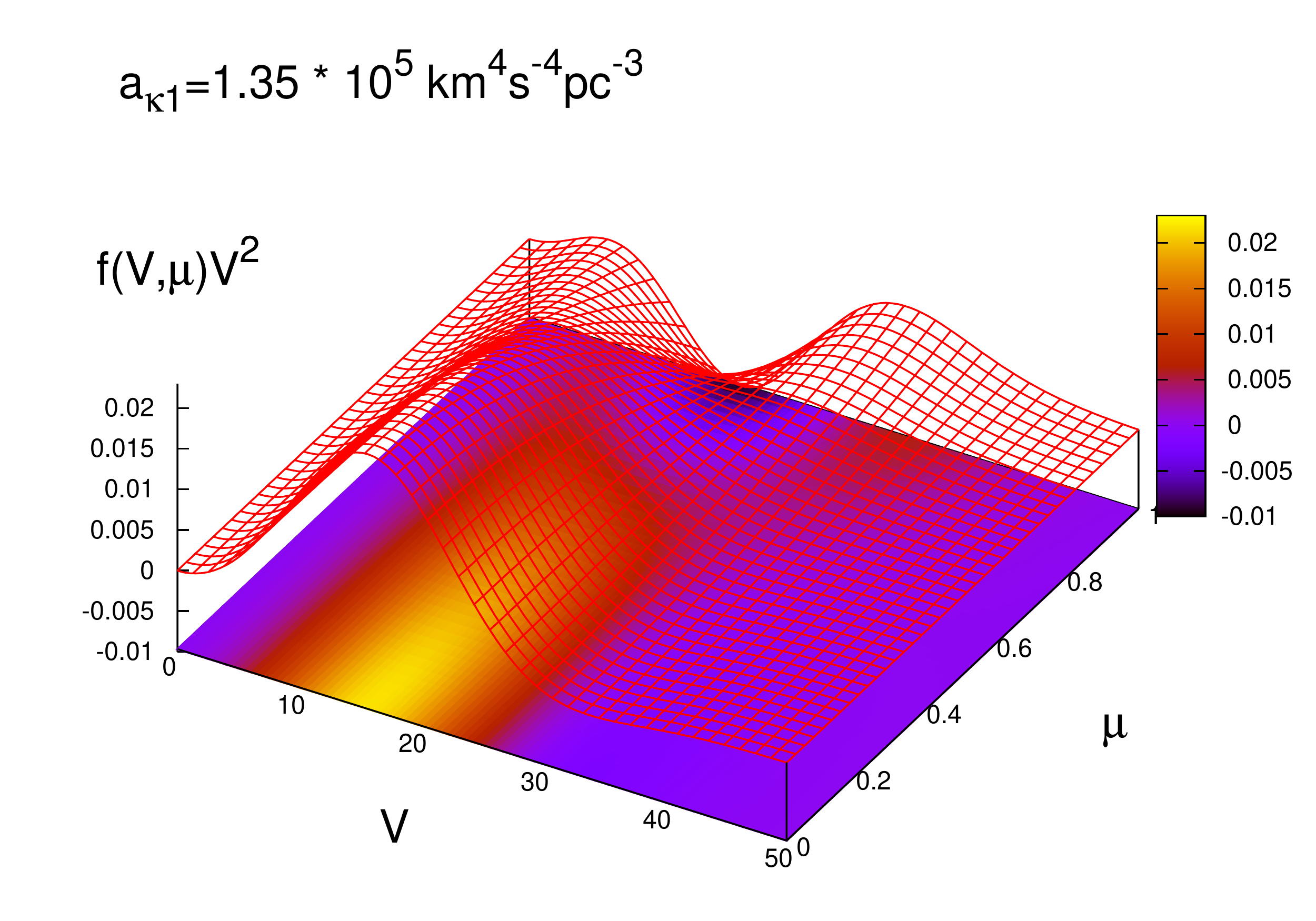}
       		\caption{VDF with two different values for the fourth 
		order anisotropy $a_{\kappa1}$. \emph{left:} The effect 
		of negative anisotropy $a_{\kappa1}$ on a MB distribution. 
		\emph{right:} The effect of a positive  anisotropy 
		$a_{\kappa1}$ on a MB distribution}
		\label{a_kappa1}
	\end{figure*}
	\begin{figure*}
		\centering
       		\includegraphics[width=8cm,clip]{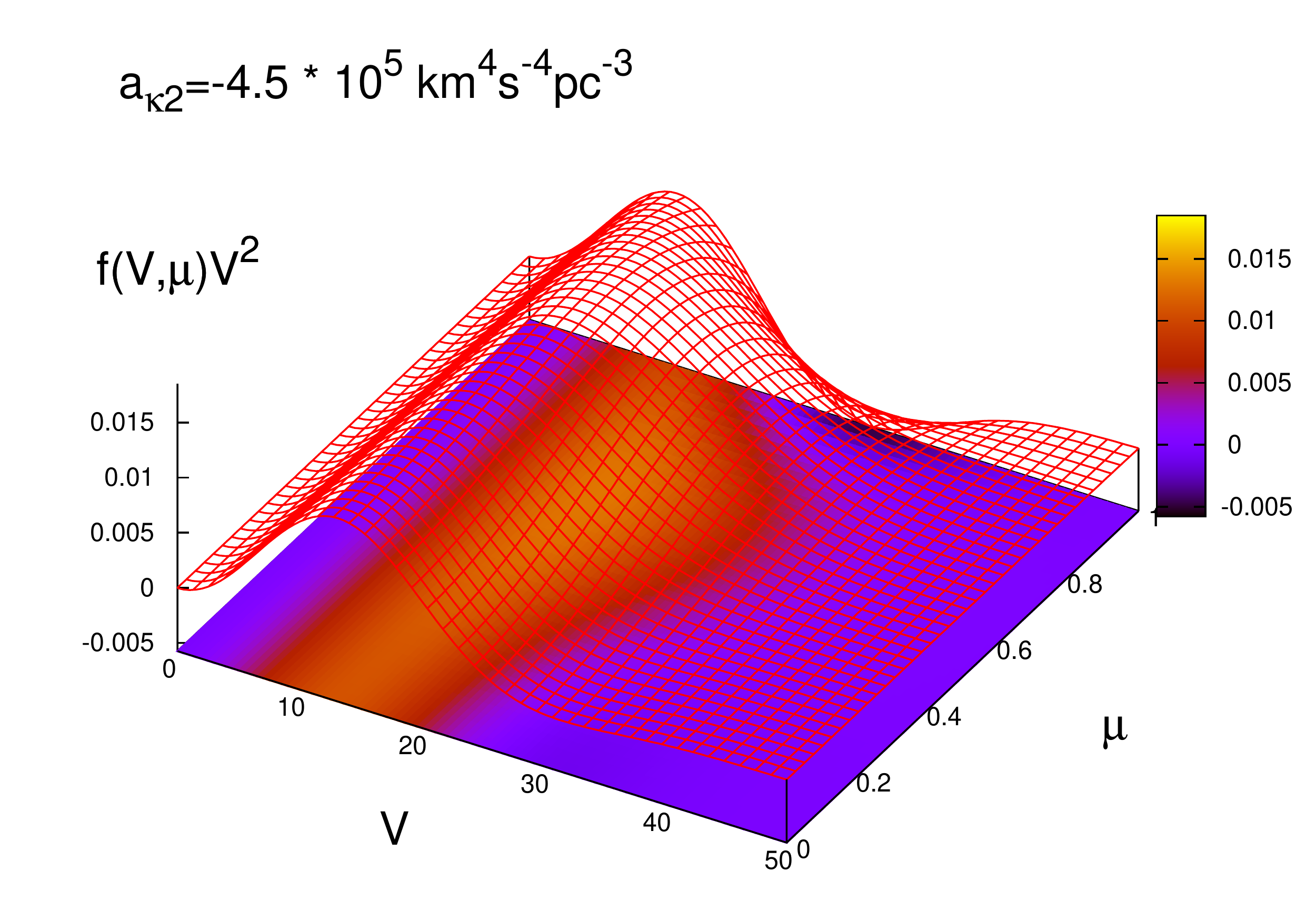}
       		\includegraphics[width=8cm,clip]{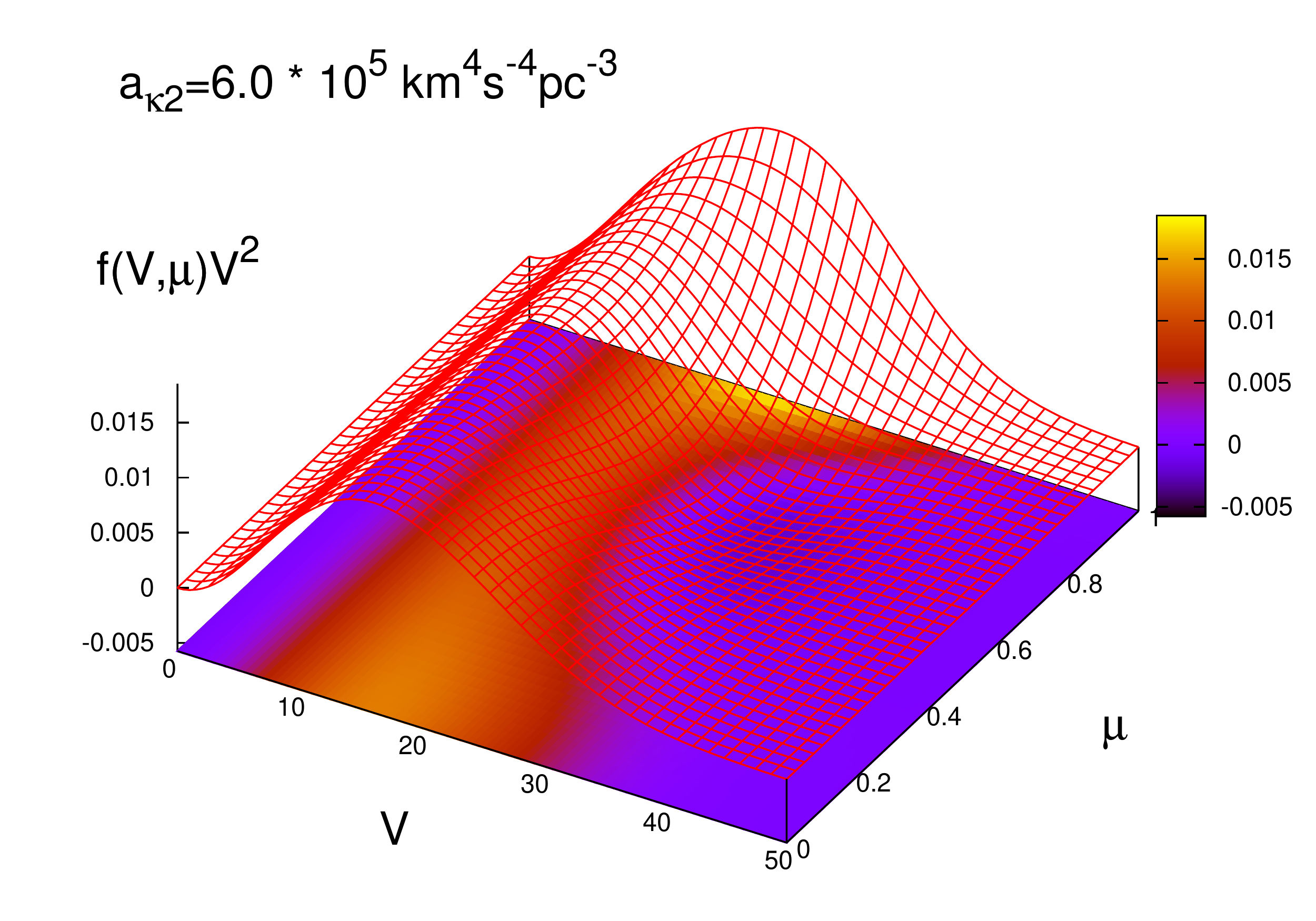}
       		\caption{VDF with two different values for the fourth order 
		anisotropy $a_{\kappa2}$. \emph{left:} The effect of negative 
		anisotropy $a_{\kappa2}$ on a MB distribution. \emph{right:} 
		The effect of of anisotropy $a_{\kappa2}$ on a MB distribution}
		\label{a_kappa2}
	\end{figure*}
	\begin{figure*}
		\centering
       		\includegraphics[width=8cm,clip]{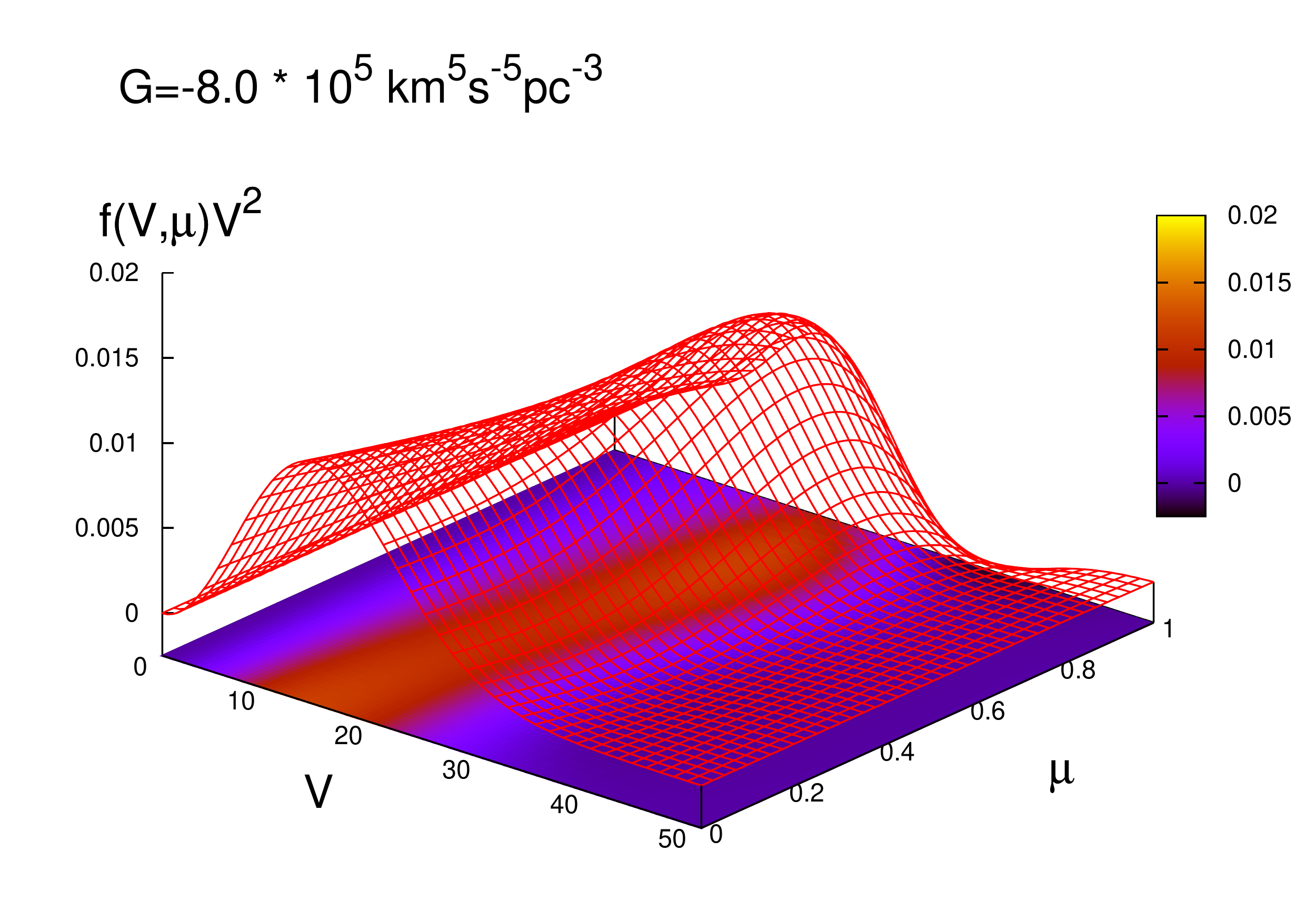}
		\includegraphics[width=8cm,clip]{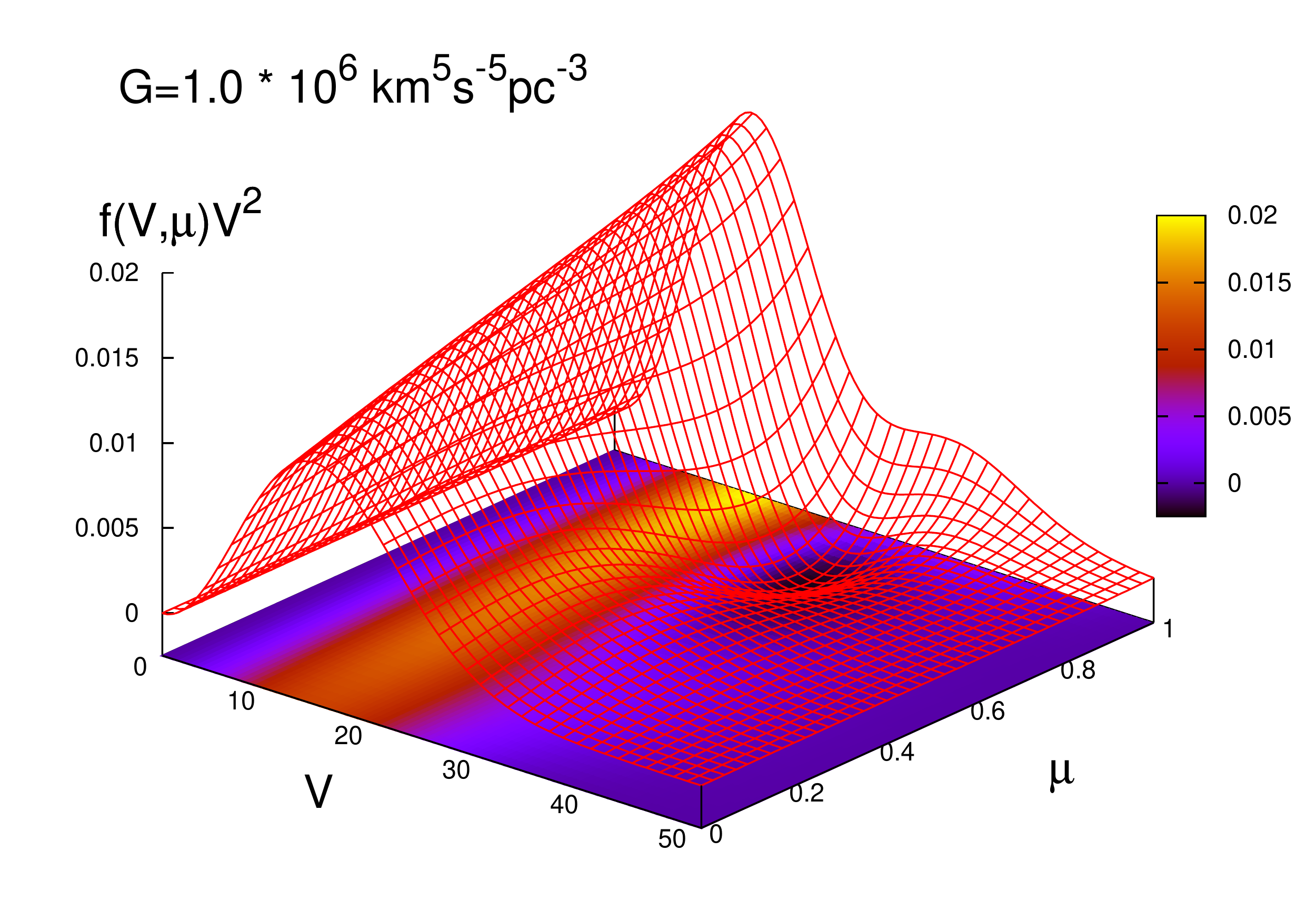}
       		\caption{VDF with two different values for the total moment $G$. 
		\emph{left:} Shows the effect of negative $G$ on a MB distribution. 
		\emph{right:} Shows the effect of positive $G$ on a MB distribution}
		\label{G}
	\end{figure*}
	\begin{figure*}
		\centering
       		\includegraphics[width=8cm,clip]{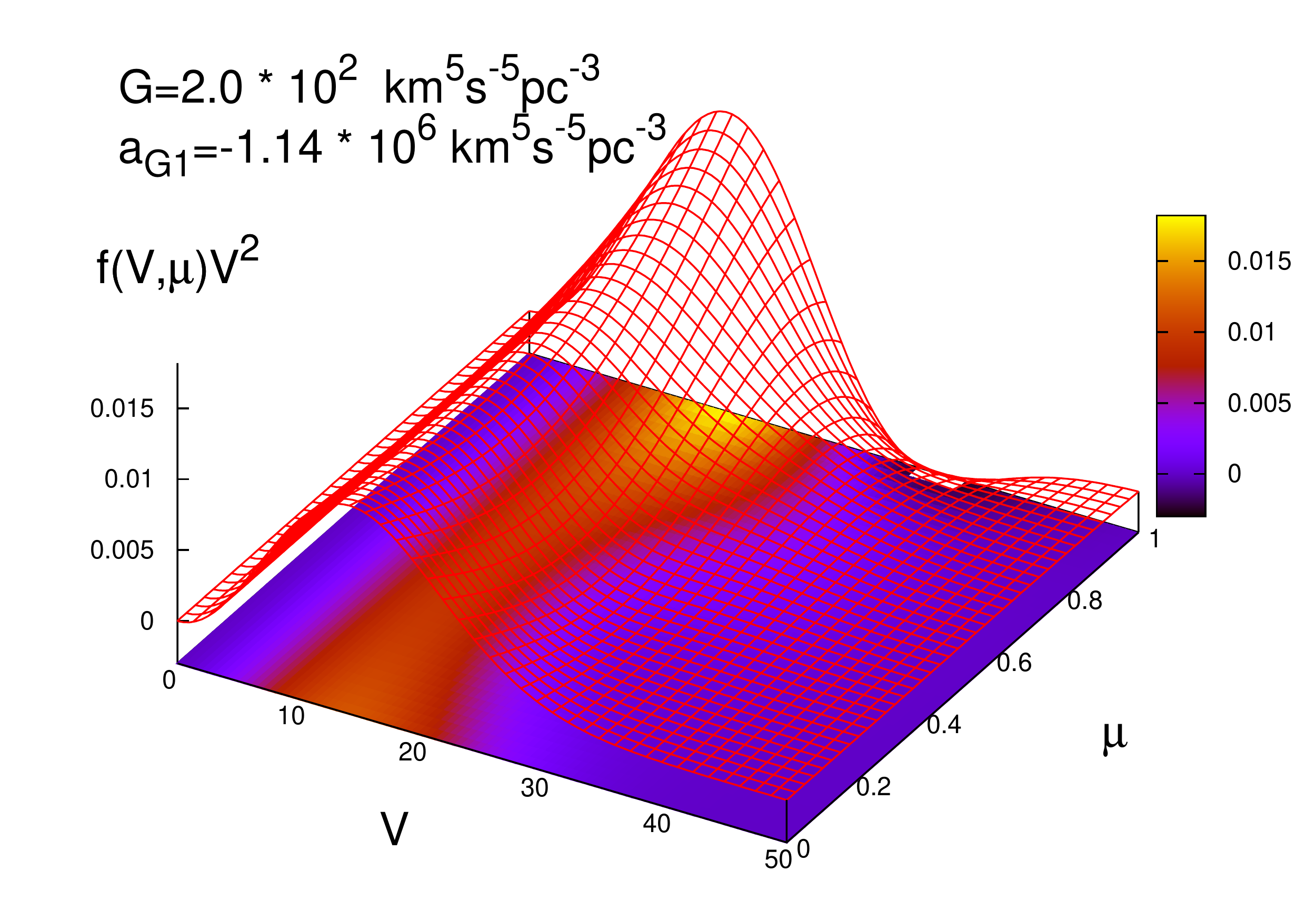}
       		\includegraphics[width=8cm,clip]{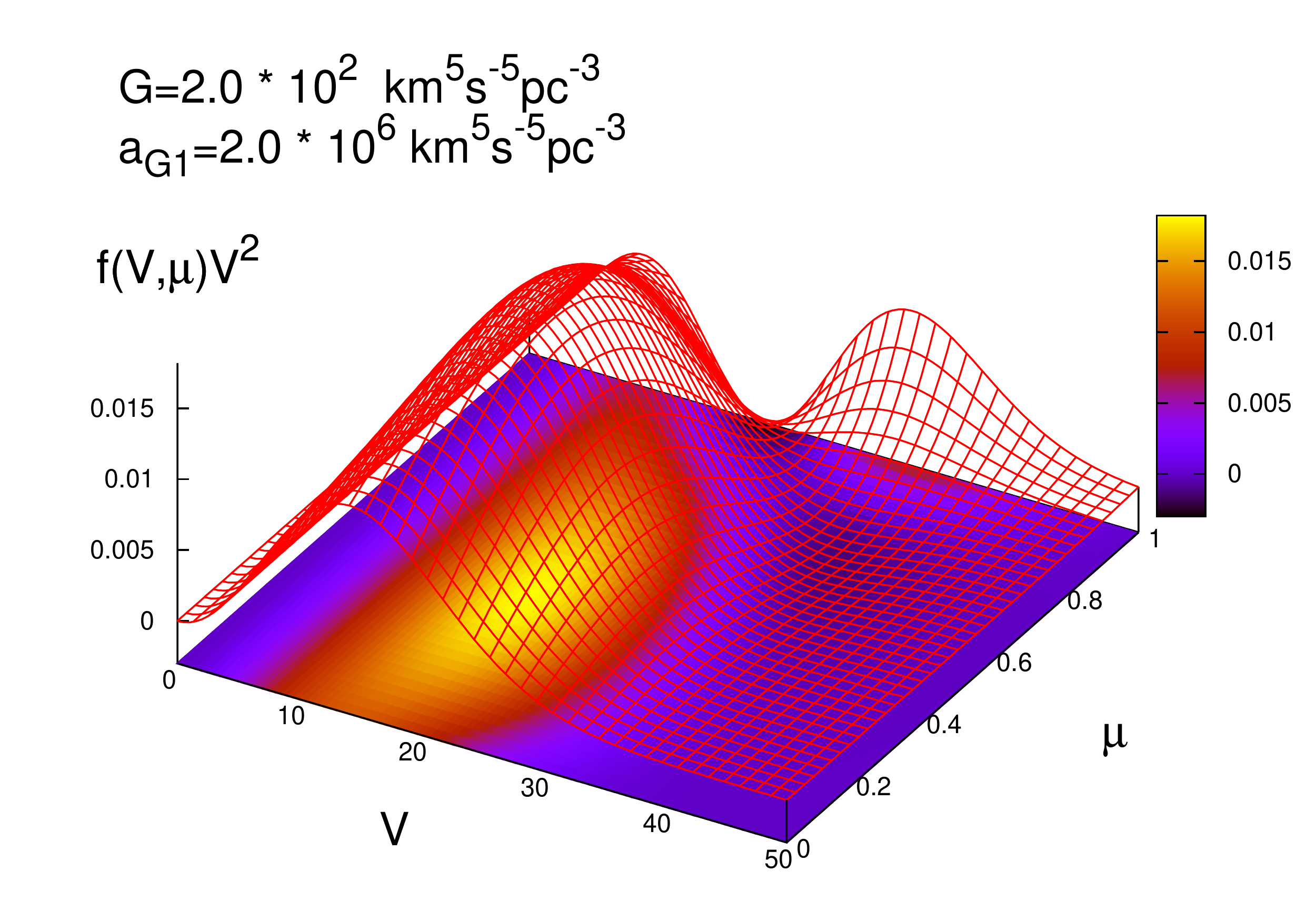}
       		\caption{VDF with two different values for the fifth order anisotropy 
		$a_{G1}$. \emph{left column:} Shows the effect of negative anisotropy 
		$a_{G1}$ on the VDF. \emph{right column:} Shows the effect of positive 
		anisotropy $a_{G1}$ on the VDF. }
		\label{aG1}
	\end{figure*}
	\begin{figure*}
		\centering
		\includegraphics[width=8cm,clip]{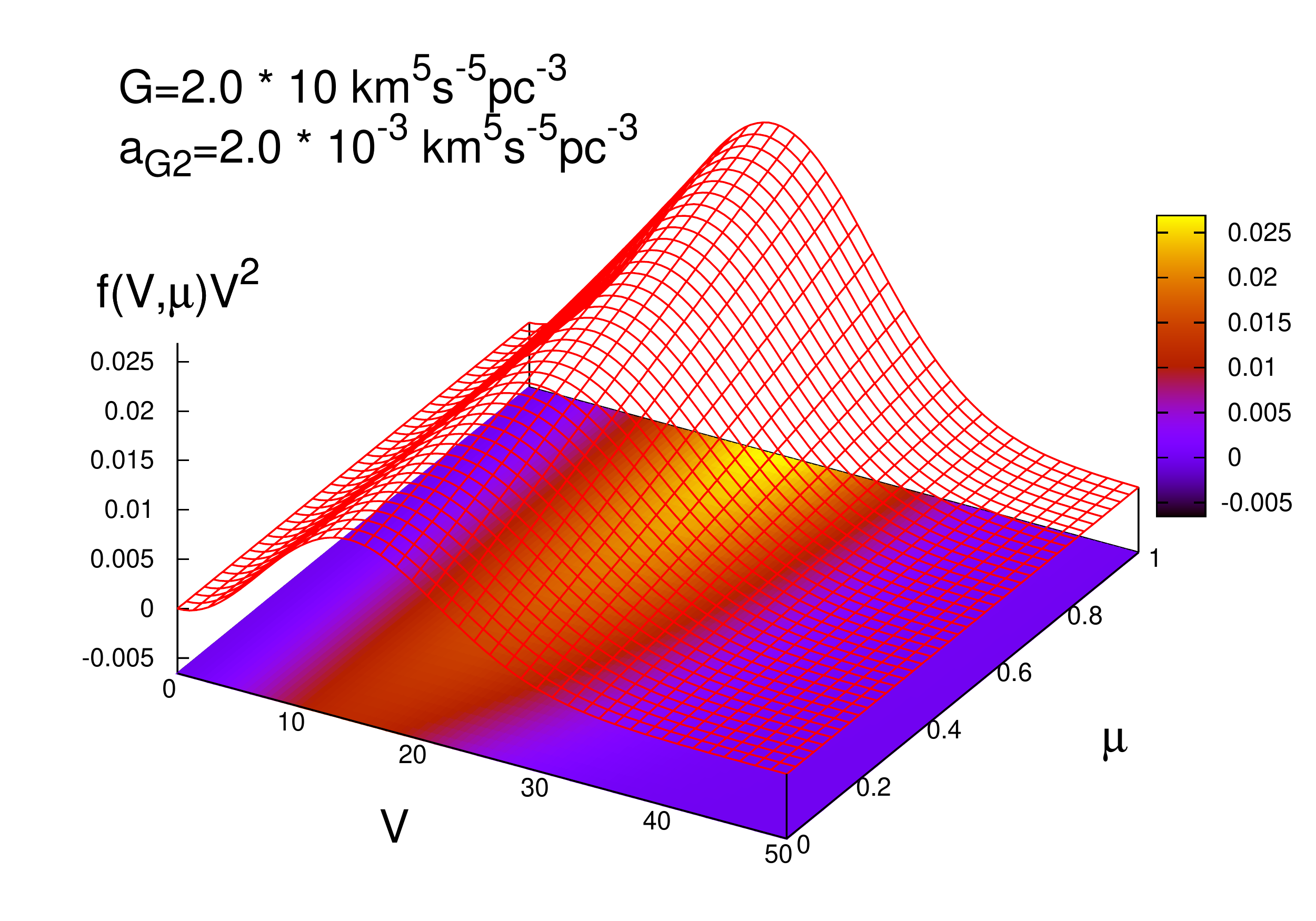}
       		\includegraphics[width=8cm,clip]{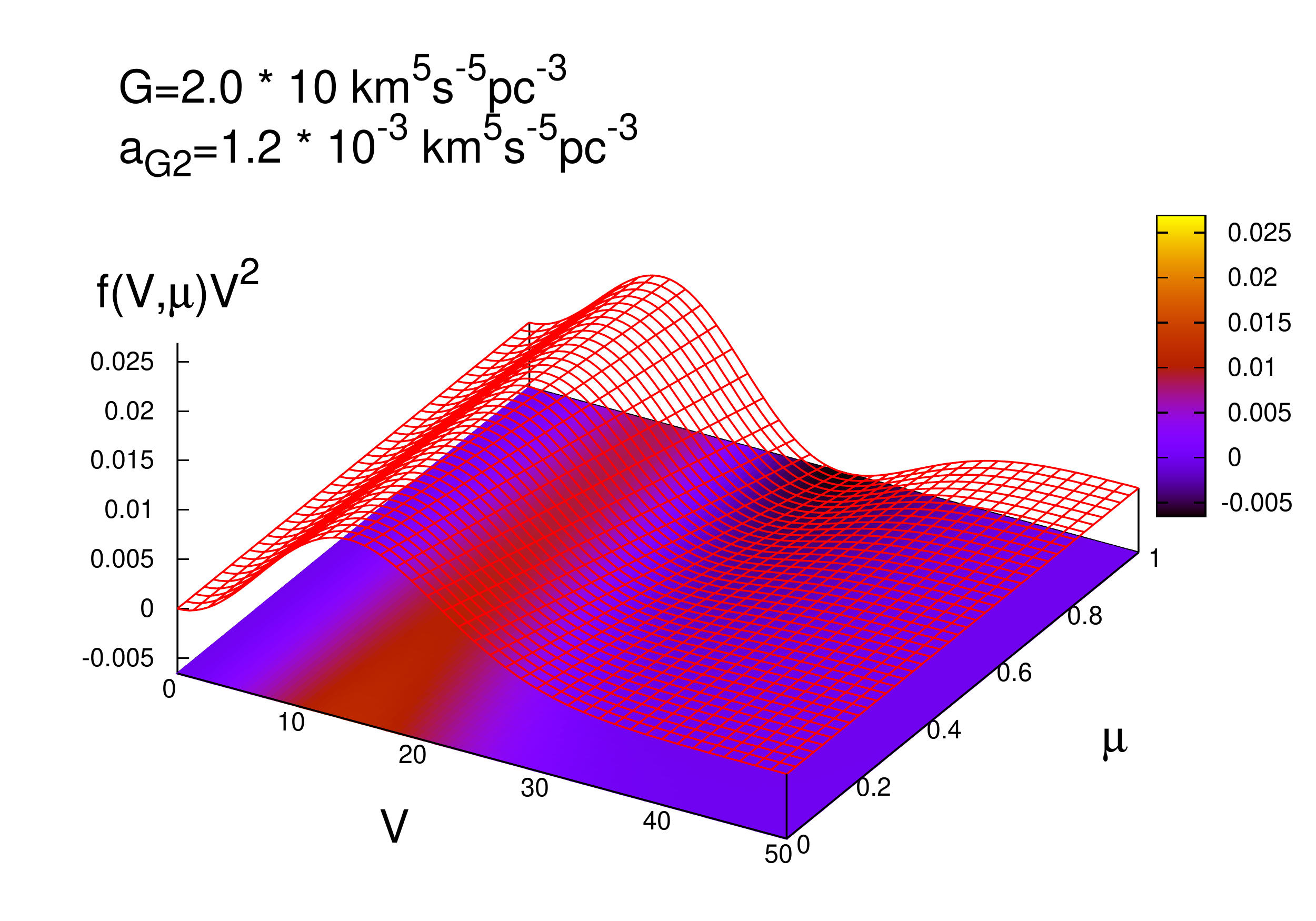}
       		\caption{VDF with two different values for the fifth order 
		anisotropy $a_{G2}$. \emph{left:} Shows the effect of negative 
		anisotropy $a_{G2}$ on the VDF. \emph{right:} The effect of 
		positive anisotropy $a_{G2}$ on the VDF. }
		\label{aG2}
	\end{figure*}

The anisotropy $a_p$ describes the difference between the second order 
moments $p_r$ and $p_t$ which represent the radial and tangential pressure 
(or equivalently energy density) respectively. In thermal equilibrium 
we have $p_r=p_t$ and thus $a_p=0$. The second order moments determine 
the width of the VDF given by the dispersion $\sigma$. 
When the anisotropy $a_p<0$ the tangential pressure exceeds 
the radial pressure. As a consequence, we observe in the left plot of 
figure \ref{ap} that for $\mu\rightarrow1$ the number of particles 
decreases whereas for $\mu\rightarrow0$ the number of particles increases. 
Since $\mu$ determines the fraction of the radial and tangential velocity 
component  this physically means that we have more particles with 
circular orbits when $a_p<0$. For $a_p>0$ we have the opposite behavior. 

A very similar effect is caused by the fourth order anisotropy $a_{\kappa1}$ 
as can be seen in figure \ref{a_kappa1}. It appears together with $P_2(\mu)$ 
and the same powers of $V$ as $a_p$. However, since $a_{\kappa1}$ and $a_p$ 
appear with a different sign they have opposite effects. Consequently,
we can assume that the fourth order anisotropy $a_{\kappa1}$ is a correction 
of the second order anisotropy $a_p$. 

The same argument holds for the third order anisotropy $a_F$ (figure \ref{aF}) 
and fifth order anisotropy $a_{G1}$ (figure \ref{aG1}). Both appear as factors 
of the Legendre polynomial $P_3(\mu)$ with the same powers of $V$, but different 
sign. Thus $a_{G1}$ can be seen as a correction of the third order anisotropy 
$a_F$. 

In equation \eqref{distr} we observe that uneven moments appear with uneven 
Legendre polynomials $P_l(\mu)$. However, these Legendre polynomials 
vanish at $\mu=0$ and thus the VDF is independent on uneven moments at 
$\mu=0$. In other words, since $\mu=0$ corresponds to stars that have 
a vanishing radial velocity component $v_r$, the distribution of 
stars that move on circular orbits are not affected by third order 
moments. This effect can be seen in in figures \ref{F}, \ref{aF}, 
\ref{G}, \ref{aG1} and \ref{aG2}.

When including third order moments the VDF depends on the total
 moment $F$ and the anisotropy $a_F$, where $F$ is related to the total 
energy flux and $a_F$ describes differences between radial and tangential 
energy fluxes. Negative values of the total moment $F$ result in 
an increase of the maximum of the VDF for $\mu\rightarrow1$ (figure 
\ref{F}, left plot). We thus find more stars with eccentric orbits for 
$F<0$. When $F>0$ the maximum of the VDF shifts to higher velocities
$V$ when $\mu\rightarrow1$ (figure \ref{F}, right plot). This means 
that positive $F$ increases the radial velocity component $v_r$, but 
leaves the tangential component constant. As a consequence it increases 
the eccentricity of orbits, but not the number of stars with eccentric 
orbits as it does for $F<0$. Note that the two plots in figure \ref{F} have 
different scaling in the $z$-axis with respect to each other in order 
to display the distinct effects for $F<0$ and $F>0$. Whereas the maximum 
of the VDF changes for $F<0$ from $\approx 0.01\text{pc}^{-3}(\text{km/s})
^{-3}$ at $\mu=0$ to $\approx 0.015\text{pc}^{-3}(\text{km/s})^{-3}$ at 
$\mu=1$ in the left plot of figure \ref{F} it roughly stays 
constant for all values of $\mu$ in the right plot corresponding to $F>0$.
The effect of the third order anisotropy $a_F$ is displayed 
in figure \ref{aF}. $a_F<0$ increases the number of stars with velocities 
above the mean and directions corresponding to $\mu\approx 0.5$ whereas 
$a_F>0$ causes an inverse effect. 

The fourth order moments are related to the kurtosis of the velocity 
distribution which gives a measure of high velocity stars as compared 
to thermal equilibrium. In figure \ref{kappa} the VDF is plotted for 
two different values of the total moment of fourth order $\kappa$. In 
the left plot $\kappa$ is chosen to be smaller than $15 \rho \sigma^4$.
 The VDF increases at its mean at the expense of high and low 
velocities. This corresponds to a deficiency of stars with high 
or low velocities but more stars with velocities near the mean,
compared to thermal equilibrium. In the right plot the value of 
$\kappa$ is chosen to be higher than $15 \rho \sigma^4$. The wing of 
the VDF towards high velocities becomes thicker whereas the maximum 
of the VDF is smaller when compared to thermal equilibrium. Here the 
number of high velocity stars increases at the expense of stars with 
intermediate velocities. 

Whereas the anisotropy $a_{\kappa1}$ should be viewed in combination 
with the second order anisotropy $a_p$ as mentioned before, the 
anisotropy $a_{\kappa2}$ gives a new characterization of anisotropy 
at fourth order, which is displayed in figure \ref{a_kappa2}. 

The effect of the total moment $G$ on the VDF is illustrated in figure 
\ref{G}. To get a physical understanding of this quantity we compare it 
to the total moment of third order $F$. $F$ is an uneven moment which 
was considered to denote the radial flux of random kinetic energy. The 
random kinetic energy density $e$ was given by the second order moments 
as $e=(p_r+2p_t)/2$. Thus the third order moment $F$ is the corresponding 
flux quantity for the second order moment $p$. Equivalently we can relate 
the fifth order moment $G$ and the fourth order moment $\kappa$. Since 
$\kappa$ is related to the number of high velocity stars $G$ can be 
considered as a measure for the flux of these stars. 
Again $a_{G1}$ (figure \ref{aG1}) should be viewed in the context of the 
third order anisotropy $a_F$ whereas $a_{G2}$ (figure \ref{aG1}) determines 
a new type of anisotropy at fifth order. 

Thus, every moment and anisotropy parameter has its own affect on the VDF. They
act on different velocity intervals and redistribute stars from distinct
orbitals. If we only include moments up to third order into our model, as it
has been done in previous studies, our VDF is strongly limited. We are then not
able to describe areas in velocity space as is possible with moments of
order $>3$.  More precisely, we obtain a much more detailed description of the
distribution of stars in velocity space for stars with high velocities and
stars which have neither radial nor tangential orbits, i.e. $0<\mu<1$. 

\section{Discussion}\label{sec:disussion}

In this work we develop two statistical moment models for dense stellar
dynamical systems. They are closed either at fifth- or at sixth-order depending
on the required accuracy. They describe in a self-consistent way (including
Fokker-Planck relaxation terms) local deviations of the velocity distribution
function from the MB distribution.  The description of the velocity
distribution function includes third- and fourth-order moments. Third-order
moments represent energy fluxes equivalent to asymmetries of the velocity
distribution around its center. Fourth order moments denote deviations from the
MB distribution at high velocities. This cannot be described by a velocity
distribution that is fully determined by its first two moments such as a
Gaussian, commonly used to fit observational data. Due to the larger
number of moments of the velocity distribution, the two models we introduce
have the potential to fit detailed star-star and integrated light observations
of globular clusters or nuclear star clusters in detail. However, they still
underly a number of approximations, such as assuming spherical symmetry, equal
stellar masses, the Fokker-Planck and the local approximation and they also
require a system with a high number of stars or high star densities so as to
justify the statistical treatment. As the model equations only account for
two-body relaxation, other mechanisms that drive the evolution of a stellar
system can be added as terms in the model equations later on (e.g. unequal
stellar masses and stellar evolution).  In this work we have focused on giving
the first complete analytical derivation of the relevant high-order moment
equations. 

One of our goals is also to improve previous models such as the AGM or the
moment model of \citet{louis1990}. For that, we achieve a more accurate
modeling by including a larger number of moments.  As we explained in section
\ref{sec:distribution_funktion}, increasing the number of moments leads to
both a more complex VDF and an increasing number of differential moment
equations (section \ref{subsec:lhs}). This argument applies to the AGM rather
than Louis' model. 

We therefore can describe the state of the system in terms of its phase space 
distribution function
more accurately. As explained previously, GCs are in dynamic equilibrium but
not in thermodynamic equilibrium. While a system in thermodynamic
equilibrium can be represented by a VDF that is fully defined by its first two
moments, the number of non-vanishing moments increases for a system which is
not in thermodynamic equilibrium. In most cases, it is impossible to exactly
compute the VDF for a system that is not in thermodynamic equilibrium. In a
stellar dynamical system such as GCs or NCs there are numerous mechanisms that
force the system away from thermodynamic equilibrium, raising the issue of
when to truncate the moment series. Mechanisms that affect the high end of the
VDF, such as the evaporation of stars from a stellar system, close three-body
encounters and mass-segregation, suggest that the inclusion of fourth- and fifth- order
moments are important. This is also fortified by observations such as in the
findings of high velocity stars in the core of Milky Way GCs. The AGM, as a 3rd
order model, does not accurately describe these mechanisms.

The correct computation of the collisional terms is still a major difficulty
and the local approximation is applied and an ansatz for the VDF is used to
handle this problem. Even so, there are evident improvements over the previous
models that stem from the use of a larger number of moments (regarding the AGM)
and a self-consistent method for the computation of the collisional terms
\cite[as compared to][]{louis1990}. In contrast to the previous models, the
collision term of a moment equation of order $n$ does not only depend on the
corresponding $n$th order anisotropy parameter but instead exhibits {more}
dependencies on anisotropy of the parameters and moments of almost all orders as
well. This leads to further coupling between the different moments. In a
comparative study between the AGM  and FP and $N$-body models,
\citet{spurzem1995} concluded that in a multi-mass model a significant fraction
of small-angle encounters, which transfer energy from the heavy to the light
stars in the core, cause the light stars to move radially outwards on elongated
orbits. As a result, the energy taken from the heavy particles is quickly
redistributed over a much larger volume than assumed by the local
approximation. Even though the local approximation is still applied in this
model, the energy transfer due to collisions should be improved due to a
stronger coupling of the moments. This will provide a better estimate for the
impact of the local approximation of the evolution of the system.

The choice of the closure relation is very important, in particular at lower
orders. In the AGM the system of equations is closed with the heat flux
equation, which relates the energy flux to the velocity dispersion. It is not
clear how well the heat conduction closure of the AGM works. It obviously
allows the model to handle heat transfer and there are certainly parallels to
gas-dynamics in GCs, but the description of energy transfer via the
gas-dynamical heat conduction equation might nevertheless be a too simple
description of this process. The heat conduction closure and the
third order differential moment equations seems to be two completely different
descriptions of a similar process. Even so, in the comparative study by
\citet{louis1991} between the AGM and the model of \citet{louis1990} reasonable
agreement in pre-core-collapse could be achieved by proper choice of the free
parameters of the AGM. However, these values of the free parameters of the AGM
are not in agreement with the values resulting from the comparative study by
\citet{giersz1994}, where the AGM was fitted and compared to FP and $N$-body
models. This indicates that Louis' model does not agree very well with FP and
$N$-body models. Furthermore, it has to be considered
that the parameter $\lambda$ determining the heat conduction in gaseous models
is just a scaling factor in isotropic gaseous models. In anisotropic gaseous
models $\lambda$ prescribes the relative speed of the two relevant processes -
the decay of anisotropy and the heat flow between warm and cold regions. With
growing $\lambda$ heat flows faster, so there is less time for gravitational
encounters to destroy anisotropy \citep{louis1991}. In our model (as in
the model of \citet{louis1990}) this free parameter is absent.
 
The closure equation in the model of \citet{louis1990} is an algebraic
relation between the flux velocities of even moments $\kappa$, $p$ and
$\rho$. It is based on the assumption that the flux velocities of
moments of order 2k increase with k. 

The closure relations we use are basically a mathematical formulation of the
fact that our model cannot describe an arbitrary degree of anisotropy, since its
description of a stellar system is bounded by the highest moment that it includes.
It also reflects the limits of variability of the VDF and, thus, is a very
natural choice. The only uncertainty of this closure relation is the error due
to the polynomial ansatz for the VDF. The closure limiting the VDF is derived
from the VDF itself. It does not stem from any other constraint that is
independent of the form of the VDF arising from the boundary conditions like
spherical symmetry and the absence of rotation. Hence, this ansatz should not
be seen as an \emph{additional} approximation, but rather as a consistency
relation.

The model equations consist of the set of equations \eqref{moment_equations_a}
and \eqref{moment_equations_b} for \emph{model a} where the right-hand sides
are given by equations \eqref{begin_fourth_order} to \eqref{end_fourth_order}
and the set of equations \eqref{moment_equations_a}, \eqref{moment_equations_b}
and \eqref{moment_equations_c} for \emph{model b} with the right-hand sides
given in equations \eqref{begin_fifth_order} to \eqref{end_fifth_order}.
Furthermore, we need the Poisson equation \eqref{Poisson} and the closure
relations \eqref{GrGrtGt} or \eqref{Hrelations} to complete the model
equations. In order to exclude errors in the computation of the collisional
terms, several measures were taken. The higher order Rosenbluth potentials were
compared to the second-order Rosenbluth potentials of \citet{giersz1994} and
showed exact agreement. Furthermore, the collisional terms for the density,
bulk velocity and energy density vanish as expected according to mass and
energy conservation and the fact that internal collisions do not disturb the
motion of the barycenter. 

Eventually, several arguments have been given that predict improvements of the
fourth and fifth order models developed in this work in comparison with its
predecessors but a final estimate of the gained accuracy can only be obtained
by means of numerical simulations and subsequent comparison with other models.
The next step will be to implement and test the model in a numerical code such
as the anisotropic gaseous
model\footnote{\url{http://www.ari.uni-heidelberg.de/gaseous-model/}}. For
that, the left-hand sides of the differential moment equations 
\eqref{moment_equations_a}, \eqref{moment_equations_b} and 
\eqref{moment_equations_c}
have to be discretized as in the appendix of \citet{pau2004}, and the
collisional terms which form the right-hand sides of the moment equations have
to be reformulated. They should be simplified and reordered to allow an
effective implementation into a numerical code.

\section*{Acknowledgments}

JS visit to the AEI and Munich have been supported by the ARI. He is
thankful to the AEI for covering some of the expenses during his visit.
He is indebted with his collegues at the ARI, in particular with
Jonathan Downing, for discussions. PAS is indebted with Dave J. Vanecek for comments on the
manuscript. This work has been partially supported by the DLR programme ``LISA Germany''.

\appendix

\section{Collisional terms}\label{app:collisional_terms}

\begin{equation}\begin{split}\label{begin_fourth_order}
	&\left(\frac{\delta p_r}{\delta t}\right)_{\text{enc}}=-2\left(\frac{\delta p_{t}}{\delta t}\right)_{\text{enc}}=\frac{1}{t_{\mathrm{rx}}}\bigg(-\frac{177 a_p}{640}\\
	&+\frac{1}{\rho\sigma^2}\bigg(\frac{39 a_p^2}{160}-\frac{27 \rho a_{\text{$\kappa $1}}}{8960}+\frac{1}{\rho\sigma^2}\bigg(\frac{9 F^2}{100}+\frac{27 F a_F}{1400}\\
	&+\frac{a_F^2}{1400}-\frac{33 \kappa  a_p}{640}-\frac{33 a_p a_{\text{$\kappa $1}}}{1568}-\frac{33 a_p a_{\text{$\kappa $2}}}{15680}\\
	&+\frac{1}{\rho\sigma^2}\bigg(\frac{3 a_{\text{$\kappa $1}}^2}{6272}+\frac{3 \kappa  a_{\text{$\kappa $1}}}{1280}+\frac{3 a_{\text{$\kappa $1}} a_{\text{$\kappa $2}}}{31360}+\frac{a_{\text{$\kappa$2}}^2}{413952}\bigg)\bigg)\bigg)\bigg)
\end{split}\end{equation}

\begin{equation}\begin{split}
	&\left(\frac{\delta F_r}{\delta t}\right)_{\text{enc}}=\frac{1}{t_{\mathrm{rx}}}\bigg(-\frac{423 F}{400}-\frac{4833 a_F}{22400}\\
	&+\frac{1}{\rho\sigma^2}\bigg(\frac{9 F a_p}{112}-\frac{9 a_F a_p}{280}+\frac{1}{\rho\sigma^2}\bigg(\frac{9 F \kappa }{400}-\frac{81 \kappa  a_F}{22400}\\
	&+\frac{9 F a_{\text{$\kappa $1}}}{1120}+\frac{9 a_F a_{\text{$\kappa $1}}}{3920}+\frac{9 F a_{\text{$\kappa $2}}}{2800}-\frac{3 a_F a_{\text{$\kappa $2}}}{78400 }\bigg)\bigg)\bigg)
\end{split}\end{equation}

\begin{equation}\begin{split}
	&\left(\frac{\delta F_t}{\delta t}\right)_{\text{enc}}=\frac{1}{t_{\mathrm{rx}}}\bigg(-\frac{141 F}{200}+\frac{4833 a_F}{22400}\\
	&+\frac{1}{\rho\sigma^2}\bigg(\frac{1203 F a_p}{2800}-\frac{81 a_F a_p}{1400}+\frac{1}{\rho\sigma^2}\bigg(\frac{3 F \kappa }{200}\\
	&+\frac{81 \kappa  a_F}{22400}-\frac{3 F a_{\text{$\kappa $1}}}{160}+\frac{9 a_F a_{\text{$\kappa $1}}}{3920}-\frac{9 F a_{\text{$\kappa $2}}}{2800}\\
	&-\frac{17 a_F a_{\text{$\kappa $2}}}{78400}\bigg)\bigg)\bigg)
\end{split}\end{equation}

\begin{equation}\begin{split}
	&\left(\frac{\delta \kappa_r}{\delta t}\right)_{\text{enc}}=\frac{1}{t_{\mathrm{rx}}}\bigg(-\frac{93 \kappa }{400}-\frac{18069 a_{\text{$\kappa $1}}}{62720}\\
	&-\frac{6773 a_{\text{$\kappa $2}}}{156800}+\frac{31 a_p^2}{32 \rho }+\sigma^2\bigg(\frac{11481 a_p}{4480}+\frac{423 \rho  \sigma^2}{160}\bigg)\\
	&+\frac{1}{\rho\sigma^2}\bigg(-\frac{2697a_p a_{\text{$\kappa $2}}}{123200}+\frac{1593 F^2}{3500}+\frac{1257 F a_F}{7000}\\
	&+\frac{81 a_F^2}{77000}-\frac{111 a_p a_{\text{$\kappa $1}}}{1120}-\frac{789 \kappa  a_p}{3200}+\frac{1}{\rho\sigma^2}\bigg(\frac{129 a_{\text{$\kappa $1}}^2}{43904}\\
	&+\frac{699 \kappa  a_{\text{$\kappa $1}}}{62720}+\frac{3 \kappa ^2}{800}-\frac{37 \kappa  a_{\text{$\kappa $2}}}{156800}+\frac{2567 a_{\text{$\kappa $1}} a_{\text{$\kappa $2}}}{2414720}\\
	&-\frac{181 a_{\text{$\kappa$2}}^2}{313913600}\bigg)\bigg)\bigg)
\end{split}\end{equation}

\begin{equation}\begin{split}
	&\left(\frac{\delta \kappa_{rt}}{\delta t}\right)_{\text{enc}}=\frac{1}{t_{\mathrm{rx}}}\bigg(-\frac{31 \kappa }{200}-\frac{6023 a_{\text{$\kappa $1}}}{125440}\\
	&+\frac{6773 a_{\text{$\kappa $2}}}{156800}-\frac{647 a_p^2}{1344 \rho}+\sigma^2\bigg(\frac{3827 a_p}{8960}+\frac{141 \rho  \sigma^2}{80}\bigg)\\
	&+\frac{1}{\rho\sigma^2}\bigg(-\frac{263 \kappa  a_p}{6400}+\frac{509 a_p a_{\text{$\kappa $1}}}{15680}-\frac{6693 a_p a_{\text{$\kappa $2}}}{1724800}\\
	&+\frac{39 F a_F}{2800}-\frac{3 a_F^2}{880}+\frac{207 F^2}{1400}+\frac{1}{\rho\sigma^2}\bigg(\frac{233 \kappa  a_{\text{$\kappa $1}}}{125440}\\
	&+\frac{\kappa ^2}{400}+\frac{37 \kappa  a_{\text{$\kappa $2}}}{156800}-\frac{53 a_{\text{$\kappa$1}}^2}{87808}+\frac{509 a_{\text{$\kappa $1}} a_{\text{$\kappa $2}}}{4829440}\\
	&-\frac{28879 a_{\text{$\kappa$2}}^2}{1883481600}\bigg)\bigg)\bigg)
\end{split}\end{equation}

\begin{equation}\begin{split}\label{end_fourth_order}
	&\left(\frac{\delta \kappa_{t}}{\delta t}\right)_{\text{enc}}=\frac{1}{t_{\mathrm{rx}}}\bigg(-\frac{31 \kappa }{50}+\frac{6023 a_{\text{$\kappa $1}}}{15680}\\
	&-\frac{6773 a_{\text{$\kappa $2}}}{156800}-\frac{53 a_p^2}{84 \rho}+\sigma^2\bigg(-\frac{3827a_p}{1120}+\frac{141 \rho  \sigma^2}{20}\bigg)\\
	&+\frac{1}{\rho\sigma^2}\bigg(\frac{263 \kappa  a_p}{800}+\frac{6393 a_p a_{\text{$\kappa $2}}}{215600}+\frac{23 a_p a_{\text{$\kappa $1}}}{980}-\frac{39 a_F^2}{5500}\\
	&-\frac{27 F^2}{875}-\frac{363 F a_F}{1750}+\frac{1}{\rho\sigma^2}\bigg(\frac{\kappa ^2}{100}+\frac{a_{\text{$\kappa $1}}^2}{5488}-\frac{233 \kappa a_{\text{$\kappa $1}}}{15680}\\
	&-\frac{37 \kappa  a_{\text{$\kappa $2}}}{156800}-\frac{769 a_{\text{$\kappa $1}} a_{\text{$\kappa $2}}}{603680}-\frac{15319 a_{\text{$\kappa $2}}^2}{470870400}\bigg)\bigg)\bigg)
\end{split}\end{equation}

And the collisional terms for the model closing at sixth order are:

\begin{equation}\begin{split}\label{begin_fifth_order}
&\left(\frac{\delta p_r}{\delta t}\right)_{\text{enc}}=-2\left(\frac{\delta p_{t}}{\delta t}\right)_{\text{enc}}=\frac{1}{t_{\text{rx}}}\bigg(-\frac{177 a_p}{640}\\
&+\frac{1}{\rho  \sigma ^2}\bigg(\frac{39 a_p^2}{160 }-\frac{27\rho  a_{\text{$\kappa $1}}}{8960 }+\frac{1}{\rho  \sigma ^2}\bigg(\frac{243 F^2}{320 }\\
&+\frac{729 F a_F}{4480}+\frac{27 a_F^2}{4480 }-\frac{33 a_p a_{\text{$\kappa $2}}}{15680
}-\frac{33 \kappa  a_p}{640 }-\frac{33 a_p a_{\text{$\kappa $1}}}{1568}\\
&+\frac{1}{\rho  \sigma ^2}\bigg(\frac{3 \kappa  a_{\text{$\kappa $1}}}{1280 }-\frac{39 F a_{\text{G1}}}{3200 }-\frac{13 a_F a_{\text{G1}}}{14400}-\frac{13 a_F a_{\text{G2}}}{177408 }\\
&+\frac{3 a_{\text{$\kappa $1}}^2}{6272 }-\frac{117 F G}{3200 }-\frac{351 G a_F}{89600 }+\frac{3 a_{\text{$\kappa $1}}
a_{\text{$\kappa $2}}}{31360}+\frac{a_{\text{$\kappa $2}}^2}{413952 }\\
&+\frac{1}{\rho  \sigma ^2}\bigg(\frac{a_{\text{G1}} a_{\text{G2}}}{177408 }+\frac{a_{\text{G2}}^2}{9225216 }+\frac{a_{\text{G1}}^2}{28800
}+\frac{27 G a_{\text{G1}}}{89600 }\\
&+\frac{81 G^2}{179200}\bigg)\bigg)\bigg)\bigg)\bigg)
\end{split}\end{equation}

\begin{equation}\begin{split}
&\left(\frac{\delta F_r}{\delta t}\right)_{\text{enc}}=\frac{1}{t_{\text{rx}}}\bigg(-\frac{171 F}{80}-\frac{5697 a_F}{17920}\\
&+\frac{1}{\rho  \sigma ^2}\bigg(\frac{1017\rho  a_{\text{G1}}}{89600}+\frac{27\rho  G}{700}+\frac{351 F a_p}{320}\\
&+\frac{1}{\rho  \sigma ^2}\bigg(\frac{9 F \kappa }{400}-\frac{1269 \kappa  a_F}{89600 }-\frac{2277 G a_p}{62720 }\\
&-\frac{a_{\text{G1}} a_p}{280}-\frac{107 a_{\text{G2}} a_p}{68992 }-\frac{99 F a_{\text{$\kappa $1}}}{4480 }+\frac{9 a_F a_{\text{$\kappa $1}}}{3920 }\\
&+\frac{81 F a_{\text{$\kappa$2}}}{11200 }-\frac{1357 a_F a_{\text{$\kappa $2}}}{3449600 }+\frac{1}{\rho  \sigma ^2}\bigg(\frac{3 \kappa  a_{\text{G1}}}{2560 }\\
&+\frac{27 G a_{\text{$\kappa $1}}}{25088 }+\frac{9 a_{\text{G2}}a_{\text{$\kappa $1}}}{137984 }-\frac{9 G a_{\text{$\kappa $2}}}{62720 }+\frac{a_{\text{G1}} a_{\text{$\kappa $2}}}{25344 }\\
&-\frac{a_{\text{G2}} a_{\text{$\kappa$2}}}{1241856}\bigg)\bigg)\bigg)\bigg)
\end{split}\end{equation}

\begin{equation}\begin{split}
&\left(\frac{\delta F_t}{\delta t}\right)_{\text{enc}}=\frac{1}{t_{\text{rx}}}\bigg(-\frac{57 F}{40}+\frac{5697 a_F}{17920}\\
&+\frac{1}{\rho  \sigma ^2}\bigg(\frac{9 \text{$\rho $G}}{350 }-\frac{1017 \text{$\rho $a}_{\text{G1}}}{89600}+\frac{1521 F a_p}{1600}-\frac{117 a_F a_p}{800 }\\
&+\frac{1}{\rho  \sigma ^2}\bigg(\frac{3 F \kappa }{200 }+\frac{1269 \kappa  a_F}{89600 }-\frac{1167 G a_p}{62720 }+\frac{11 a_{\text{G1}} a_p}{1120}\\
&+\frac{107a_{\text{G2}} a_p}{68992 }-\frac{201 F a_{\text{$\kappa $1}}}{4480 }+\frac{99 a_F a_{\text{$\kappa $1}}}{15680 }-\frac{81 F a_{\text{$\kappa $2}}}{11200}\\
&-\frac{6269 a_F a_{\text{$\kappa $2}}}{10348800 }+\frac{1}{\rho  \sigma ^2}\bigg(\frac{117 G a_{\text{$\kappa $1}}}{125440 }-\frac{3 \kappa  a_{\text{G1}}}{2560 }-\frac{a_{\text{G1}}a_{\text{$\kappa $1}}}{2240 }\\
&-\frac{9 a_{\text{G2}} a_{\text{$\kappa $1}}}{137984 }+\frac{9 G a_{\text{$\kappa $2}}}{62720 }+\frac{23 a_{\text{G1}}a_{\text{$\kappa $2}}}{532224 }-\frac{a_{\text{G2}} a_{\text{$\kappa $2}}}{338688 }\bigg)\bigg)\bigg)\bigg)
\end{split}\end{equation}

\begin{equation}\begin{split}
&\left(\frac{\delta \kappa_r}{\delta t}\right)_{\text{enc}}=\frac{1}{t_{\text{rx}}}\bigg(-\frac{93 \kappa }{400}-\frac{18069 a_{\text{$\kappa $1}}}{62720}-\frac{6773 a_{\text{$\kappa $2}}}{156800}\\
&+\frac{31a_p^2}{32 \rho }+\sigma ^2\bigg(\frac{423 \rho  \sigma ^2}{160}+\frac{11481\text{  }a_p}{4480}\bigg)\\
&+\frac{1}{\rho  \sigma ^2}\bigg(\frac{27189 F^2}{8000 }-\frac{111 a_p a_{\text{$\kappa $1}}}{1120 }-\frac{2697 a_p a_{\text{$\kappa $2}}}{123200}\\
&+\frac{18861 F a_F}{16000}+\frac{2713 a_F^2}{176000}-\frac{789 \kappa  a_p}{3200 }+\frac{1}{\rho  \sigma ^2}\bigg(\frac{3 \kappa ^2}{800}\\
&-\frac{5319 F G}{31360}-\frac{537 G a_F}{17920 }-\frac{1061 F a_{\text{G1}}}{13440 }-\frac{643
a_F a_{\text{G1}}}{221760 }\\
&+\frac{1499 F a_{\text{G2}}}{517440 }-\frac{139 a_F a_{\text{G2}}}{149760}+\frac{699 \kappa  a_{\text{$\kappa $1}}}{62720}+\frac{129 a_{\text{$\kappa $1}}^2}{43904}\\
&-\frac{37 \kappa  a_{\text{$\kappa $2}}}{156800}+\frac{2567 a_{\text{$\kappa $1}} a_{\text{$\kappa$2}}}{2414720}-\frac{181a_{\text{$\kappa $2}}^2}{313913600}\\
&+\frac{1}{\rho  \sigma ^2}\bigg(\frac{2889 G^2}{1254400}+\frac{587 G a_{\text{G1}}}{268800}+\frac{413 a_{\text{G1}}^2}{2851200}\\
&-\frac{97G a_{\text{G2}}}{2069760 }+\frac{7643 a_{\text{G1}} a_{\text{G2}}}{103783680}-\frac{53 a_{\text{G2}}^2}{264176640}\bigg)\bigg)\bigg)\bigg)
\end{split}\end{equation}

\begin{equation}\begin{split}
&\left(\frac{\delta \kappa_{rt}}{\delta t}\right)_{\text{enc}}=\frac{1}{t_{\text{rx}}}\bigg(-\frac{31 \kappa }{200}-\frac{6023 a_{\text{$\kappa $1}}}{125440}\\
&+\frac{6773 a_{\text{$\kappa $2}}}{156800}-\frac{647a_p^2}{1344 \rho }+\sigma ^2\bigg(\frac{141 \rho  \sigma ^2}{80}+\frac{3827a_p}{8960}\bigg)\\
&+\frac{1}{\rho  \sigma ^2}\bigg(\frac{3627 F^2}{3200 }+\frac{291 F a_F}{6400}-\frac{3799 a_F^2}{211200}-\frac{263 \kappa  a_p}{6400}\\
&+\frac{509 a_p a_{\text{$\kappa $1}}}{15680}-\frac{6693 a_p a_{\text{$\kappa $2}}}{1724800}+\frac{1}{\rho  \sigma ^2}\bigg(-\frac{17937 F G}{313600}\\
&+\frac{\kappa ^2}{400}+\frac{807 G a_F}{1254400}-\frac{691 F a_{\text{G1}}}{134400}+\frac{2179 a_F a_{\text{G1}}}{950400}\\
&-\frac{1499 F a_{\text{G2}}}{517440}-\frac{67289 a_F a_{\text{G2}}}{484323840}+\frac{233 \kappa  a_{\text{$\kappa $1}}}{125440}-\frac{53a_{\text{$\kappa $1}}^2}{87808}\\
&+\frac{37 \kappa  a_{\text{$\kappa $2}}}{156800}+\frac{509 a_{\text{$\kappa $1}} a_{\text{$\kappa $2}}}{4829440}-\frac{28879 a_{\text{$\kappa $2}}^2}{1883481600}\\
&+\frac{1}{\rho  \sigma ^2}\bigg(\frac{1971 G^2}{2508800}-\frac{14947 a_{\text{G2}}^2}{17435658240}+\frac{701 a_{\text{G1}} a_{\text{G2}}}{88957440}\\
&-\frac{G a_{\text{G1}}}{76800}+\frac{97G a_{\text{G2}}}{2069760}-\frac{1279 a_{\text{G1}}^2}{17107200}\bigg)\bigg)\bigg)\bigg)
\end{split}\end{equation}

\begin{equation}\begin{split}
&\left(\frac{\delta \kappa_t}{\delta t}\right)_{\text{enc}}=\frac{1}{t_{\text{rx}}}\bigg(-\frac{31 \kappa }{50}+\frac{6023 a_{\text{$\kappa $1}}}{15680}-\frac{6773 a_{\text{$\kappa $2}}}{156800}\\
&-\frac{53a_p^2}{84 \rho }+ \sigma ^2\bigg(\frac{141 \rho  \sigma ^2}{20}-\frac{3827 a_p}{1120}\bigg)+
\frac{1}{\rho  \sigma ^2}\bigg(\frac{9 F^2}{2000 }\\
&-\frac{5079 F a_F}{4000 }-\frac{8927 a_F^2}{264000 }+\frac{263 \kappa  a_p}{800}+\frac{23
a_p a_{\text{$\kappa $1}}}{980}\\
&+\frac{6393 a_p a_{\text{$\kappa $2}}}{215600}+\frac{1}{\rho  \sigma ^2}\bigg(\frac{\kappa ^2}{100}-\frac{207 F G}{39200}+\frac{4497 G a_F}{156800}\\
&+\frac{1499 F a_{\text{G1}}}{16800}+\frac{943 a_F a_{\text{G1}}}{237600}+\frac{1499 F a_{\text{G2}}}{517440}\\
&-\frac{233 \kappa  a_{\text{$\kappa $1}}}{15680}+\frac{73013 a_F a_{\text{G2}}}{60540480}+\frac{a_{\text{$\kappa$1}}^2}{5488}-\frac{37 \kappa  a_{\text{$\kappa $2}}}{156800}\\
&-\frac{769 a_{\text{$\kappa $1}} a_{\text{$\kappa $2}}}{603680}-\frac{15319a_{\text{$\kappa $2}}^2}{470870400}+\frac{1}{\rho  \sigma ^2}\bigg(\frac{9 G^2}{62720}\\
&-\frac{29 G a_{\text{G1}}}{13440}-\frac{19 a_{\text{G1}}^2}{171072}-\frac{97G a_{\text{G2}}}{2069760}-\frac{6959 a_{\text{G1}} a_{\text{G2}}}{77837760}\\
&-\frac{251 a_{\text{G2}}^2}{136216080}\bigg)\bigg)\bigg)\bigg)
\end{split}\end{equation}

\begin{equation}\begin{split}
&\left(\frac{\delta G_r}{\delta t}\right)_{\text{enc}}=\frac{1}{t_{\text{rx}}}\bigg(-\frac{25009 a_{\text{G1}}}{161280}-\frac{35521 a_{\text{G2}}}{1241856}\\
&+\frac{a_p}{\rho }\bigg(\frac{3231F }{448}+\frac{a_F}{168}\bigg)-\sigma ^2\bigg(\frac{1683 F }{560}+\frac{1013 a_F}{2560}\bigg)\\
&-\frac{1161 G}{7840}+\frac{1}{\rho  \sigma ^2}\bigg(-\frac{351 F \kappa }{2800}-\frac{8207 \kappa  a_F}{89600}\\
&-\frac{16119 G a_p}{62720}-\frac{181 a_{\text{G1}}a_p}{3465}-\frac{45 F a_{\text{$\kappa $1}}}{896}+\frac{59 a_F a_{\text{$\kappa $1}}}{4312}\\
&-\frac{164105 a_{\text{G2}} a_p}{8072064}+\frac{4689 Fa_{\text{$\kappa $2}}}{61600}-\frac{326083 a_F a_{\text{$\kappa $2}}}{44844800}\\
&+\frac{1}{\rho  \sigma ^2}\bigg(\frac{9 G \kappa }{1120}+\frac{1247 \kappa  a_{\text{G1}}}{161280}-\frac{\kappa  a_{\text{G2}}}{25344}+\frac{1143G a_{\text{$\kappa $1}}}{175616 }\\
&+\frac{11 a_{\text{G1}} a_{\text{$\kappa $1}}}{7056}+\frac{97805 a_{\text{G2}} a_{\text{$\kappa $1}}}{113008896}-\frac{3303G a_{\text{$\kappa $2}}}{2414720}\\
&+\frac{49523 a_{\text{G1}} a_{\text{$\kappa $2}}}{80720640}-\frac{4027 a_{\text{G2}} a_{\text{$\kappa $2}}}{113008896}\bigg)\bigg)\bigg)
\end{split}\end{equation}

\begin{equation}\begin{split}
&\left(\frac{\delta G_{rt}}{\delta t}\right)_{\text{enc}}=\frac{1}{t_{\text{rx}}}\bigg(\frac{25009 a_{\text{G1}}}{1612800}+\frac{35521 a_{\text{G2}}}{1241856}\\
&+\frac{a_p}{\rho}\bigg(\frac{9669 F}{4480}-\frac{3229 a_F}{6720}\bigg)-\sigma ^2\bigg(\frac{1683 F }{1400}-\frac{1013 a_F}{25600}\bigg)\\
&-\frac{1161 G}{19600}+\frac{1}{\rho  \sigma ^2}\bigg(-\frac{351 F \kappa }{7000}+\frac{8207 \kappa  a_F}{896000}-\frac{1557 G a_p}{25088}\\
&+\frac{5017 a_{\text{G1}}a_p}{221760}+\frac{210101 a_{\text{G2}} a_p}{80720640}-\frac{12573 F a_{\text{$\kappa $1}}}{313600}\\
&+\frac{34381 a_F a_{\text{$\kappa $1}}}{1724800}-\frac{130341 F a_{\text{$\kappa $2}}}{8624000}-\frac{1572817 a_F a_{\text{$\kappa $2}}}{448448000}\\
&+\frac{1}{\rho  \sigma ^2}\bigg(\frac{9 G \kappa }{2800}-\frac{1247 \kappa  a_{\text{G1}}}{1612800}+\frac{\kappa  a_{\text{G2}}}{25344}+\frac{189857 a_{\text{G1}} a_{\text{$\kappa $2}}}{807206400}\\
&-\frac{311 a_{\text{G1}} a_{\text{$\kappa $1}}}{282240}-\frac{4483 a_{\text{G2}} a_{\text{$\kappa $1}}}{32288256}+\frac{309 G a_{\text{$\kappa $2}}}{985600}+\frac{459 G a_{\text{$\kappa $1}}}{250880}\\
&-\frac{731 a_{\text{G2}} a_{\text{$\kappa $2}}}{32288256}\bigg)\bigg)\bigg)
\end{split}\end{equation}

\begin{equation}\begin{split}\label{end_fifth_order}
&\left(\frac{\delta G_t}{\delta t}\right)_{\text{enc}}=\frac{1}{t_{\text{rx}}}\bigg(-\frac{387 G}{4900}+\frac{25009 a_{\text{G1}}}{201600}-\frac{35521 a_{\text{G2}}}{1241856}\\
&+\frac{a_p}{\rho }\bigg(\frac{1069 F}{560}-\frac{533 a_F}{1680}\bigg)-\sigma ^2\bigg(\frac{561 F }{350}-\frac{1013 a_F}{3200}\bigg)\\
&+\frac{1}{\rho  \sigma ^2}\bigg(-\frac{117 F \kappa }{1750}+\frac{8207 \kappa  a_F}{112000}-\frac{2229 G a_p}{78400}+\frac{2581 a_{\text{G1}}a_p}{277200}\\
&+\frac{76303 a_{\text{G2}} a_p}{5045040}-\frac{3141 F a_{\text{$\kappa $1}}}{39200}+\frac{5669 a_F a_{\text{$\kappa $1}}}{431200}-\frac{197889 F a_{\text{$\kappa $2}}}{4312000}\\
&-\frac{505481 a_F a_{\text{$\kappa $2}}}{168168000}+\frac{1}{\rho  \sigma ^2}\bigg(\frac{3 G \kappa }{700}-\frac{1247 \kappa  a_{\text{G1}}}{201600}-\frac{\kappa  a_{\text{G2}}}{25344}\\
&+\frac{309 G a_{\text{$\kappa $1}}}{219520}-\frac{31 a_{\text{G1}} a_{\text{$\kappa $1}}}{70560}-\frac{8303 a_{\text{G2}} a_{\text{$\kappa $1}}}{14126112}+\frac{17889G a_{\text{$\kappa $2}}}{24147200}\\
&+\frac{63001 a_{\text{G1}} a_{\text{$\kappa $2}}}{302702400}-\frac{3119 a_{\text{G2}} a_{\text{$\kappa $2}}}{169513344}\bigg)\bigg)\bigg)
\end{split}\end{equation}

\bibliographystyle{mn}

\label{lastpage}
\end{document}